\documentclass[article]{aa}
\usepackage{natbib}
\usepackage{graphicx}
\usepackage{color}

\definecolor{remgray}{rgb}{0.5,0.5,0.5}

\begin{document}{}
 
  \title{Internetwork magnetic field as revealed by 2D inversions}
   \titlerunning{2D Inversions of Quiet Sun}
   \author{S.~Danilovic \inst{1} \and M.~van~Noort \inst{1} \and M.~Rempel\inst{2}}

   \institute{Max-Planck-Institut f\"ur Sonnensystemforschung, Justus-von-Liebig-Weg 3
37077 G\"ottingen, Germany  \and 
 High Altitude Observatory, NCAR, P.O. Box 3000, Boulder, Colorado 80307, USA}

   \date{\today}

\abstract
 {Properties of magnetic field in the internetwork regions are still fairly unknown due to rather weak spectropolarimetric signals.}
  {We address the matter by using the 2D inversion code that is able to retrieve the information on smallest spatial scales, up to the diffraction limit, while being less susceptible to noise than most of the previous methods used.}
  {Performance of the code and the impact of the various effects on the retrieved field distribution is tested first on the realistic MHD simulations. The best inversion scenario is then applied to the real Hinode/SP data.}
  {Tests on simulations show: (1) the best choice of node position ensures a decent retrieval of all parameters, (2) code performs well for different configurations of magnetic field, (3) slightly different noise level or slightly different defocus included in the spatial PSF produces no significant effect on the results and (4) temporal integration shifts the field distribution to the stronger, more horizontally inclined field.}
  {Although the contribution of the weak field is slightly overestimated due to noise, the 2D inversions are able to recover well the overall distribution of the magnetic field strength. Application of the 2D inversion code on the Hinode/SP internetwork observations reveals a monotonic field strength distribution. The mean field strength at optical depth unity is $\sim 130$~G. At the higher layers, field strength drops as the field becomes more horizontal. Regarding the distribution of the field inclination, tests show that we cannot directly retrieve it with the observations/tools at hand, however the obtained distributions are consistent with those expected from simulations with a quasi-isotropic field inclination after accounting for observational effects.}
\keywords{Sun: photospheric magnetic field --- techniques: polarimetric --- techniques: spectroscopic}

\maketitle

\section{Introduction}

Determining the magnetic properties of the internetwork has always been an important task \citep{Almeida:Marian:2011}, because it carries a substantial fraction of the solar magnetic
flux and has a large impact on the energy budget in the solar atmosphere. Since decades, it has been clear that the internetwork is filled with magnetic field elements of opposite polarities organized on small-scales \citep{Livingston:Harvey:1975,Livi:etal:1985,Martin:1988}. Yet, whether their field strength is predominantly hG or kG, i.e. how much flux they carry, has been the subject of debate \citep{Keller:etal:1994,Lin:1995,Almeida:Lites:2000,Socas:Lites:2004,Cerdena:2006,Marian:2006}. 

As the polarimetric sensitivity and spatial resolution of the observations increased, it became possible to characterize the full magnetic field vector. The debate then further expanded on the inclination of the magnetic field. Currently, three hypothesis remain: that internetwork magnetic field is predominantly horizontal \citep{david07a,david07b,lites08}, predominantly vertical \citep{reza2009,stenflo10,ryuko11} or quasi-isotropic \citep{asensio2009,asensio2014}.

The basic problem lies in retrieving information on magnetic field based on the rather weak internetwork spectropolarimetric signals. The influence of noise on the result is overwhelming and leads to a systematic overestimation of the inclination of the magnetic field vector \citep{Borrero:Kobel:2011}. One way of getting round that problem is to limit the analysis only to pixels where the signal to noise ratio is high enough. However, these selection criteria tend to exclude a significant portion of the internetwork surface and to bias the retrieved distributions of magnetic field strength and inclination in different ways \citep{Borrero:Kobel:2012}.

In this paper, we address the issue by using the 2D inversion technique \citep{Michiel2012} that accounts for spatial coupling between the neighbouring pixels and simultaneously and self-consistently fits the observed spectra, which makes it less susceptible to noise than most of the previous methods used. The code is first tested on 'synthesized observations' produced from realistic  magnetohydrodynamic simulations and then the best inversion strategy is applied to real observations. We limit our study to the general properties of the distributions of the magnetic field strength and inclination.

\section{2D inversions}

To invert a Hinode SP map of the quiet Sun, we use the SPINOR inversion code \citep{frutigerthesis, frutiger00}. This inversion code fits a strongly simplified atmosphere using one or more spectral lines, with the assumption of Local Thermodynamic Equilibrium (LTE). The atmosphere is described by the values of a selectable number of atmospheric quantities, such as temperature, line of sight velocity, magnetic field strength and direction, and microturbulent broadening, specified at a selectable number of optical heights, the so called ``node'' positions.  The code is used in its spatially coupled mode \citep{Michiel2012}, in which the accurately known spatial PSF of the Hinode SOT is taken into account when the observed data are fitted.

To be able to reach the diffraction limit of Hinode/SOT, we oversampled all Stokes maps by a factor two, to $0.08\arcsec$, as was done in \cite{Michiel2013}. Taken that the current version of the code can only use shared memory, the size of the input map is limited to $200\times 200$ pixels. The final maps are then collected as mosaics of individually inverted smaller maps.
  
Since the model atmosphere used in previous studies \citep{Michiel2013,Tino2013,Tiwari2013,Andreas2014,David2015} proved to provide a good fit even to very complex profiles, it is also used here. The model describes a height dependent atmosphere at three nodes in optical depth, with the following free parameters: temperature, magnetic field strength, magnetic field inclination with respect to the (LOS), azimuth of the magnetic field vector, LOS velocity, and a microturbulent velocity. 
The inversion strategy is also the same as in the previous studies: the code was run for 10 iteration steps, followed by a spatial smoothing of the fitted atmosphere, the result of which is then used as the input for the next 10 iteration steps. This cycle is repeated 8 times before the result is considered fully converged. To make sure that the result is really converged, the cycle was repeated up to 20 times for certain maps, but these additional cycles brought no significant improvements to the mean $\chi^{2}$ value. 
\begin{figure*}
  \centering
  \includegraphics[angle=90,width=0.3\linewidth,trim= 3.2cm 8cm 1.5cm 0cm,clip=true]{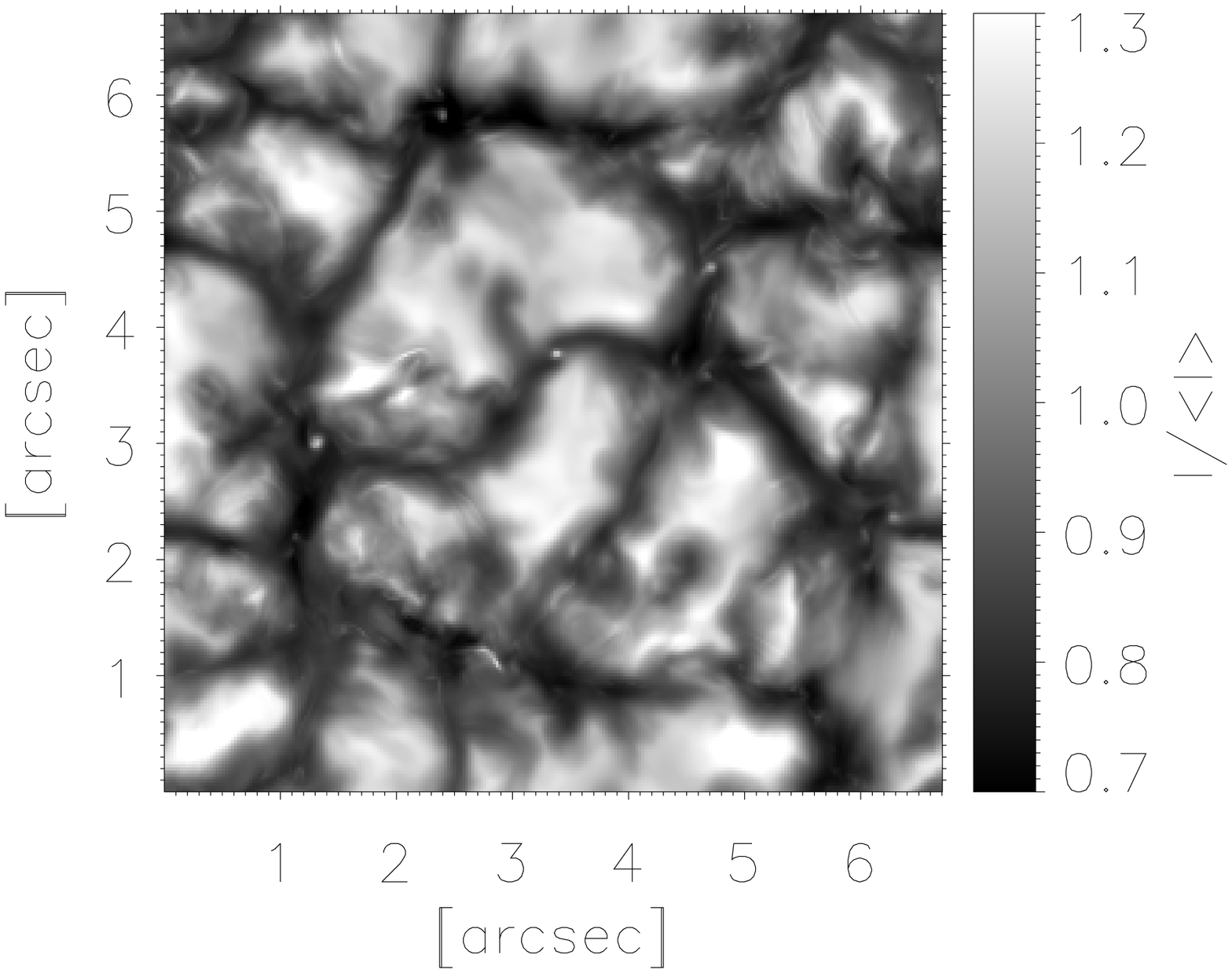}
  \includegraphics[angle=90,width=0.3\linewidth,trim= 3.2cm 8cm 1.5cm 0cm,clip=true]{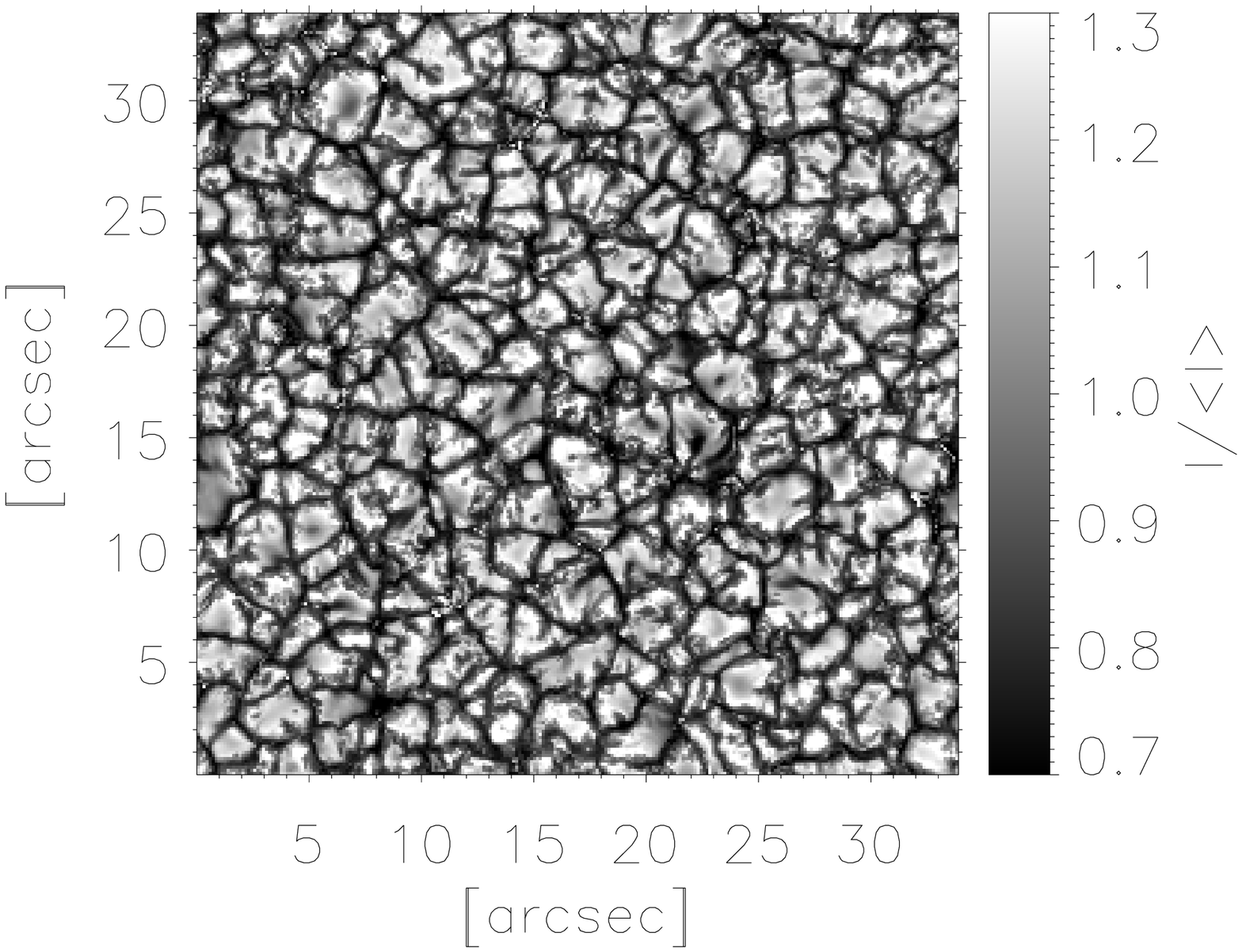}
  \includegraphics[angle=90,width=0.38\linewidth,trim= 3.2cm 3cm 1.5cm 0cm,clip=true]{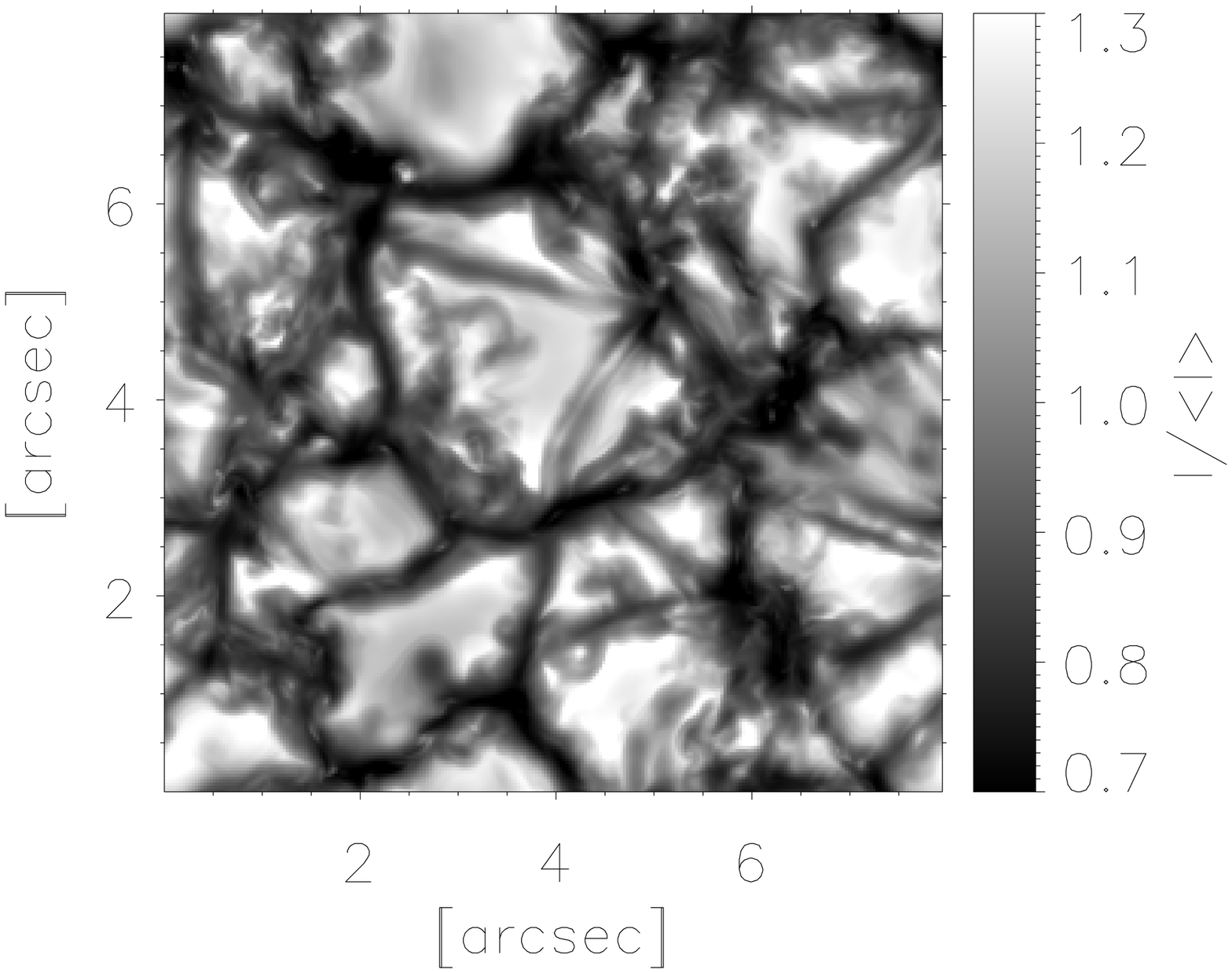}
    \includegraphics[angle=90,width=0.3\linewidth,trim= 3.2cm 8cm 1.5cm 0cm,clip=true]{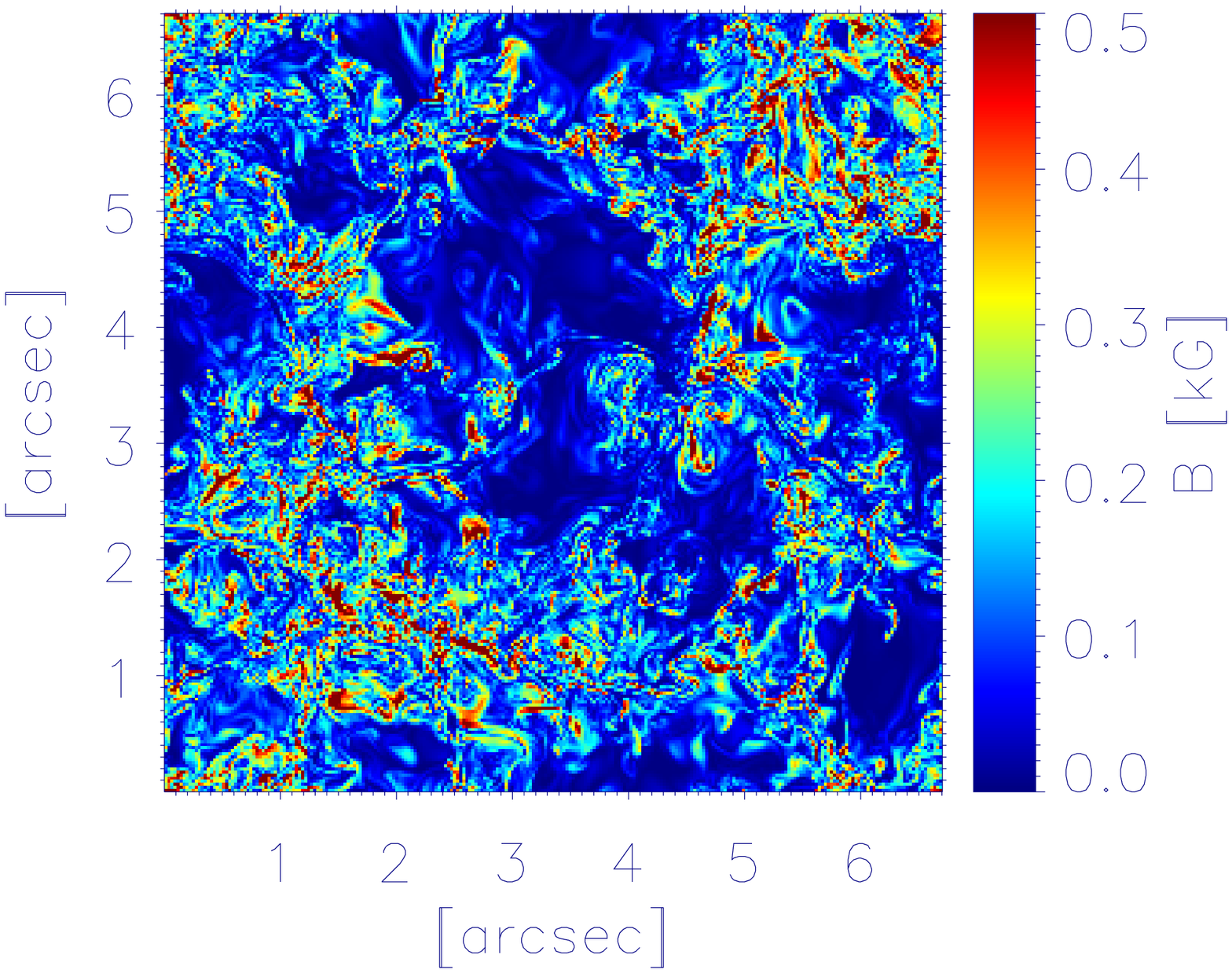}
  \includegraphics[angle=90,width=0.3\linewidth,trim= 3.2cm 8cm 1.5cm 0cm,clip=true]{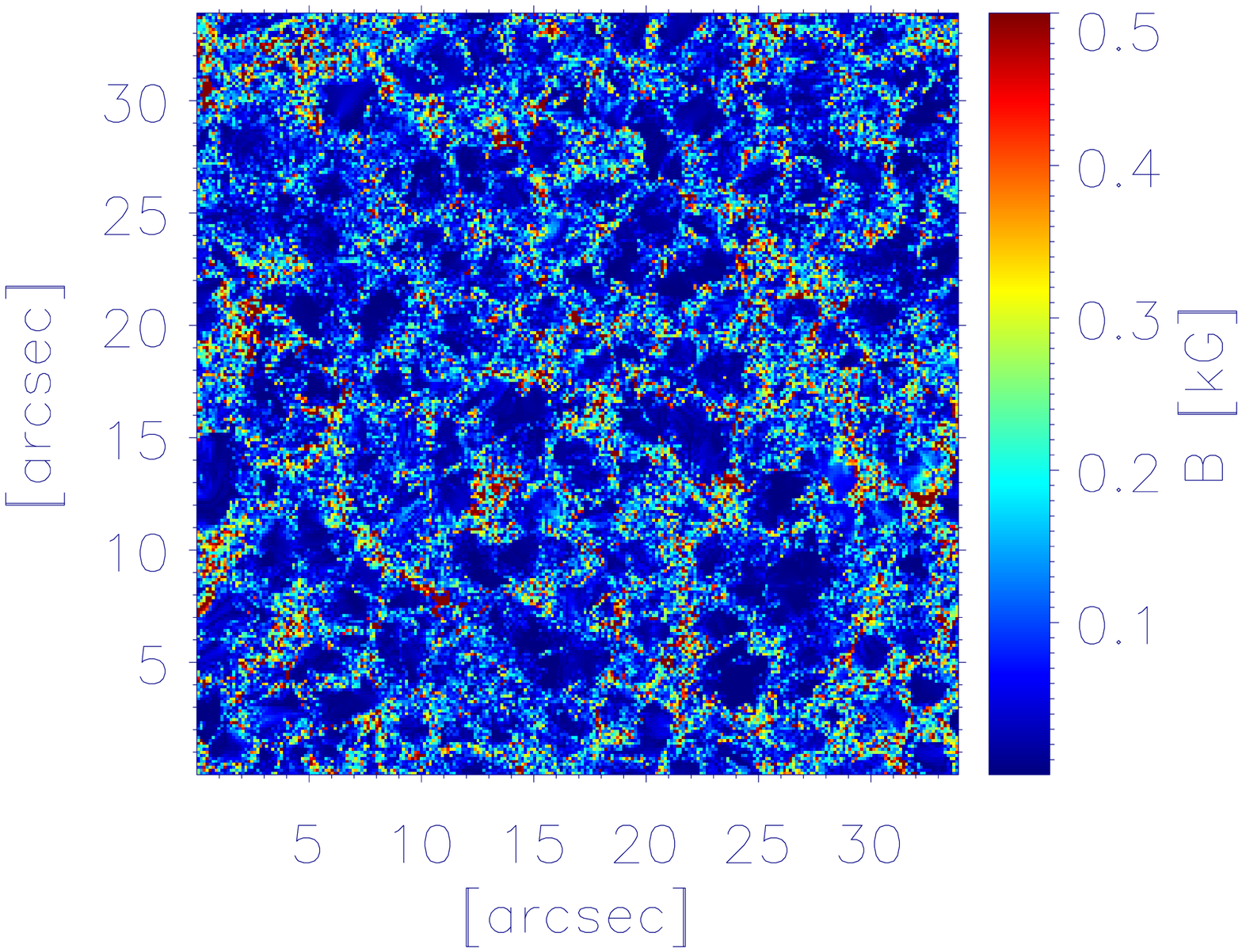}
  \includegraphics[angle=90,width=0.38\linewidth,trim= 3.2cm 3cm 1.5cm 0cm,clip=true]{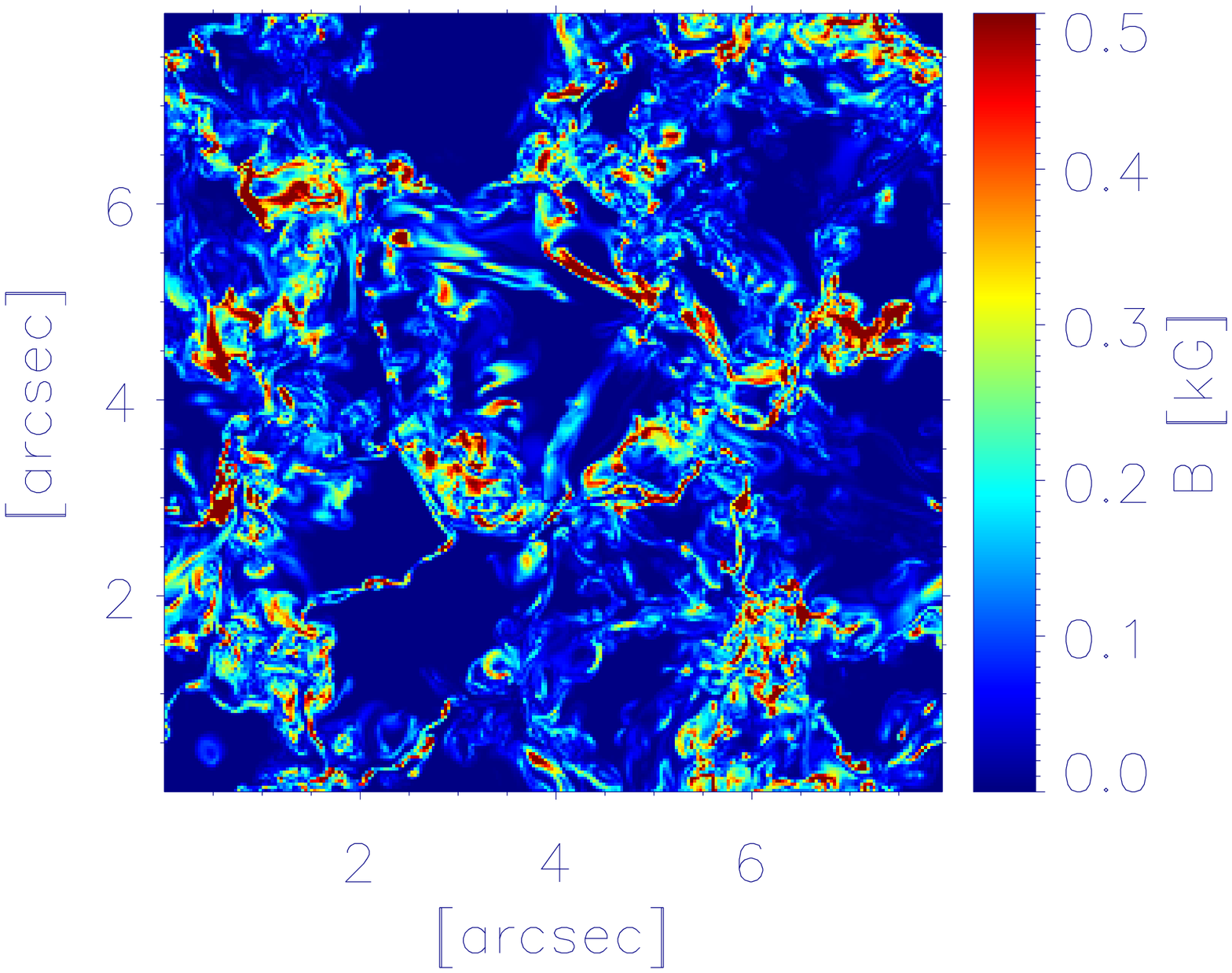}
      \includegraphics[angle=90,width=0.3\linewidth,trim= 0.5cm 8cm 1.5cm 0cm,clip=true]{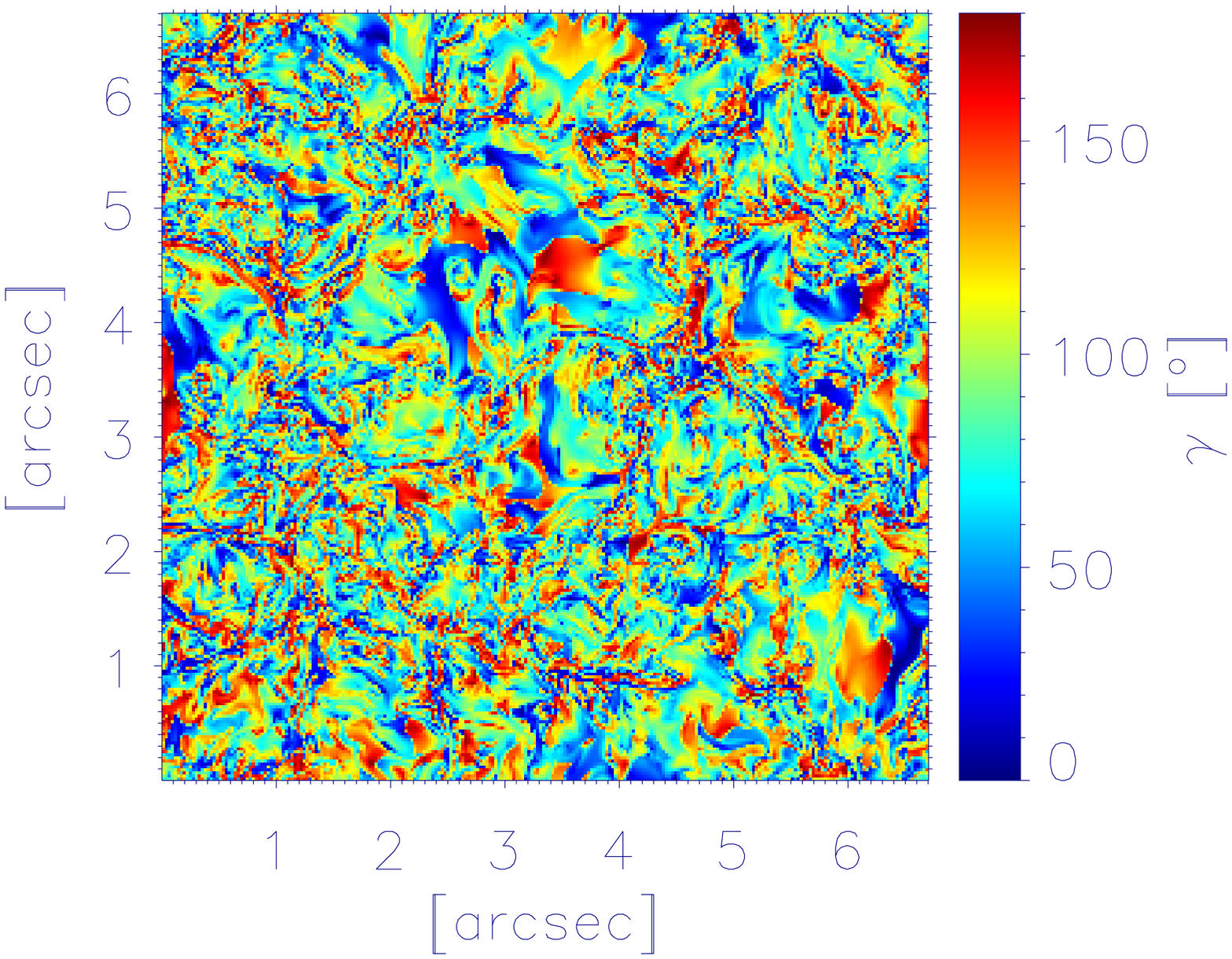}
  \includegraphics[angle=90,width=0.3\linewidth,trim= 0.5cm 8cm 1.5cm 0cm,clip=true]{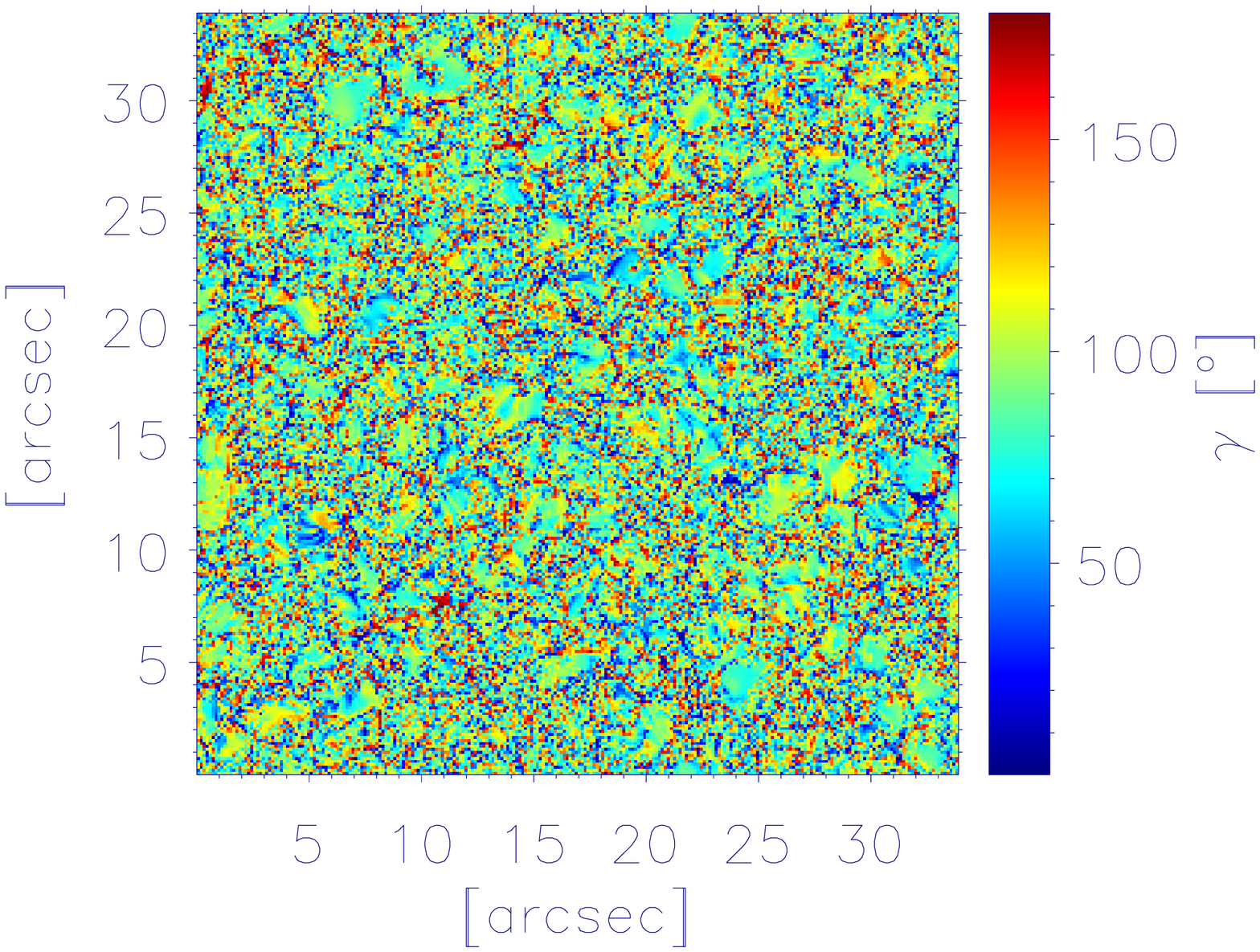}
  \includegraphics[angle=90,width=0.38\linewidth,trim= 0.5cm 3cm 1.5cm 0cm,clip=true]{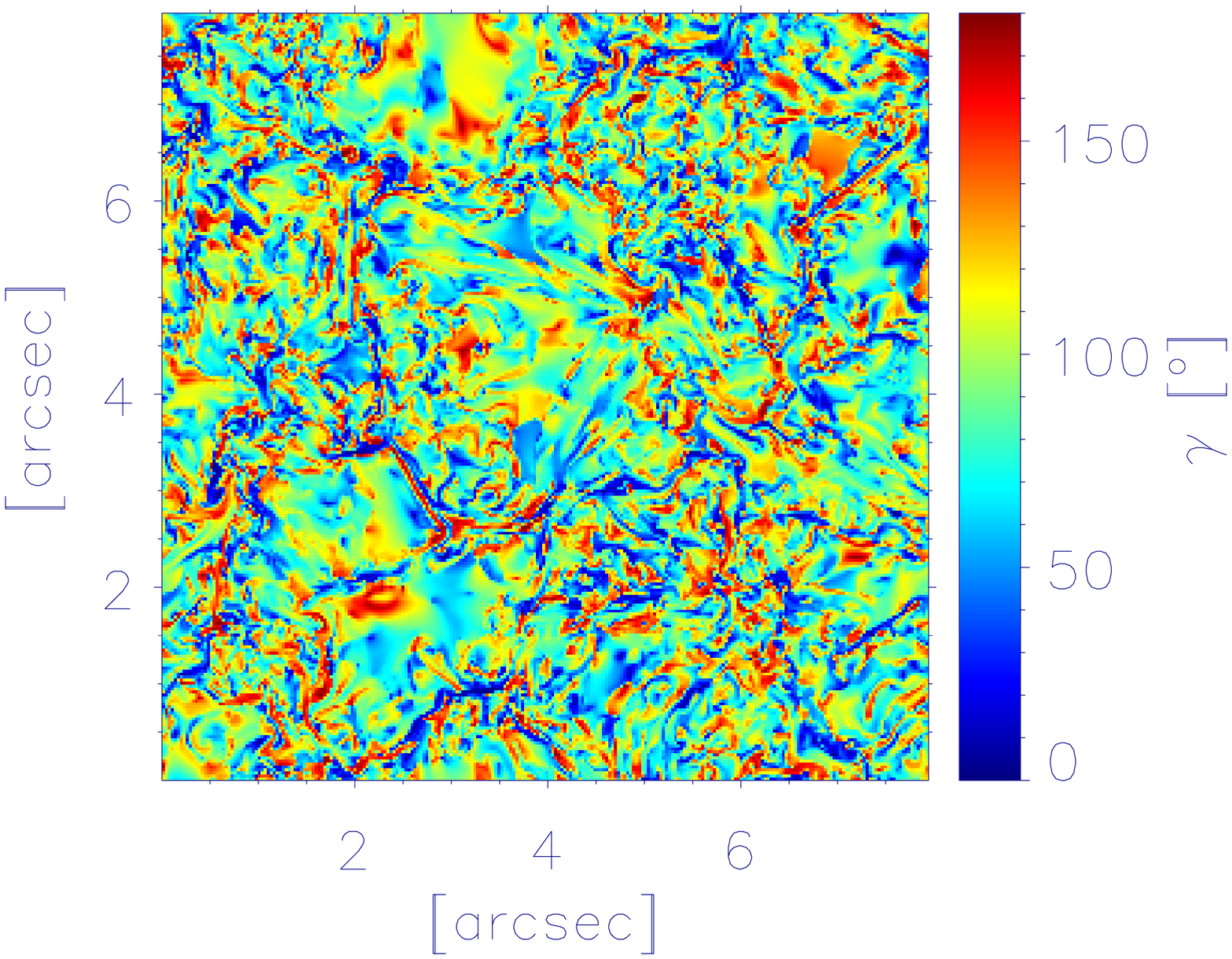}
  \caption{Overview of the MHD snapshots used for testing 2D inversions: Intensity (top), magnetic field strength (middle) and inclination (bottom) in Sim~1 (left), Sim~2 (middle) and Sim~3 (right) at $\log  \tau =0$. }
\label{maps_sim_all_bg}
\end{figure*}

\section{Simulations}

To test our inversions, we use three snapshots from three different runs produced with the MURaM code \citep{vogler05}. Since we are interested in finding out what the inversion process does with weak quiet Sun signals, we choose snapshots that produce spectropolarimetric signals of similar strength. Given that the origin of the quiet Sun field is still a subject of debate, we take two snapshots from the local dynamo runs and one where pure flux emergence is taking place. The field configuration in all three simulations are predominantly horizontal. Although some studies suggest the real quiet Sun may have predominantly vertical field \citep{stenflo10,ryuko11}, we are not able to produce a snapshot with such field configuration that matches the observed level of linear polarization as well as circular polarization signals. Hence, we do not include this type of simulations in the test. The main properties of the chosen snapshots are given in Table~\ref{sim_overview}. The maps of the intensity and magnetic field parameters in all three snapshots are given in Fig.~\ref{maps_sim_all_bg}.
\begin{figure*}
  \centering
  \includegraphics[angle=0,width=\linewidth]{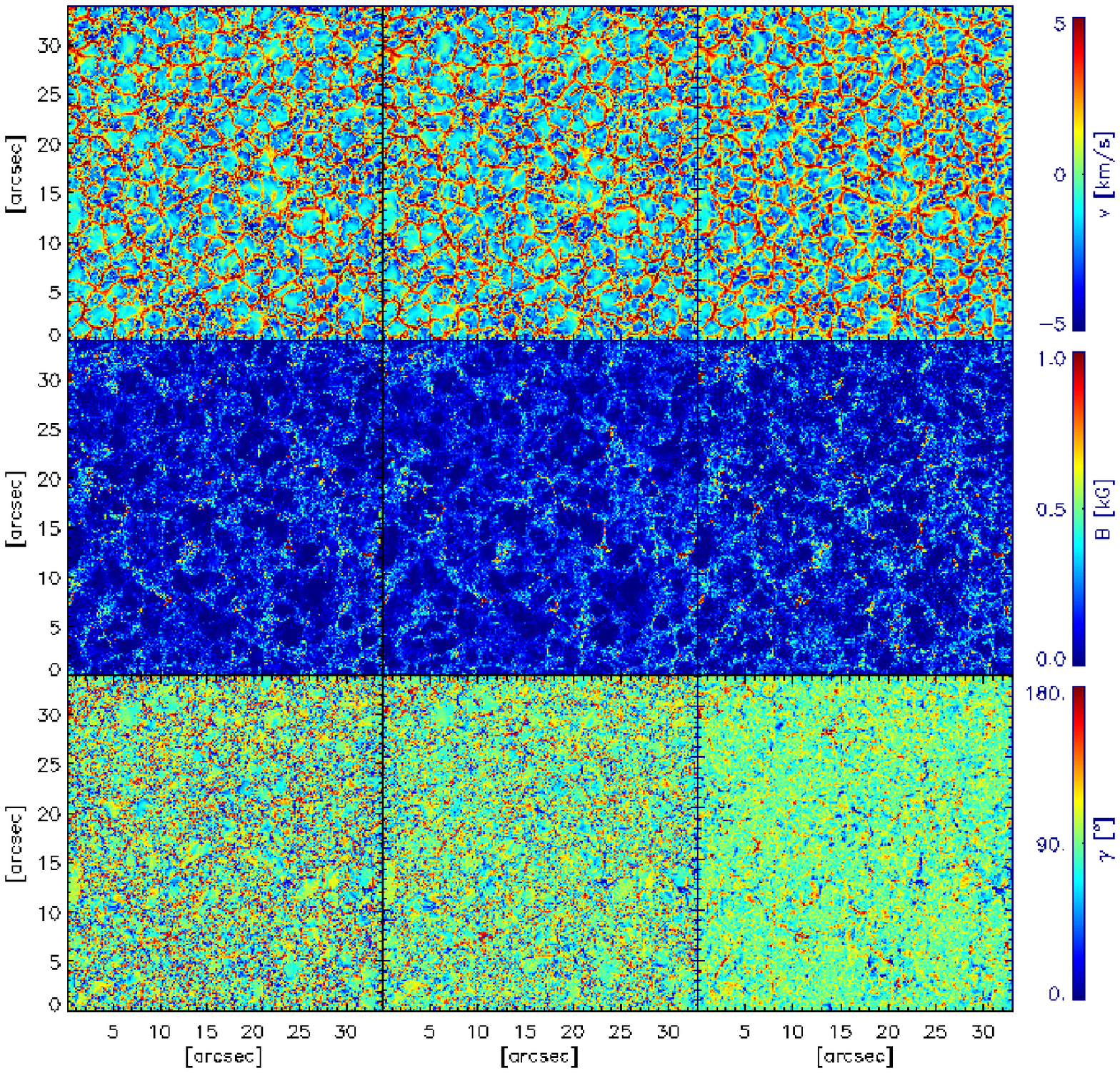}
  \caption{Results of 2D inversions applied to simulations - comparison of maps of velocity (top row), magnetic field strength (middle row) and inclination (bottom row) at $\log  \tau =0$ in Sim~2. Three columns, from left to right, show the original unsmeared maps from the snapshot, the same maps after the highest spatial frequencies are filtered out and the results from the inversions.}
\label{maps_sim2_bg}
\end{figure*}

\begin{table*}
\caption{Main characteristics of the MHD snapshots used to test the inversion code. Mean unsigned longitudinal and transverse flux densities are calculated after all the instrumental effects are taken into account.}
   \centering                                            
   \begin{tabular}{c l l l l}                              
   \hline                                                
   \noalign{\smallskip}
   Name & Dimension    & Resolution & Longitudinal & Transverse   \\
       &  &       & flux density      & flux density  \\   
   \noalign{\smallskip}                                                                                           
              &  [Mm]   &  [km]   & [G]      & [G]   \\
   \hline                                                                                                       
   \noalign{\smallskip}                                                                                              
   Sim~1      & $4.86\times4.86\times1.4$   &  $5\times5\times7$   &      4.2     &   54.2  \\
   Sim~2      & $24.6\times24.6\times7.68$   &  $16\times16\times16$   &      10.5      &   52.8  \\
   Sim~3      &  $6\times6\times1.68$  &   $10\times10\times14$  &     6.8     &    53.1    \\
\noalign{\smallskip}
\hline                                                
\noalign{\smallskip}
\end{tabular}
\label{sim_overview}
\end{table*} 

The first snapshot, Sim~1, is a snapshots from the saturated phase of Run~G ($R_m \sim 5200$) from \cite{danilovic10a}. A detailed description of these dynamo simulations is given in \cite{vogler07} and \cite{pietarilagraham10}. To get the observed level of spectropolarimetric signals, we multiplied the field strength with a factor of 2, as it is done in \cite{danilovic10a}. 
Sim~2 is a snapshot from a more realistic dynamo simulations that tries to take into account the contribution coming from the large scale dynamo by introducing an open bottom boundary. Snapshot is a non-grey version of the run O16bM from \cite{Rempel14}. The photosphere is located about 1.5~Mm beneath the top boundary. The open bottom boundary allows the for presence of small-scale horizontal magnetic field in upflow regions in an attempt to mimic a deep magnetized convection zone \citep{Rempel14}.
 The last snapshot, Sim~3, is taken from a run that simulates pure flux emergence with no local dynamo action present. The run is similar to the run used in \cite{danilovic14}, with the difference that a horizontal flux sheet is introduced into a purely hydrodynamic run.  The field strength in the flux sheet is set to vary across the cross section as a Gaussian with a width of  $110$~km and a maximum value of $300$~G. The initial position of the sheet is some $300$~km below the height where the mean optical depth is unity. We take a snapshot some $15$~min into the run.

All three snapshots are treated in the same way to produce synthetic observations. Hinode spectropolarimetric observations are simulated by convolving with an appropriate spectral and spatial PSF \citep{danilovic08}. Also Gaussian noise at the level of $8\times10^{-4}$~I$_{c}$ is added to simulate deep magnetogram observations.  Table~\ref{sim_overview} shows the longitudinal and transverse flux densities for every snapshot calculated by application of the solarsoft routines \citep{lites13} on the artificial data, after the convolutions were applied. The values are very close to the ones found in observations \citep{danilovic10a}\footnote{The values for Sim~1 are obtained after the field strength was scaled.}. The final preparatory step is the interpolation of such obtained maps to a finer sampling, as is done for real observations.

\section{Tests on simulated Hinode data}

\cite{Michiel2012} showed that the 2D inversions are able to retrieve all physical parameters when the input data are spectra computed from simplified  atmospheres. Here, we determine how far the final result is off if we feed the inversions with spectral profiles calculated from the complex original stratification of MHD simulations. We are not interested in the accuracy of the retrieved values of individual pixels, but in the overall distribution of the key parameters: velocity, magnetic field strength and inclination.
To make a quantitative measure of the ability of the inversion code to retrieve these physical parameters, one needs first to smear the original simulation somehow to the resolution of the simulated observations. Since the 2D inversions  should retrieve only spatial frequencies lower than the diffraction limit of the Hinode/SOT, we degrade the original maps for every parameter at a constant optical depth with a low-pass filter with a cut-off at $3.85$~arcsec$^{-1}$ and then rebin to the pixel size of the oversampled observations ($0.08\arcsec$). The inclination maps are treated differently because the spatial averaging in this case is weighted with the field strength. Since Stokes V signal scales linearly with longitudinal component B$_{long}$, and Q and U scale quadratically with transverse component B$_{tra}$ of the magnetic field, smearing in Stokes space will make the derived fields more horizontal due to the different scaling. We thus apply the filter to B$_{long}$ and B$_{tra}^{2}$ separately, rebin them and then compute the inclination for every resized pixel. Treating original maps in this way enables us to make a pixel-to-pixel comparison of the retrieved and original simulated parameter as a function of optical depth. Figure~\ref{maps_sim2_bg} shows the example for a snapshot from Sim~2. The spatial filtering seems to change the appearance of the inclination map the most.

The parameter that is retrieved most robustly in Fig.~\ref{maps_sim2_bg} is the velocity. The inverted velocity map retains all the details visible in the original maps, i.e. the small downflow in the center of the granule at [23\arcsec , 17\arcsec] or the strong upflow at the edge of granule at [5\arcsec , 21\arcsec]. Apart from being noisy, the inverted map of magnetic field strength looks very much like the original. The strong field features together with patches of no field are well recovered. In places, though, the field strength is a bit higher, probably due to the noise included in the simulated observations. The inclination, on the other hand, seems to reveal more horizontal field than is really there. The areas where the inversion code misses on the small scale mixed polarity correspond to regions where the field strength is very weak. The inversion code, however, retrieves the inclination of the stronger magnetic features well.

Coming to this result took some testing which we present below. We investigate how the results depend on the setup of the node positions in the inversions, the input configuration of the magnetic field, the choice of a wrong PSF, the noise level and the evolution of the underlying solar scene.

\begin{figure*}
  \centering
  \includegraphics[angle=90,width=0.37\linewidth ,trim=3cm 1cm 1cm 0.5cm,clip=true]{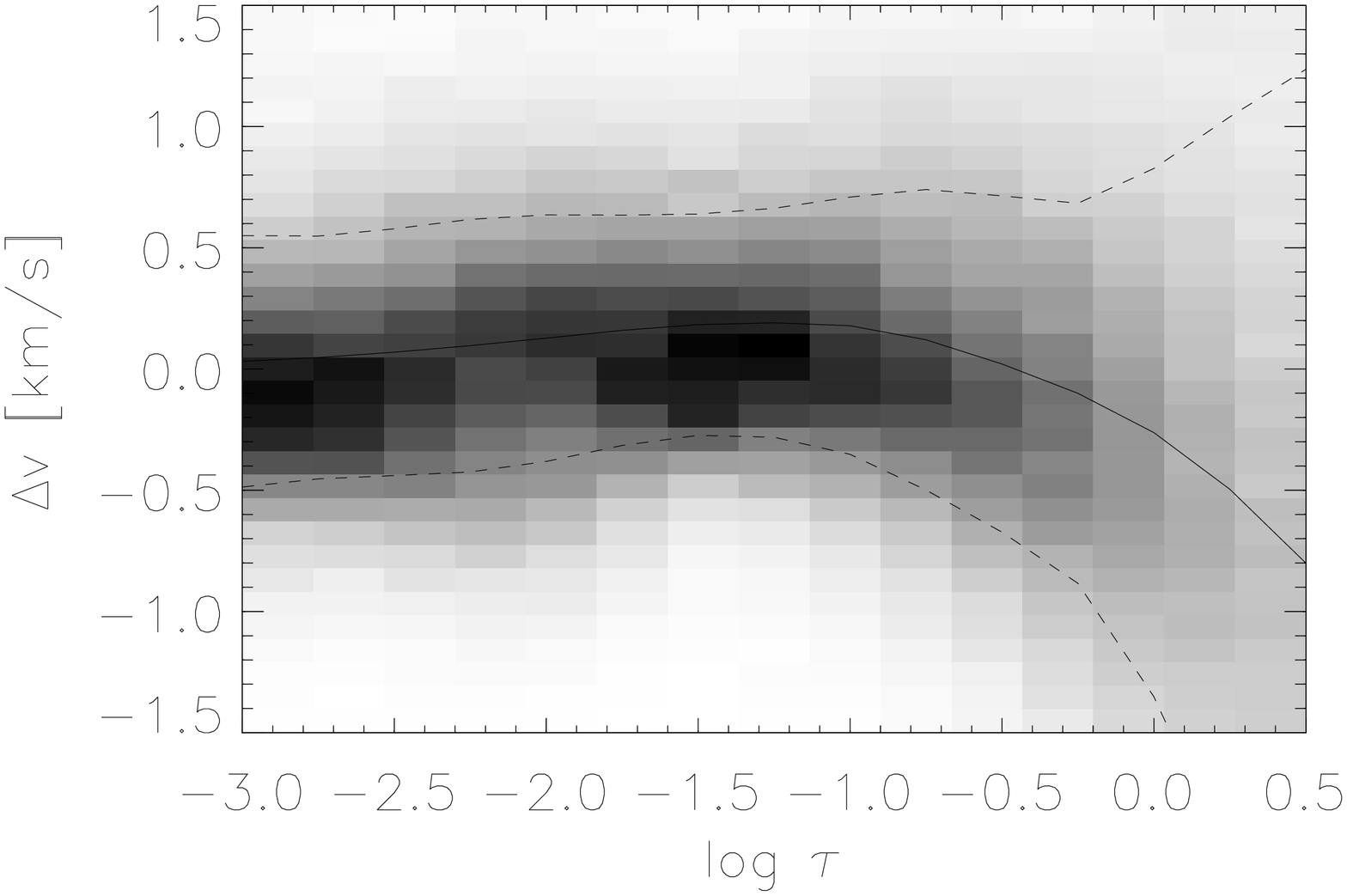}
  \includegraphics[angle=90,width=0.3\linewidth ,trim=3cm 1cm 1cm 5cm,clip=true]{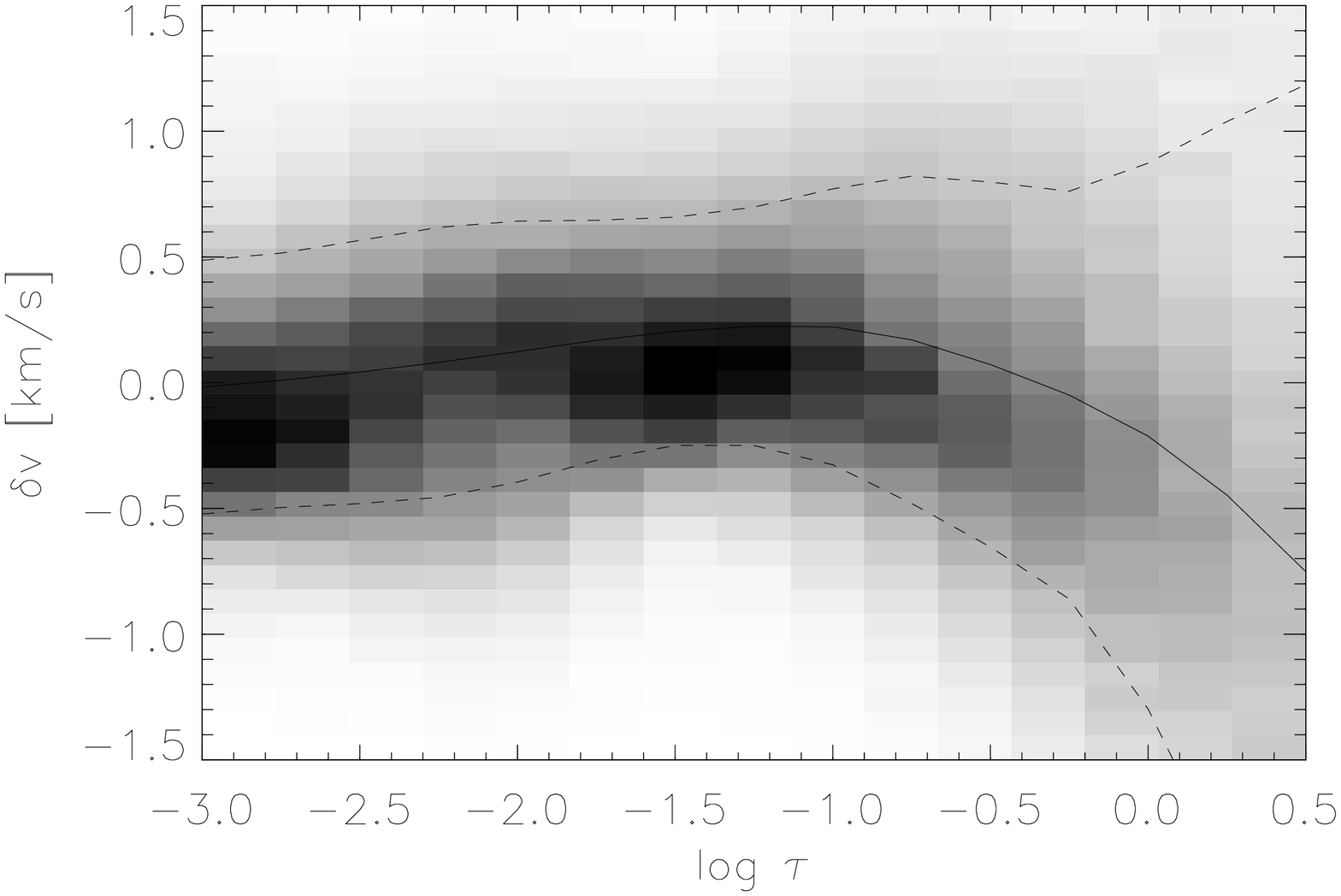}
  \includegraphics[angle=90,width=0.3\linewidth ,trim=3cm 1cm 1cm 5cm,clip=true]{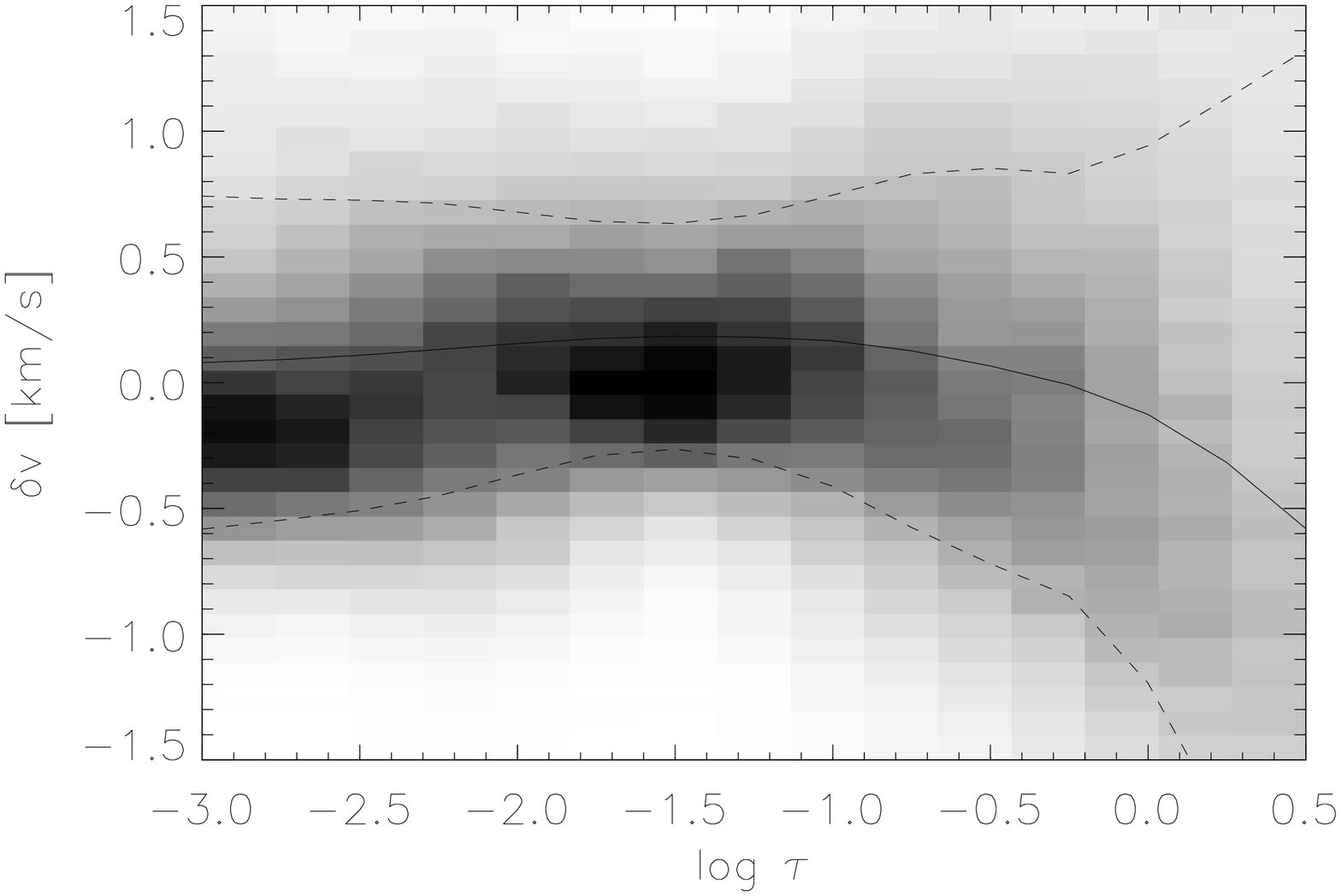}  
  \includegraphics[angle=90,width=0.37\linewidth ,trim=3cm 1cm 1cm 0.5cm,clip=true]{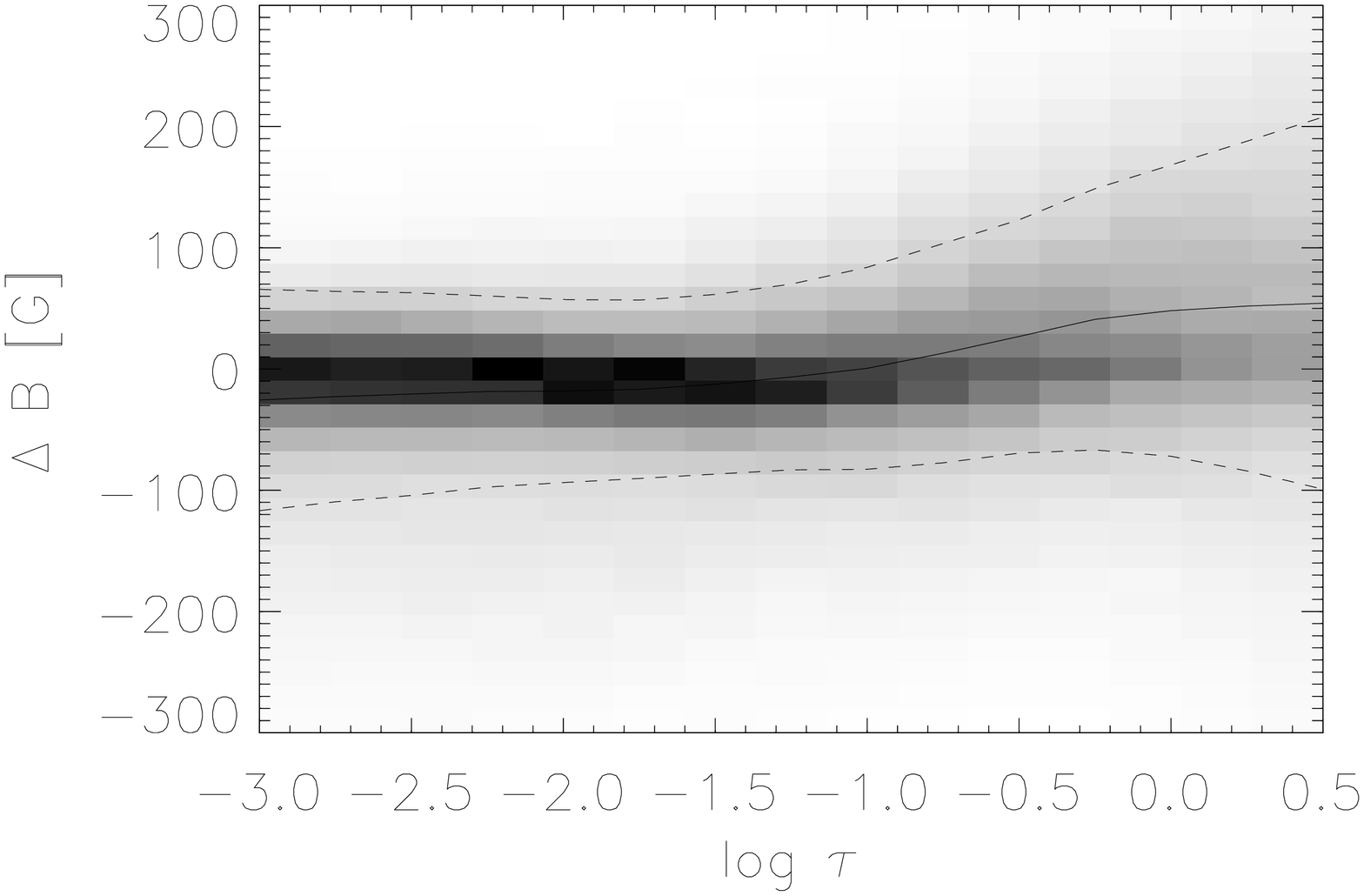}
   \includegraphics[angle=90,width=0.3\linewidth ,trim=3cm 1cm 1cm 5cm,clip=true]{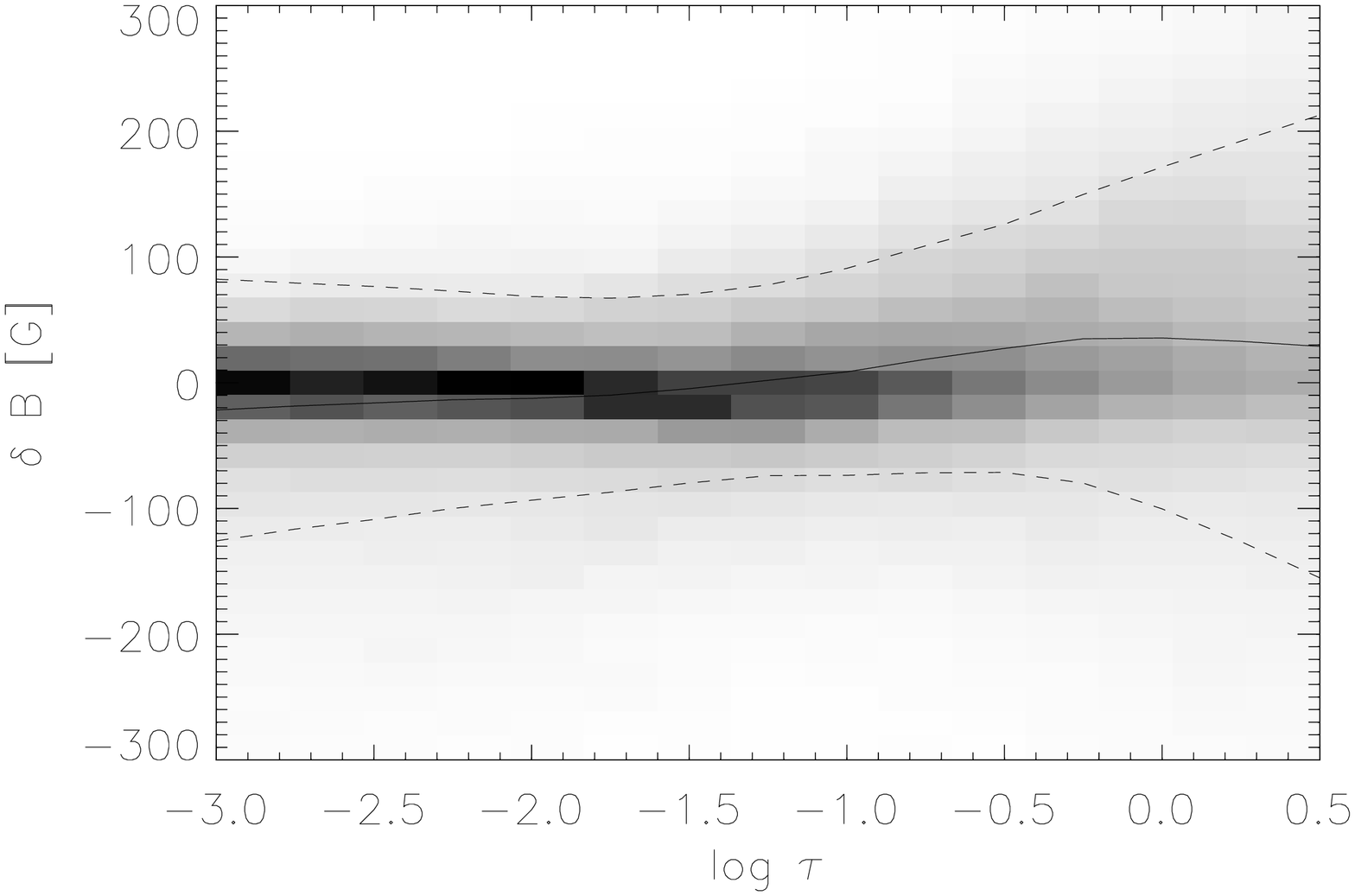}
   \includegraphics[angle=90,width=0.3\linewidth ,trim=3cm 1cm 1cm 5cm,clip=true]{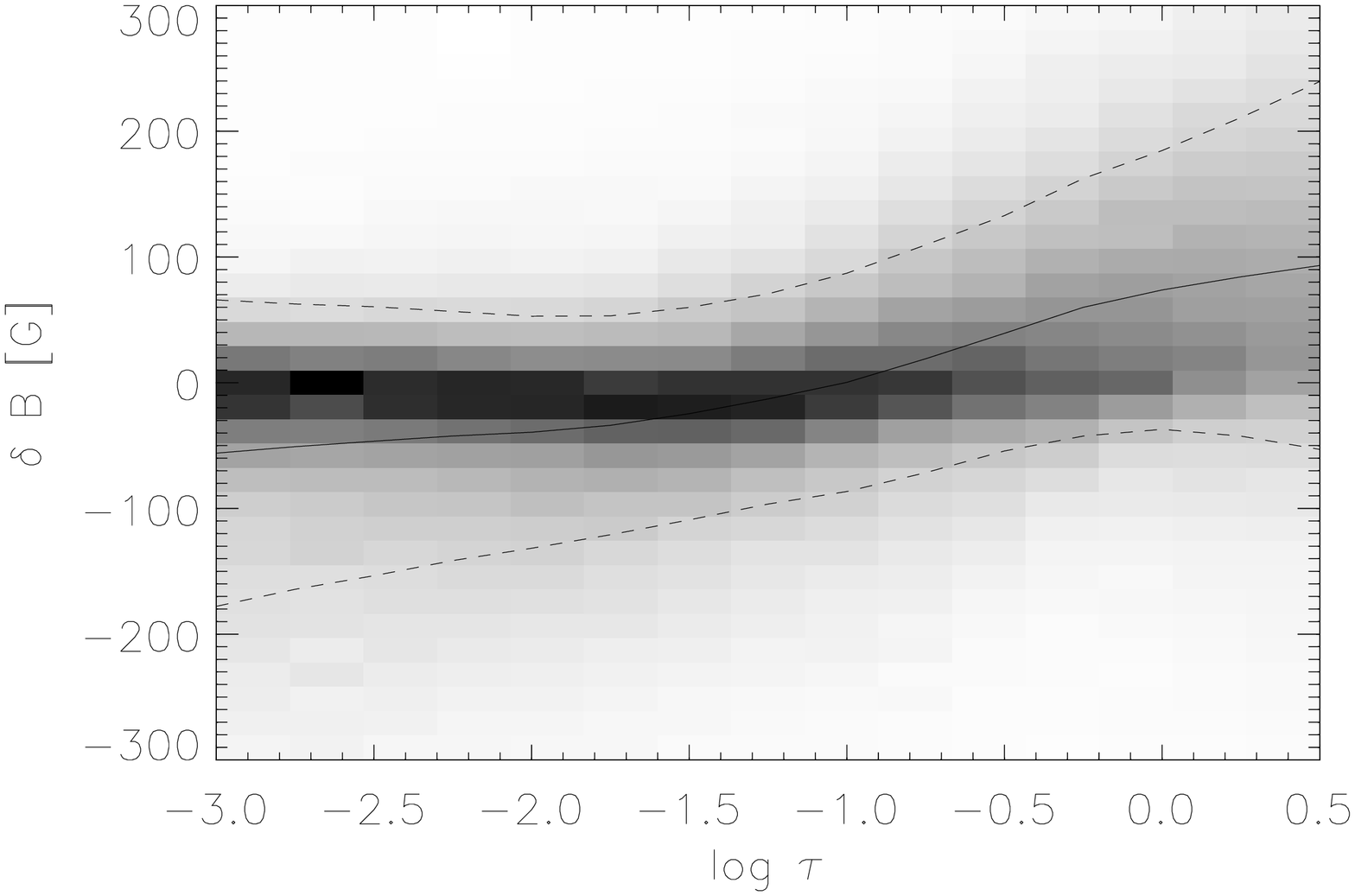}
   \includegraphics[angle=90,width=0.37\linewidth ,trim=0.5cm 1cm 1cm 0.5cm,clip=true]{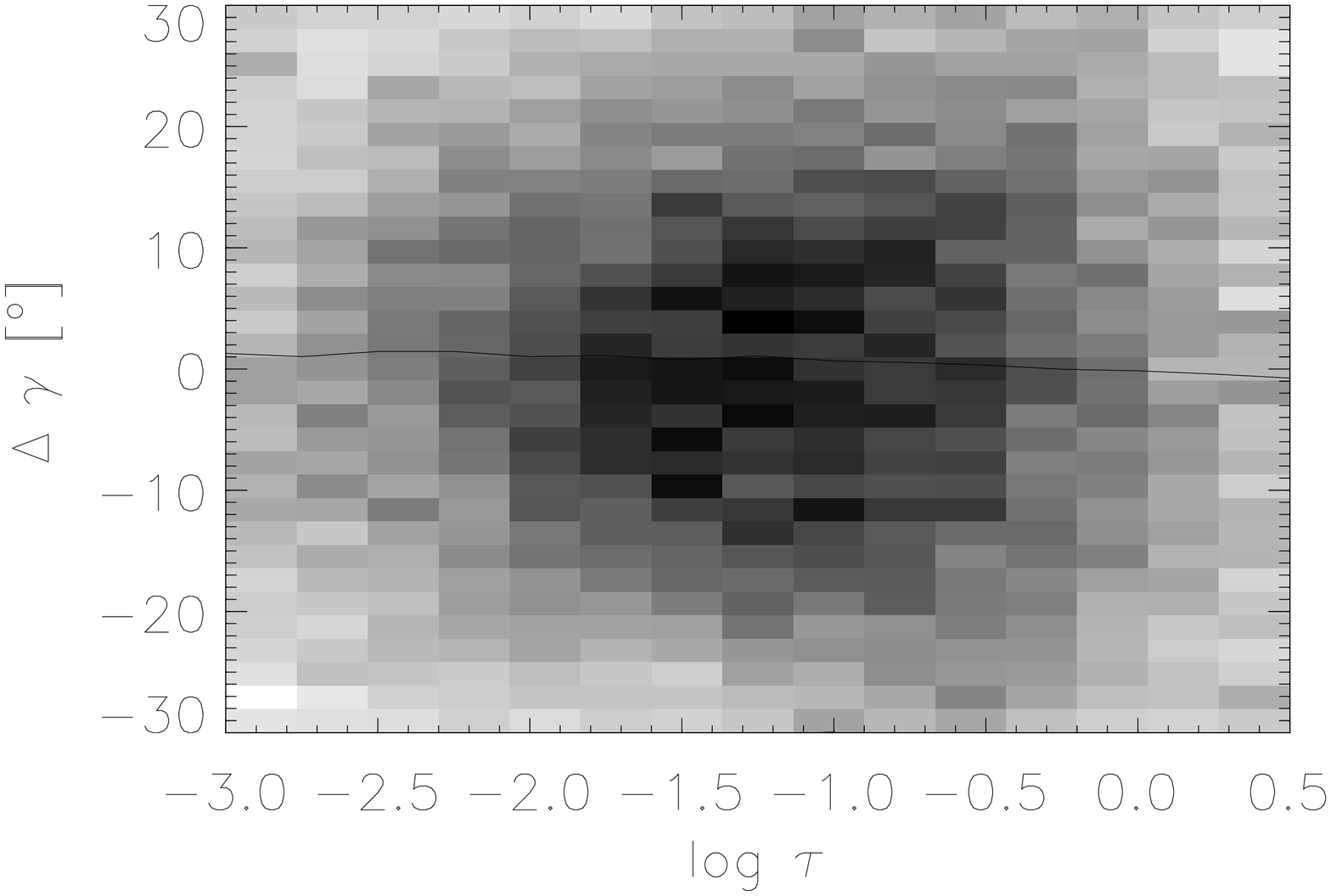}
   \includegraphics[angle=90,width=0.3\linewidth ,trim=0.5cm 1cm 1cm 5cm,clip=true]{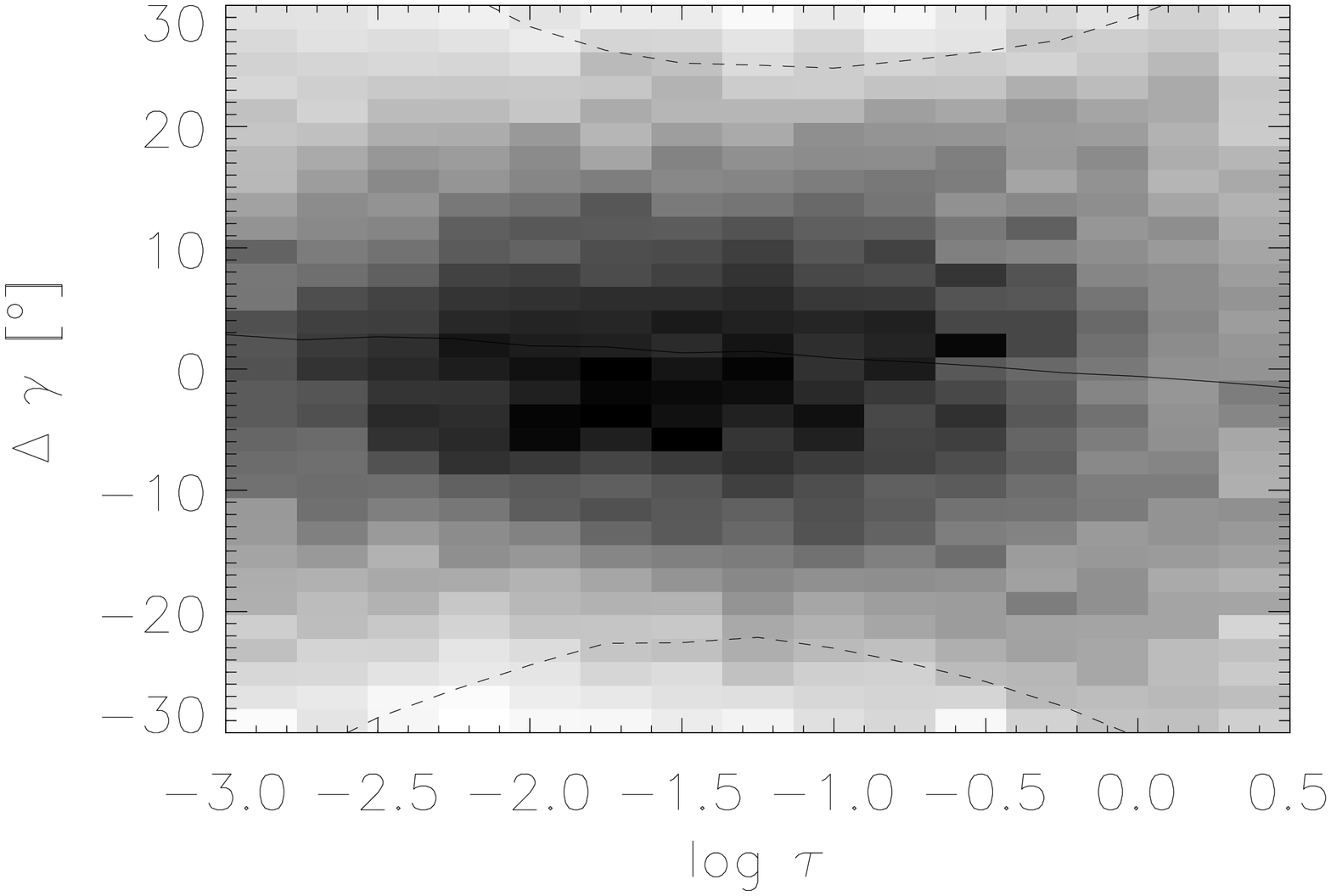}
   \includegraphics[angle=90,width=0.3\linewidth ,trim=0.5cm 1cm 1cm 5cm,clip=true]{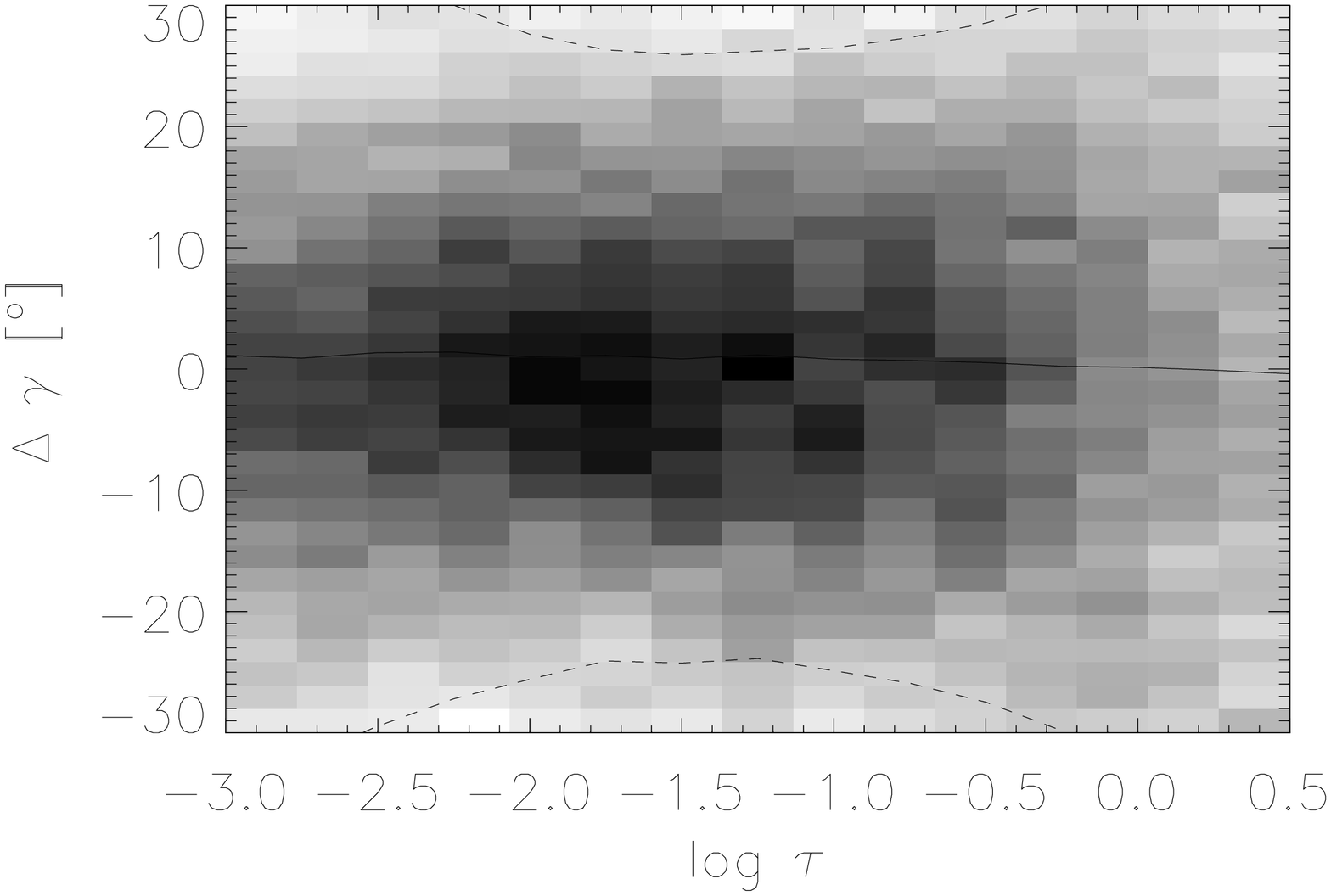}
  \caption{Results of 2D inversions applied to simulations - test of node position in Sim~1 when the top node is placed at $\log  \tau =-1.5$ (left column), $\log  \tau =-2.0$ (middle column) and $\log  \tau =-2.5$ (right column): 2D histograms of the difference of the original smeared and inverted quantities: magnetic field (top), LOS velocity (middle) and inclination (bottom) as a function of optical depth. The mean difference is marked by solid lines and the standard deviation of the scatter by the dashed.}
\label{tauplot_diffnodes}
\end{figure*}

\subsection{Node position setup}

Our tests show that the distribution of inverted parameters varies the most with the changes in inversion node positions. Different combinations are tried of which only extreme cases are shown here to illustrate the amount of scatter. Figure~\ref{tauplot_diffnodes} shows the difference between the input and the inverted values as a function of optical depth, and demonstrates how much systematic error is introduced when only the top node is shifted from $\log  \tau =-1.5$ to $\log  \tau =-2.5$. The lower two nodes are kept fixed at $\log  \tau =0$ and $-0.8$.

As shown by \cite{Michiel2012}, the velocity is a pretty robust quantity, whose determination is not very sensitive to noise since it can be basically inferred directly from the intensity. It is well recovered in all layers of the atmosphere, with a mean error that is always smaller than $20$~m/s. The best match to the original value, in the lower as well as in the upper atmosphere, is obtained when the top node is placed at $\log  \tau =-2.0$. Shifting the top node upwards or downwards gives underestimated or overestimated velocities at the surface respectively. Velocities in the upper atmosphere are underestimated in both cases.

Although the determination of the magnetic field strength is more prone to noise induced errors, Fig.~\ref{tauplot_diffnodes} shows that a pixel-to-pixel comparison yields an accumulation of points around 0. The mean error in the top layers goes from close to $100$~G when the top node is at $\log  \tau =-2.5$ to less than $20$~G when we put it at $\log  \tau =-2.0$. All three curves come the closest to zero around the middle node. This is not a surprise given that the inversions are most sensitive and accurate at this height when the Fe~I~$630$~nm lines are fitted \citep{cs2005}.  

A pixel-to-pixel comparison of the original and inverted values shows the largest scatter for the magnetic field inclination. The standard deviation curves exceed $30^{o}$ in all three cases, peaking when the top node is at $\log  \tau =-1.5$. 

Shifting the middle node from $\log  \tau =-0.8$ to $-1$ gives a larger mean error at higher layers of the atmosphere. It is even doubled for some quantities.

For further tests, we use the optimum combination of node positions at $\log  \tau =0,-0.8$ and $-2.0$.

\subsection{Difference in the magnetic field configuration}

Since they all give similar amount of spectropolarimetric signal, as shown in Table~\ref{sim_overview}, all three snapshots have a similar field strength distribution. A slight difference exists in the upper atmosphere where Sim~1 has significantly lower field strength than the other two snapshots. This is because this early dynamo run contains predominantly small scale magnetic field loops that close low down \citep{manfred08}, so the height profile of the mean field strength in this run drops off more rapidly than in the other two cases. Another difference is that Sim~2 is the only snapshot that shows a large scale magnetic field distribution, and hence shows more strong field features at the solar surface than the other two examples.

The main difference is, however, the field entanglement, which is the main reason why these runs are chosen. The complexity of the field configuration increases with the resolution of the simulations. So, while Sim~3 has well resolved magnetic features, Sim~1 and 2 show small-scale salt and pepper pattern of opposite polarities characteristic of all local dynamo runs. Therefore, this test determines how much detail our 2D simulations are able to retrieve.

\begin{figure*}
  \centering
     \includegraphics[angle=90,width=0.37\linewidth ,trim=3cm 1cm 1cm 0.5cm,clip=true]{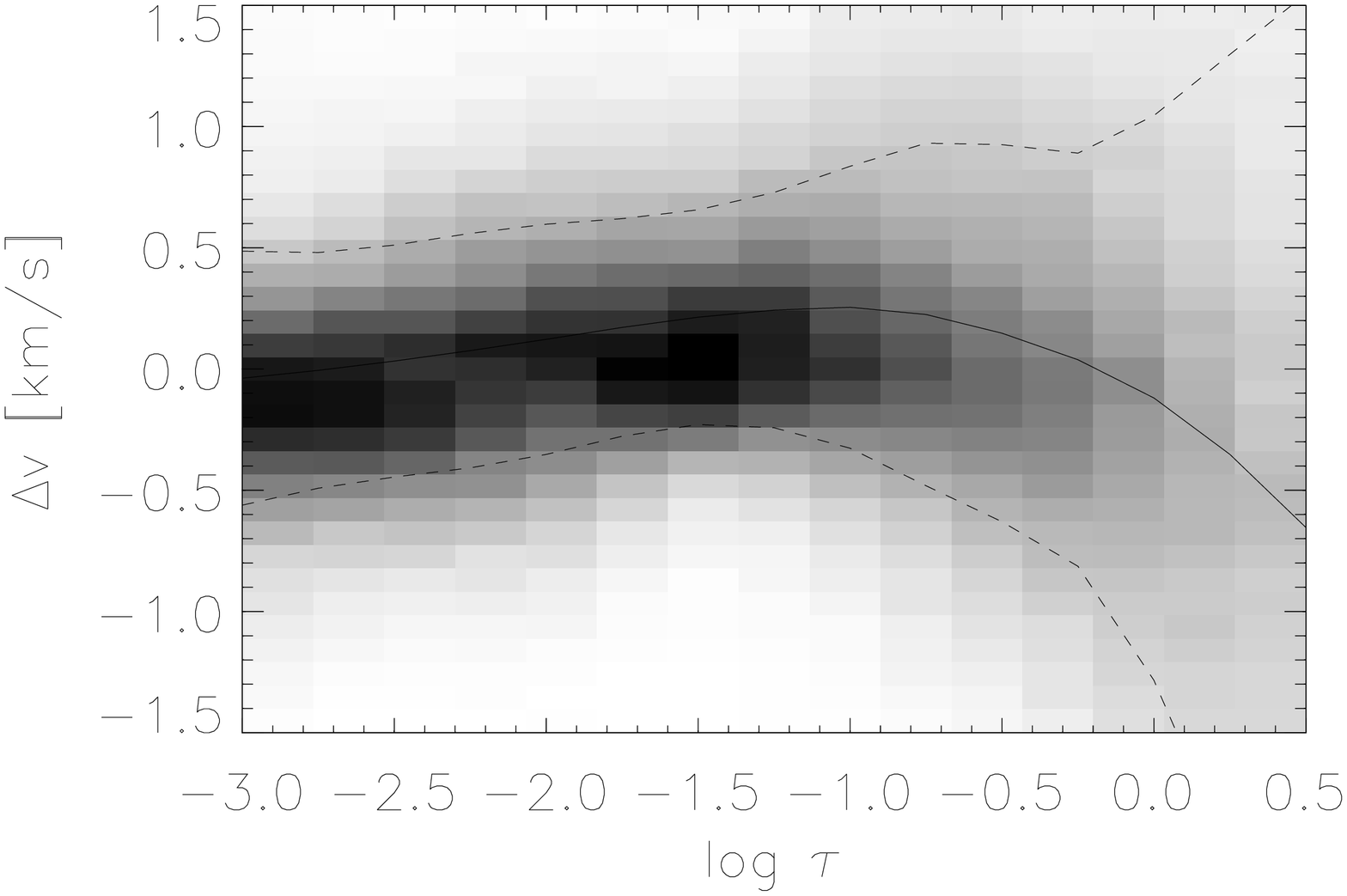}
     \includegraphics[angle=90,width=0.3\linewidth ,trim=3cm 1cm 1cm 5cm,clip=true]{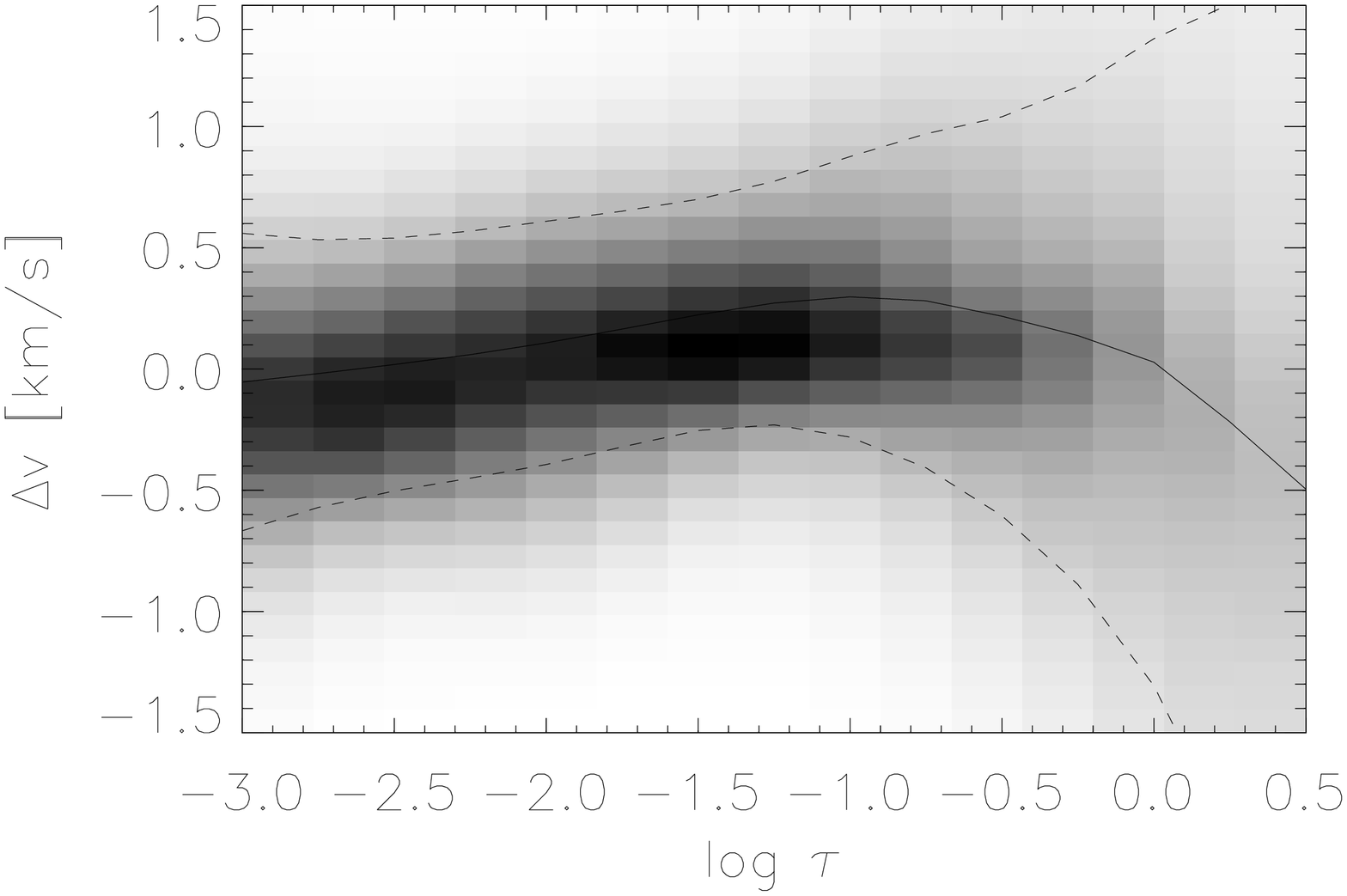}   
    \includegraphics[angle=90,width=0.3\linewidth ,trim=3cm 1cm 1cm 5cm,clip=true]{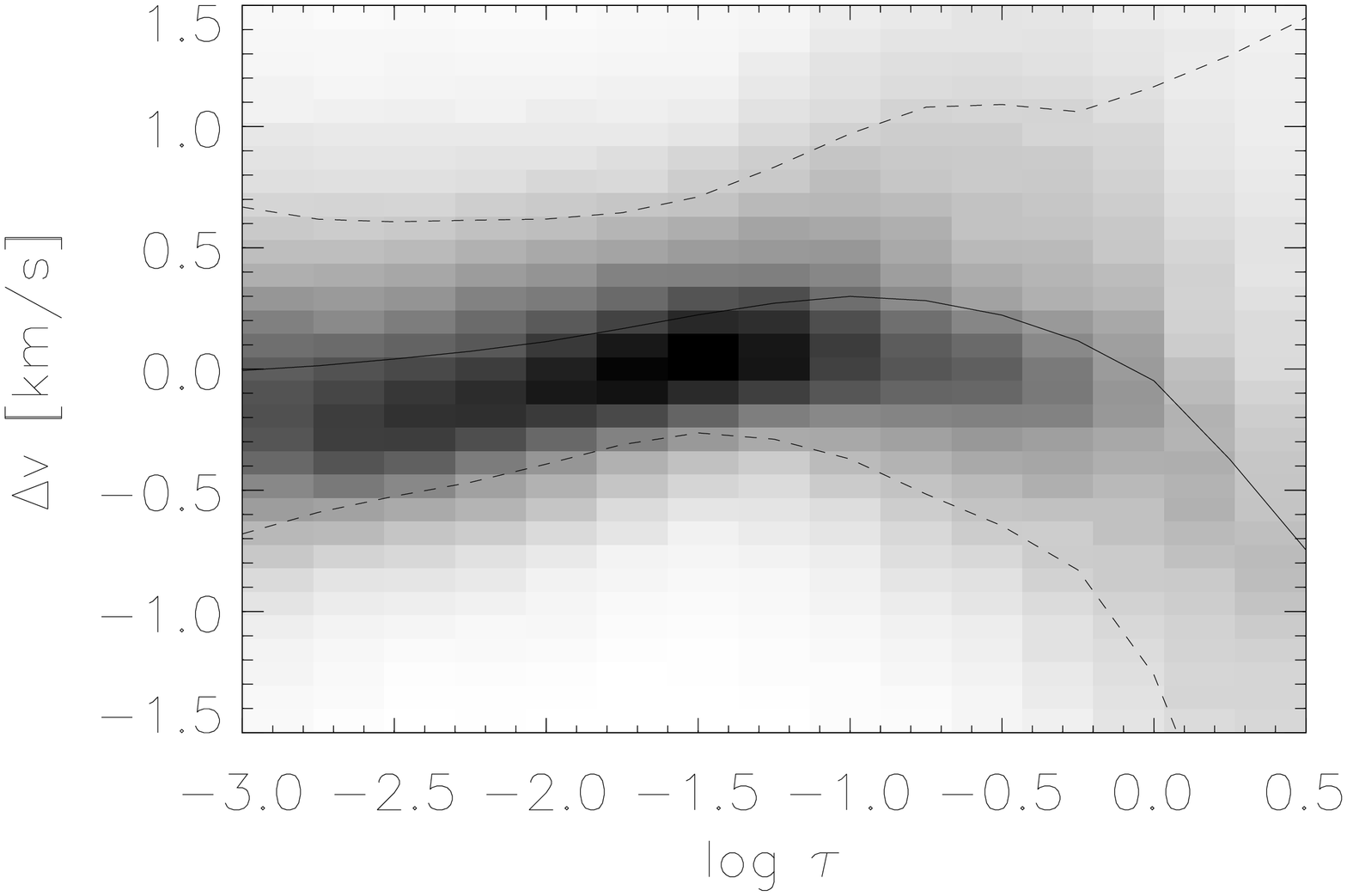}
  
  \includegraphics[angle=90,width=0.37\linewidth ,trim=3cm 1cm 1cm 0.5cm,clip=true]{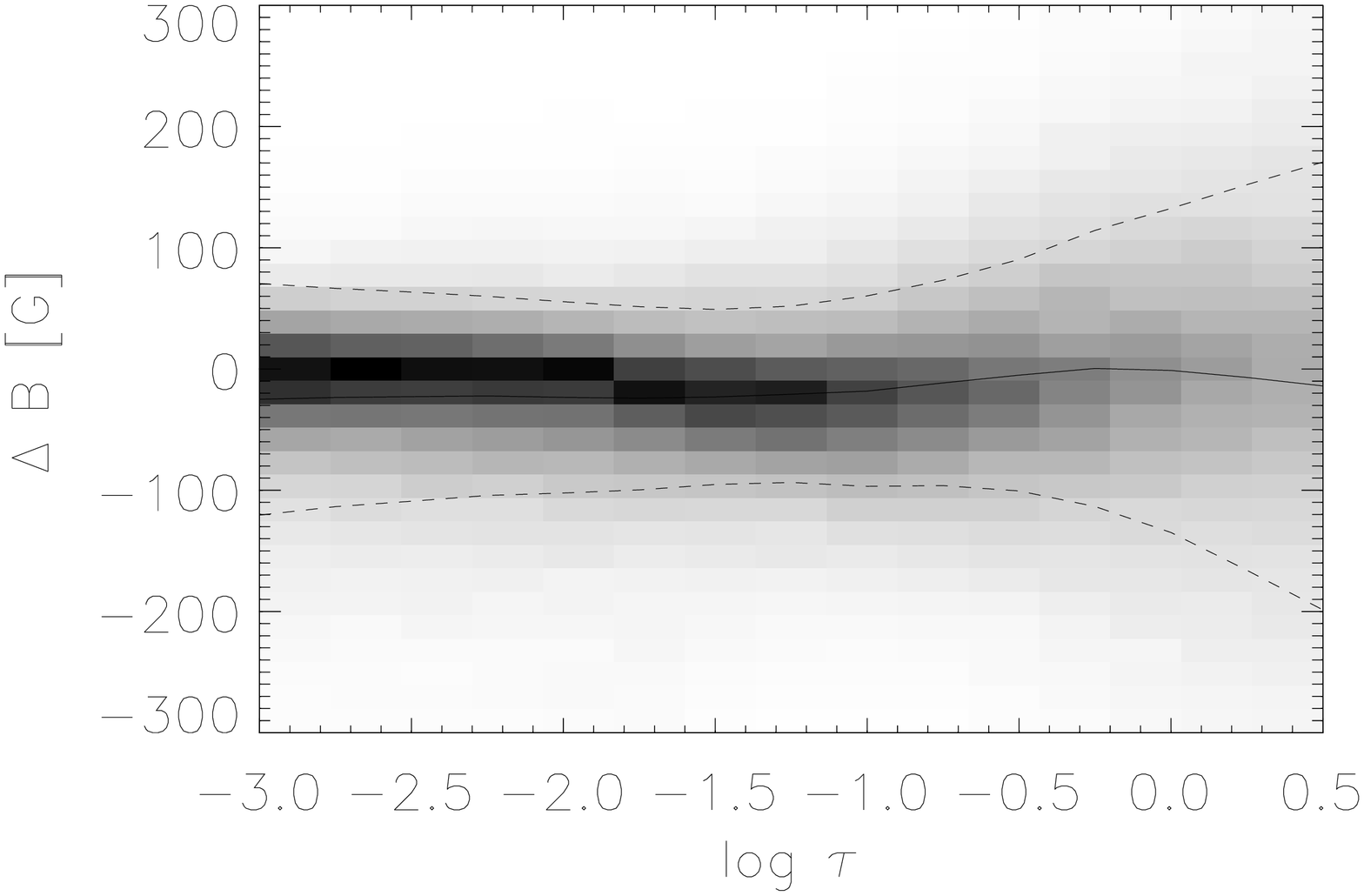}
  \includegraphics[angle=90,width=0.3\linewidth ,trim=3cm 1cm 1cm 5cm,clip=true]{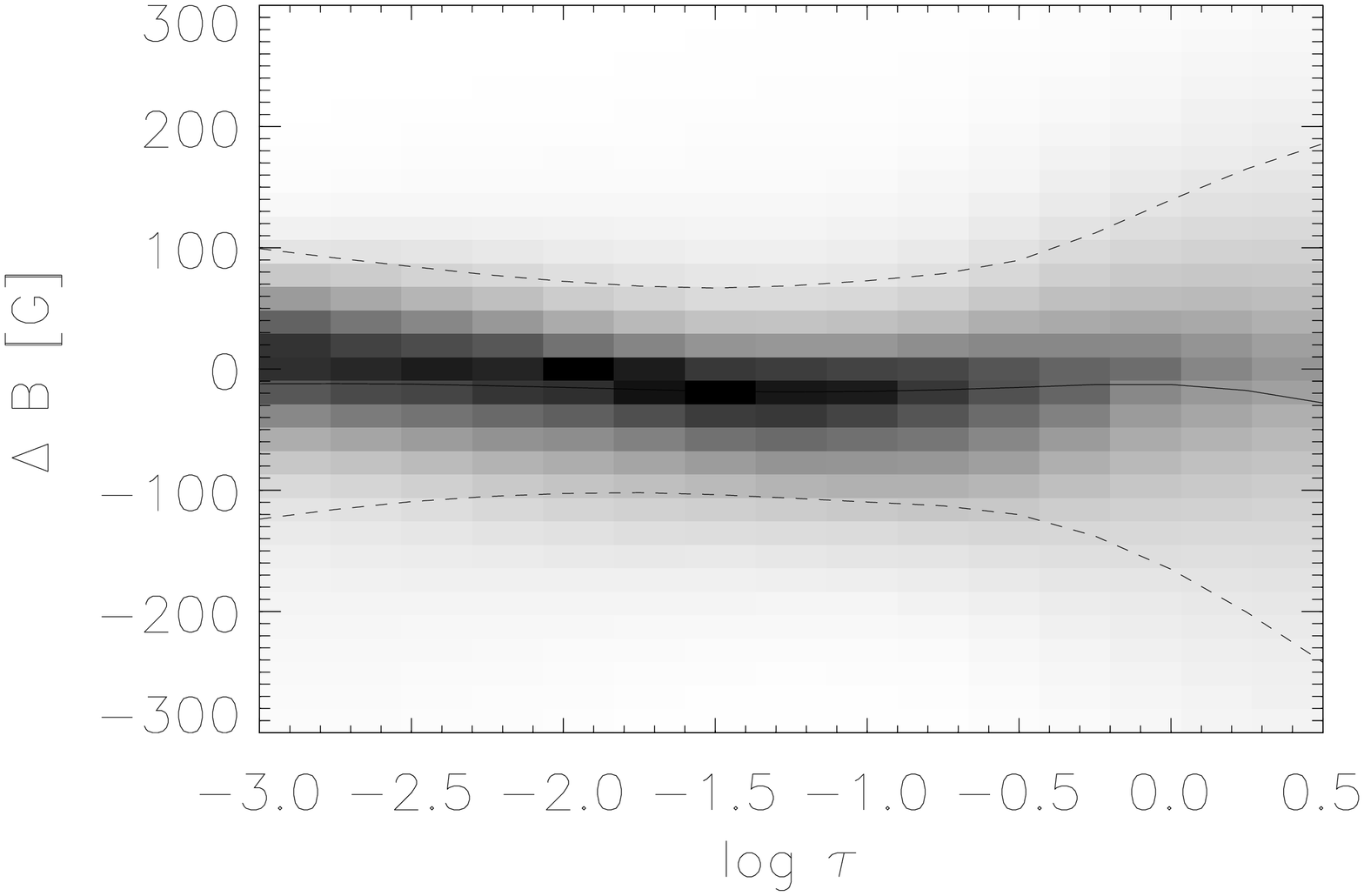}
  \includegraphics[angle=90,width=0.3\linewidth ,trim=3cm 1cm 1cm 5cm,clip=true]{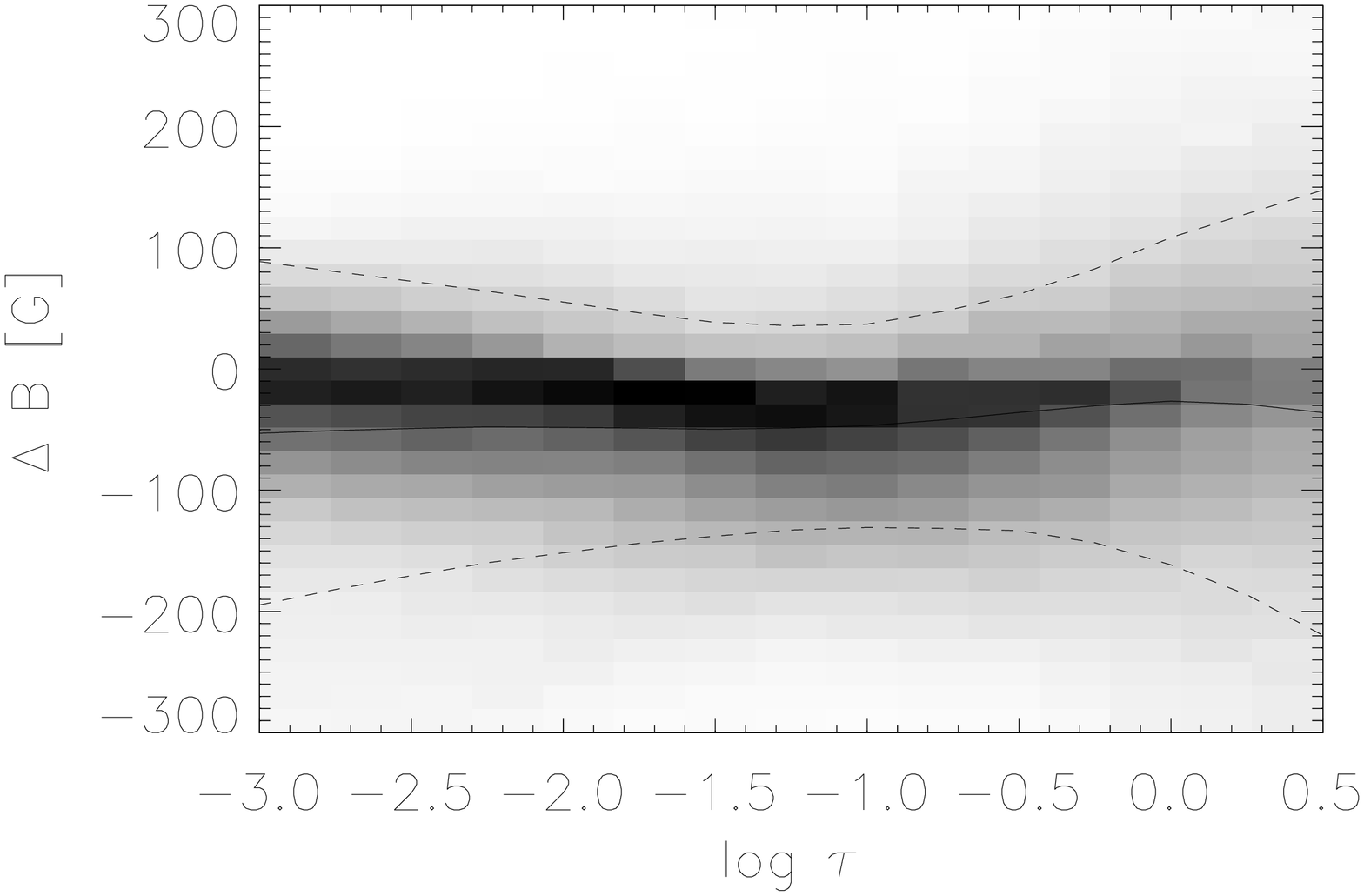}
   
   \includegraphics[angle=90,width=0.37\linewidth ,trim=0.5cm 1cm 1cm 0.5cm,clip=true]{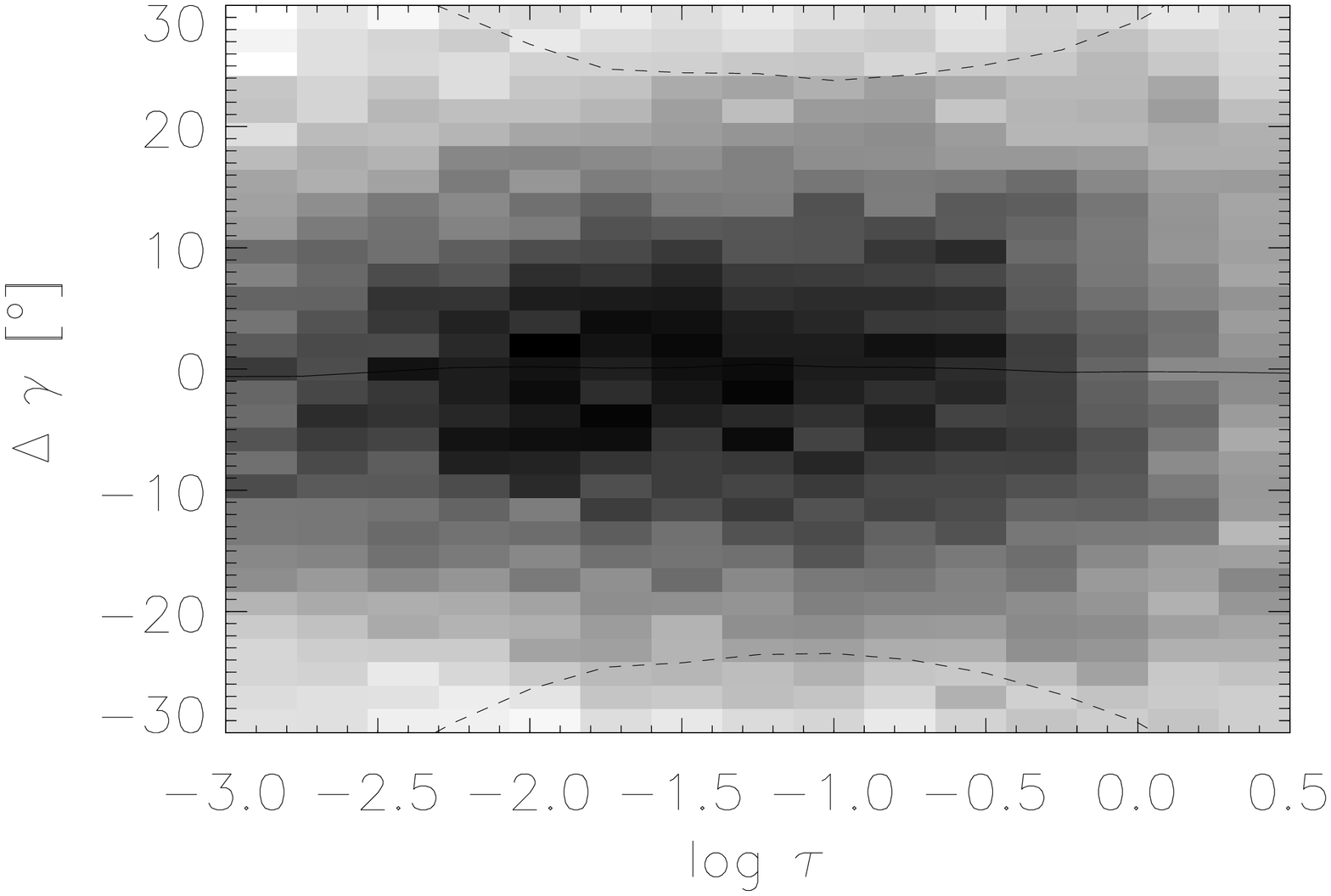}
  \includegraphics[angle=90,width=0.3\linewidth ,trim=0.5cm 1cm 1cm 5cm,clip=true]{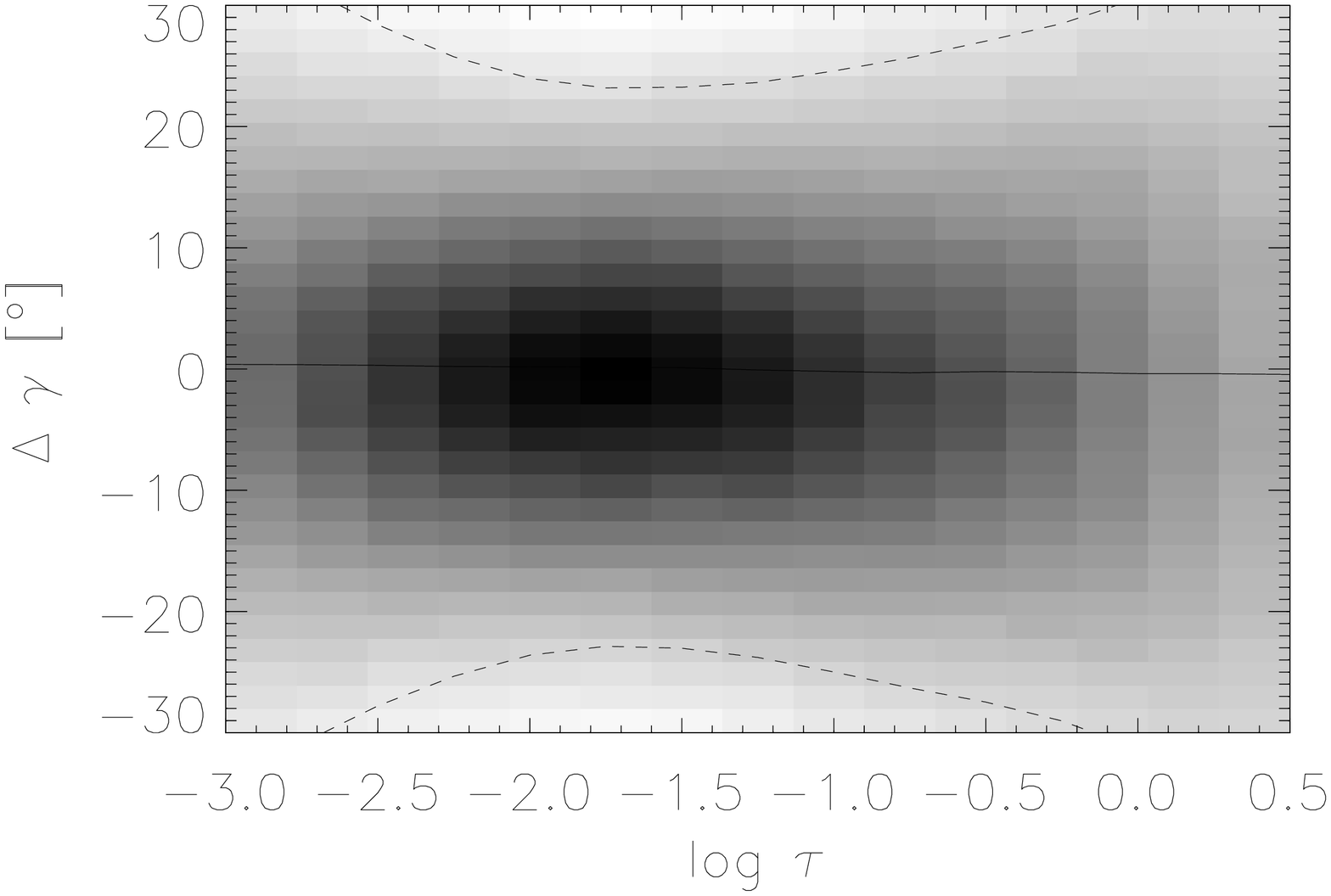}
  \includegraphics[angle=90,width=0.3\linewidth ,trim=0.5cm 1cm 1cm 5cm,clip=true]{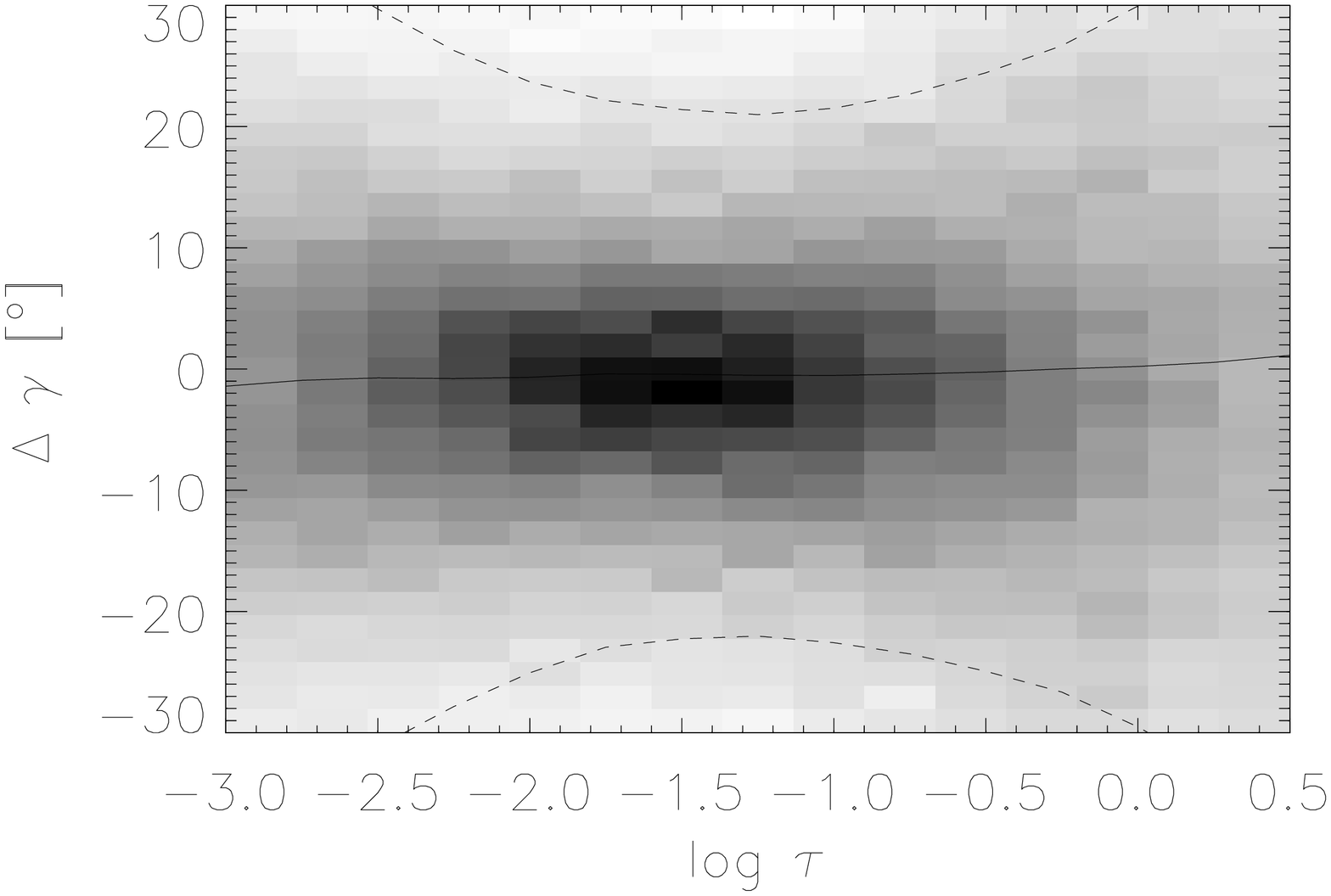}
  \caption{Results of 2D inversions applied to simulations - test on different simulations: difference of original and inverted velocity (top row), magnetic field strength (middle row) and inclination (bottom row) as a function of optical depth for Sim~1,2 and 3 from left to right. }
\label{tauplot_diffsim}
\end{figure*}

Figure~\ref{tauplot_diffsim} shows again the difference of initial and retrieved parameters as a function of optical depth for all three cases. The 2D histograms look smoother for Sim~2 because of the larger domain which gives better statistics. The absolute, and not relative values are given here, since the quantities do not differ significantly in magnitude for all three simulation. The figure demonstrates that the inversions behave similarly in all cases. The systematic errors are very similar for all three parameters. 

\begin{figure*}
  \centering
  \includegraphics[angle=90,width=0.355\linewidth ,trim=0.5cm 1cm 1.5cm 1.5cm,clip=true]{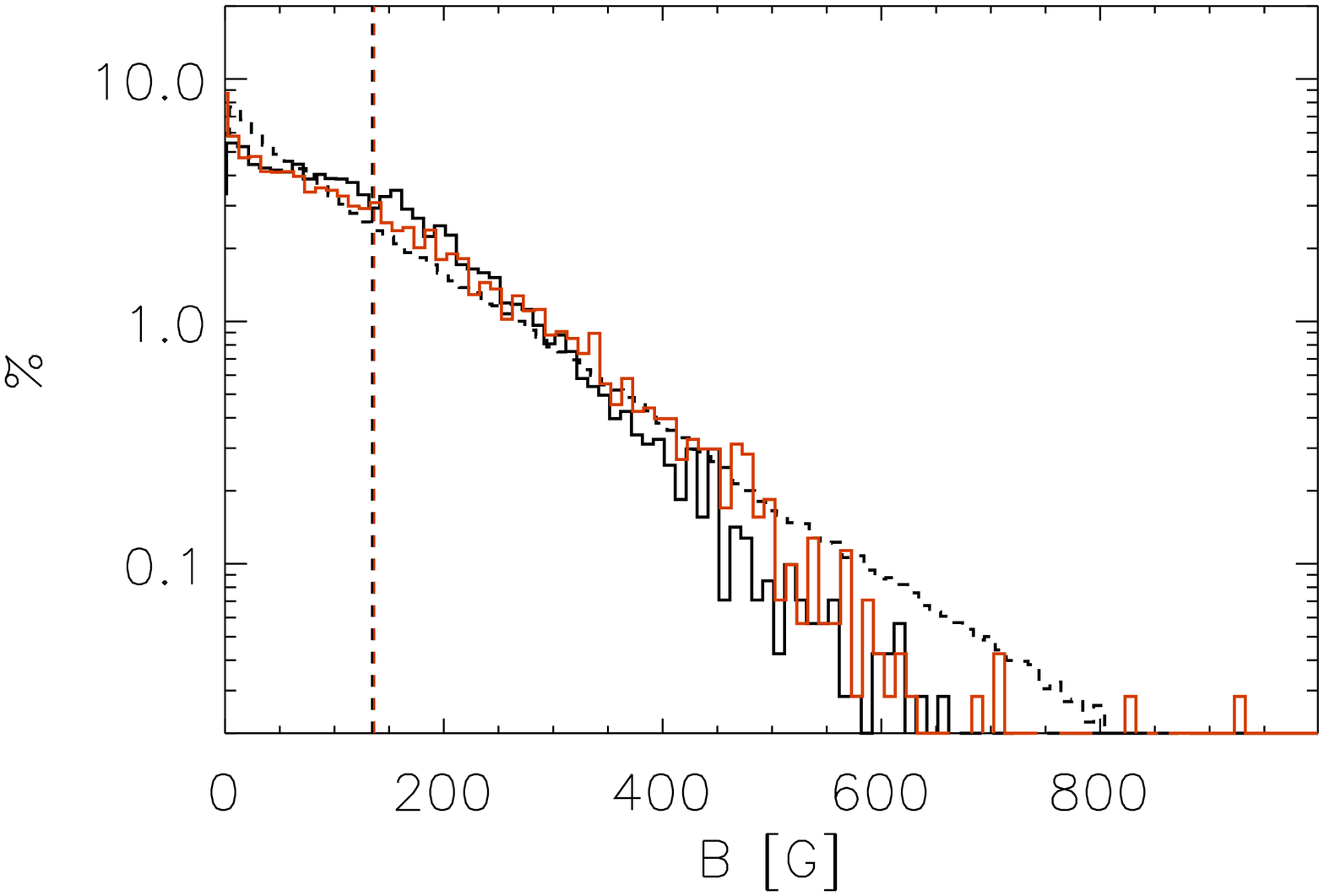}
  \includegraphics[angle=90,width=0.3\linewidth ,trim=0.5cm 1cm 1cm 5.2cm,clip=true]{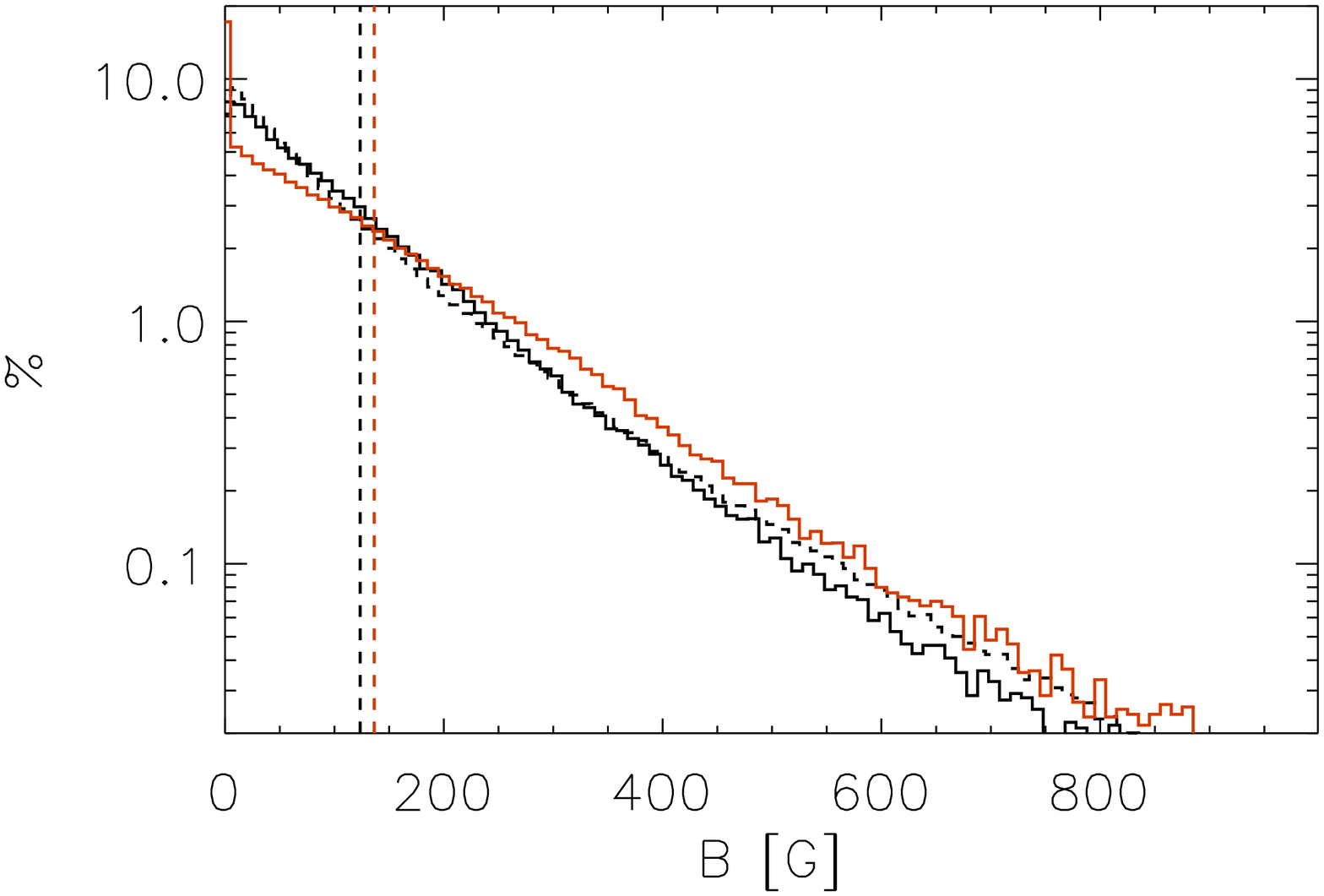}
  \includegraphics[angle=90,width=0.3\linewidth ,trim=0.5cm 1cm 1cm 5.2cm,clip=true]{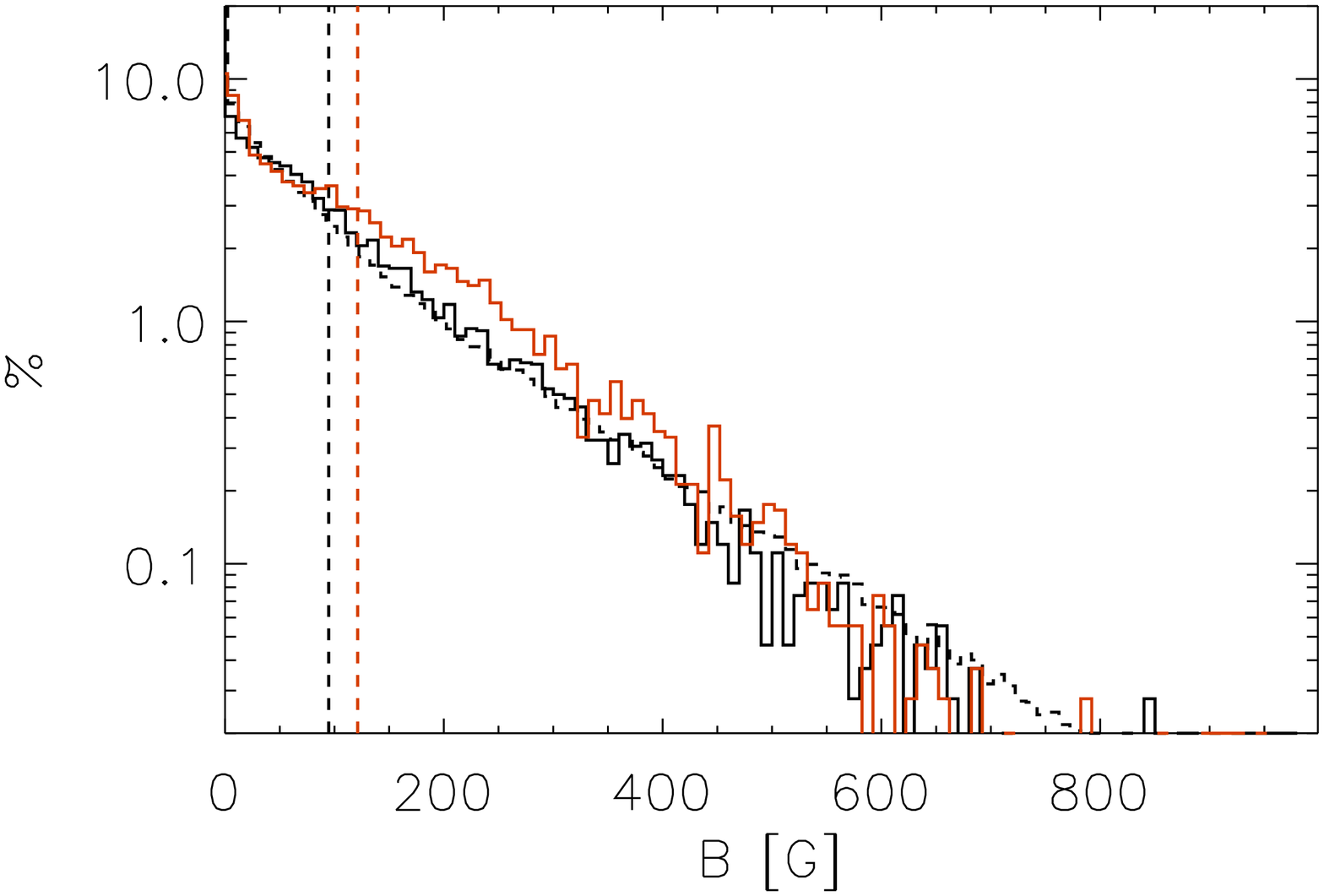}
  \includegraphics[angle=90,width=0.355\linewidth ,trim=0.5cm 1cm 1.5cm 1.5cm,clip=true]{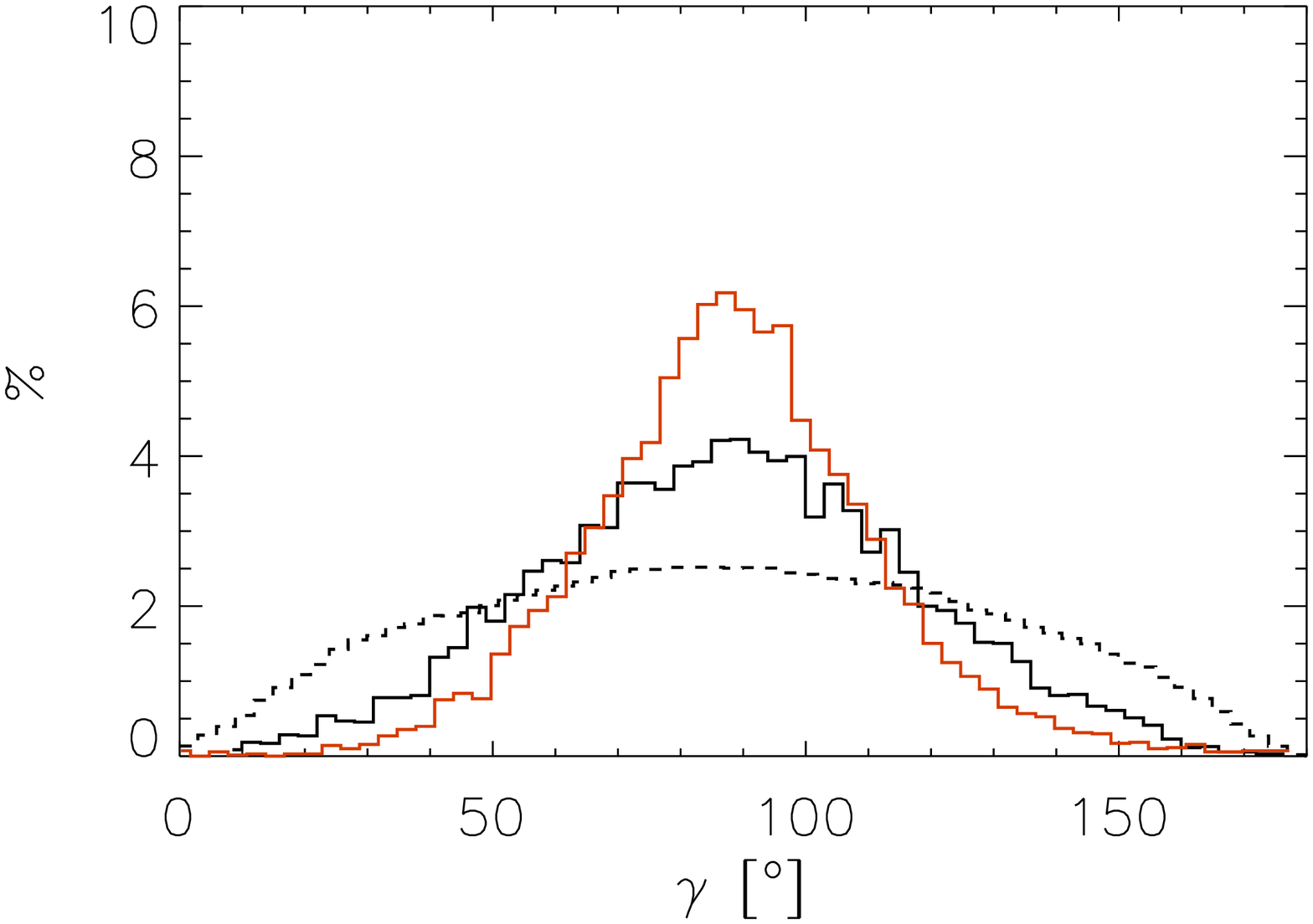}
  \includegraphics[angle=90,width=0.3\linewidth ,trim=0.5cm 1cm 1cm 5.2cm,clip=true]{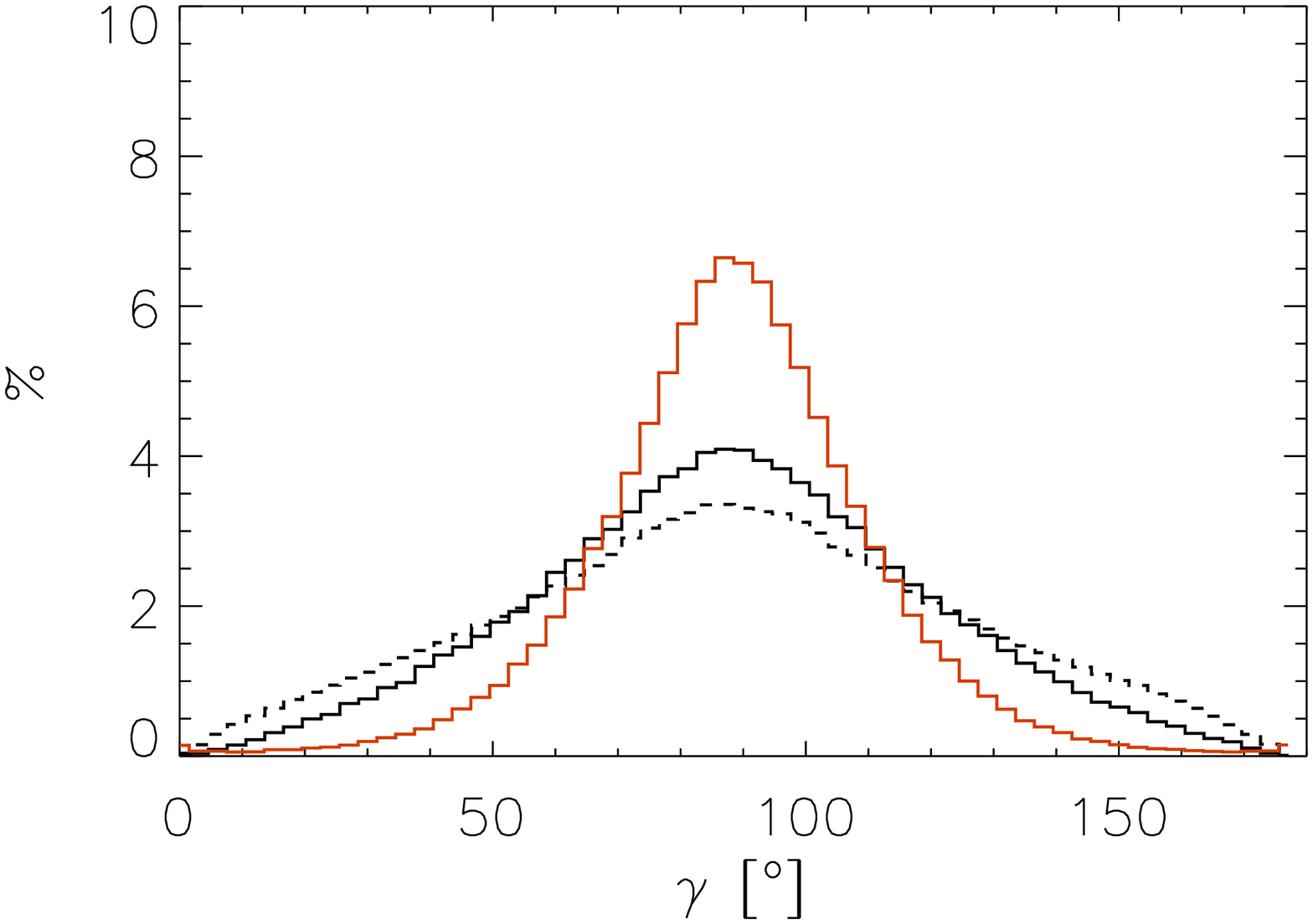}
  \includegraphics[angle=90,width=0.3\linewidth ,trim=0.5cm 1cm 1cm 5.2cm,clip=true]{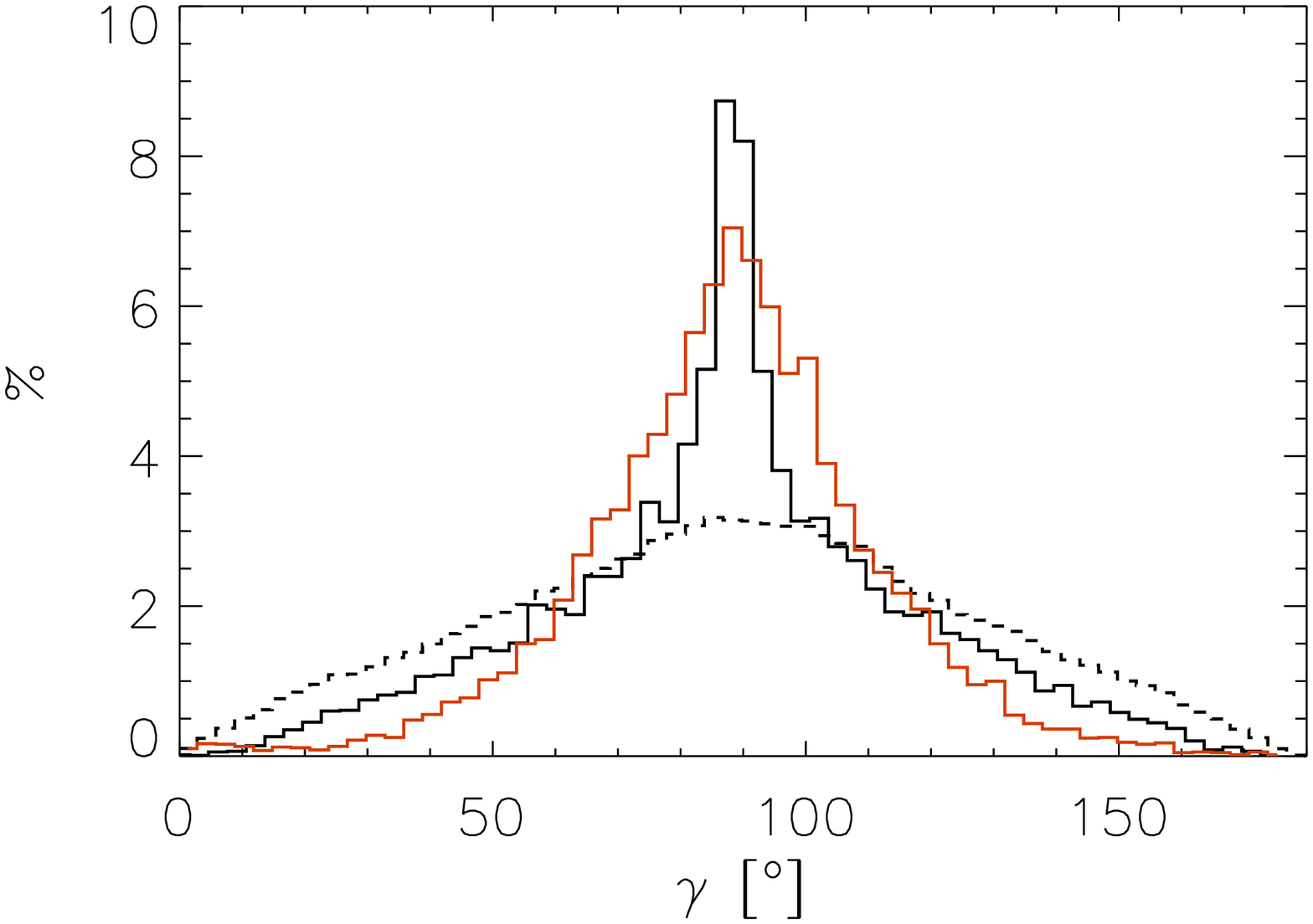}
  \caption{Results of 2D inversions applied to simulations - test on different simulations: Histograms of magnetic field strength (top row) and inclination (bottom row) at $\log  \tau =0$ for Sim~1,2,3 from left to right. Black dashed and solid lines mark the original distributions before and after the spatial smearing, respectively. Red line shows distributions retrieved with 2D inversions. Bin sizes are $10$~G and $3^{o}$. Vertical lines mark the corresponding mean field strengths at $\log  \tau =0$: from the original when the highest spatial frequencies are filtered out (black dashed line) and retrieved with inversions (red dashed line).}
\label{hist_diffsim}
\end{figure*}

What do the retrieved distributions look like? Figure~\ref{hist_diffsim} shows histograms of the original and retrieved magnetic field strength and inclination for all three snapshots. For clarity, we choose to give here only distributions at optical depth unity. This height is also the most interesting one, since the parameters change on the smallest scales. 

In the case of magnetic field strength, the retrieved distributions are very close to the original ones generated after the spatial filtering is applied and the distribution tail at the highest field strengths is cut off. At the smallest values, the peaks at $5-20$~G, visible in the original distributions, are lost after the 2D inversions because the inversions tend to set the field strength to 0 if no significant signal is found. Also, due to noise, the inversions tend to overestimate the field strength for a large number of pixels, which is most obvious in the case of Sim~2, where the sample is the largest (largest field of view). This will be further demonstrated later in Fig.~\ref{hist_diffnoise}.

Vertical lines in Figure~\ref{hist_diffsim} show the mean values for the corresponding field strengths. Retrieved numbers are fairly close to the original after the spatial filtering is applied. Because of the effects of noise, which tend to overestimate the weak field, the mean inverted values are slightly larger for all simulations. The same trend is present for all node positions. The deference is always the largest for Sim~3 where rapid emergence is simulated over whole simulation domain and the line-of-sight atmospheres are quite complicated with large gradients in the field vector and velocity.

Inspection of the inclinations shows that the spatial smearing also removes a significant number of pixels with inclined or completely vertical field. After the inversions, the distributions are additionally modified to show more horizontal field, especially in the regions where the field strength is very weak, as shown in Fig.~\ref{maps_sim2_bg}. The change in the distribution is somewhat less severe  for Sim~3 which does not show salt and paper pattern  visible in the local dynamo simulations. 

\begin{figure*}
  \centering
  \includegraphics[angle=90,width=0.45\linewidth ,trim=0cm 1cm 0.5cm 1.3cm,clip=true]{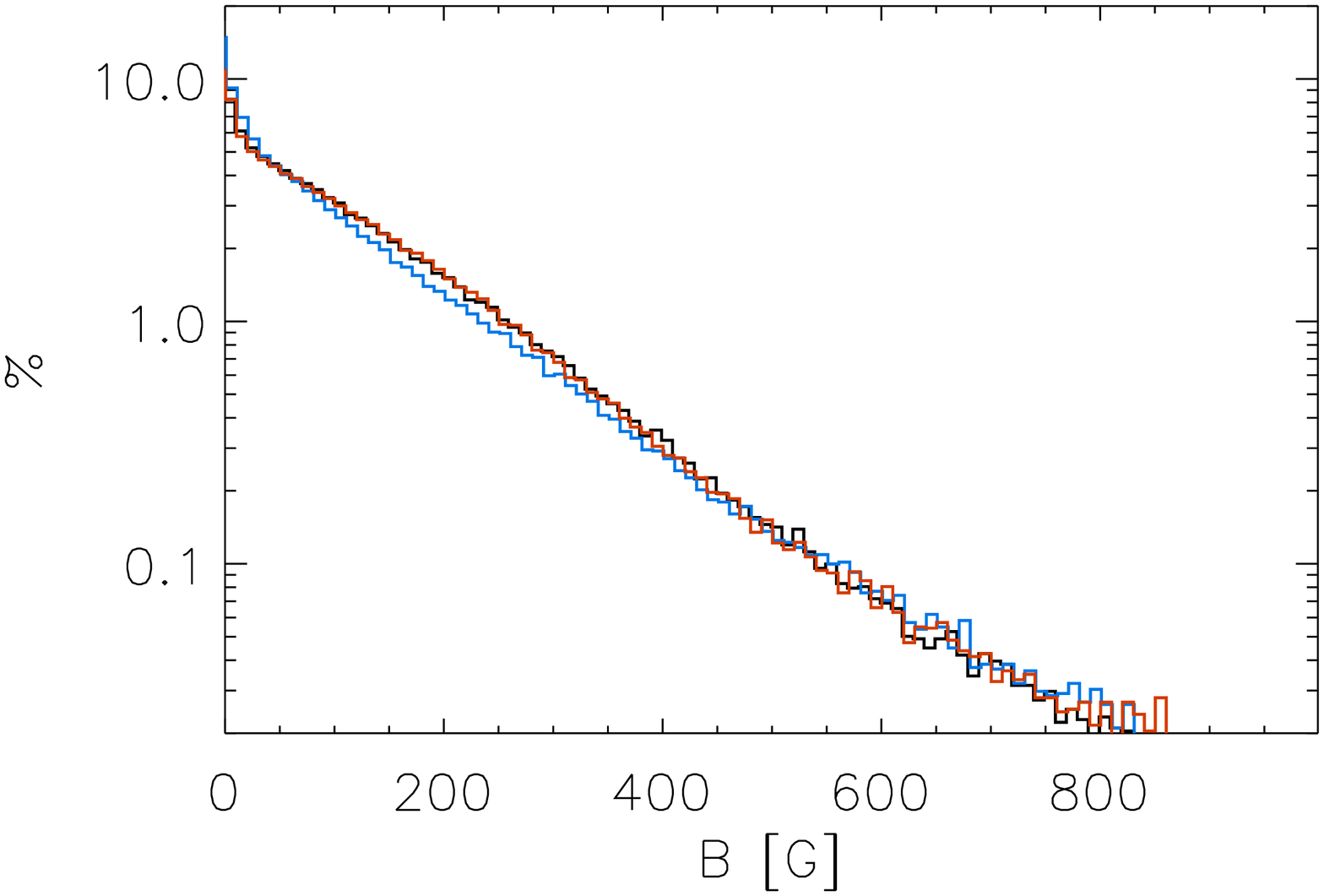}
  \includegraphics[angle=90,width=0.45\linewidth ,trim=0cm 1cm 0.5cm 1.3cm,clip=true]{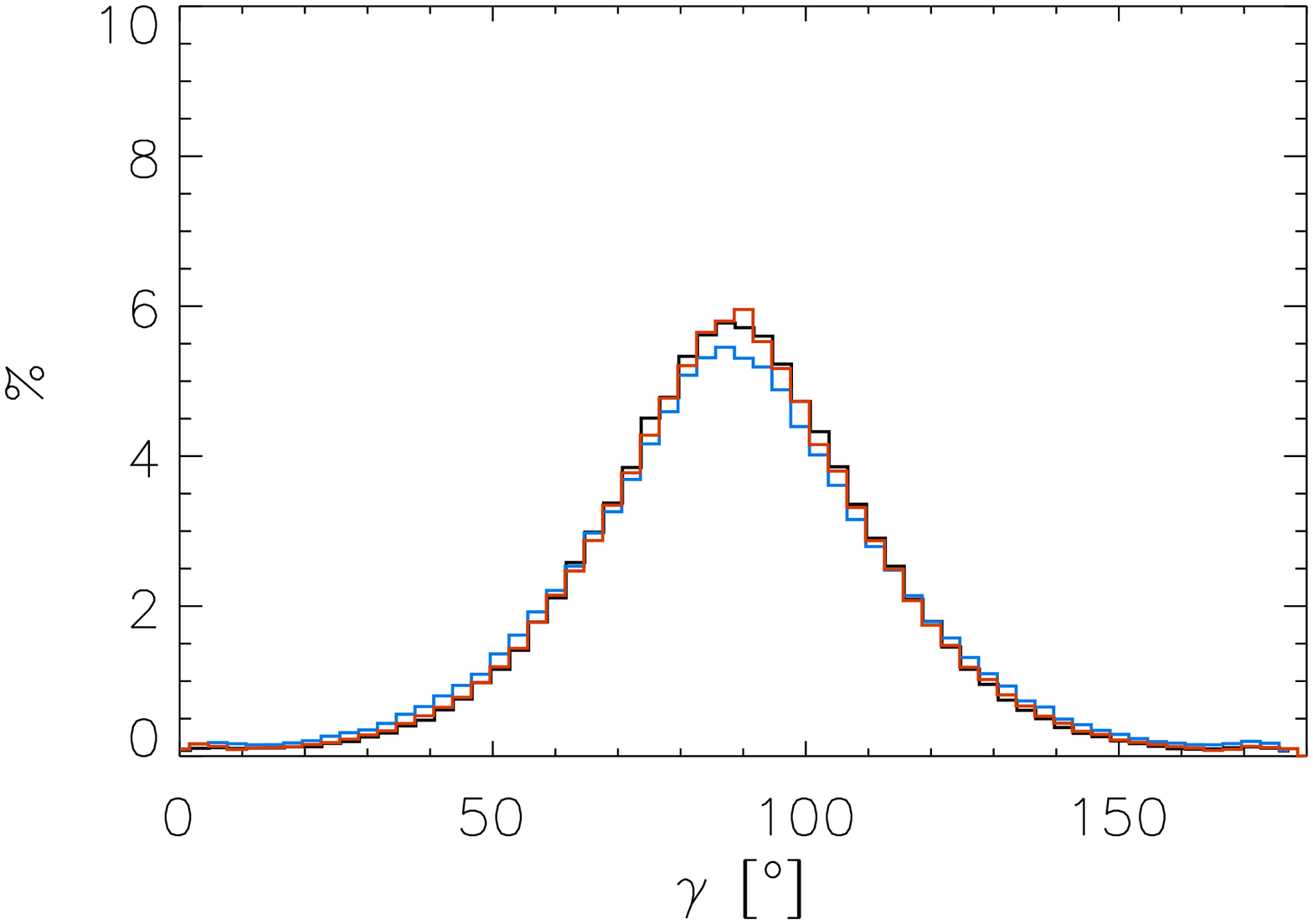}
  \caption{Results of 2D inversions applied to simulations - effect of used PSF: Histograms of magnetic field strength (left) and inclination (right) at $\log  \tau =0$ for Sim~2 when different PSFs are used in 2D inversions. Black line marks the distributions when the same PSF is used to smear and retrieve the parameters. Blue and red lines show what happens when PSF used by 2D inversions assumes too little or too much of the defocus, respectively. Bin sizes are $10$~G and $3^{o}$.}
\label{hist_diffpsf}
\end{figure*}

\begin{figure*}
  \centering
  \includegraphics[angle=90,width=0.45\linewidth ,trim=0cm 1cm 0.5cm 1.3cm,clip=true]{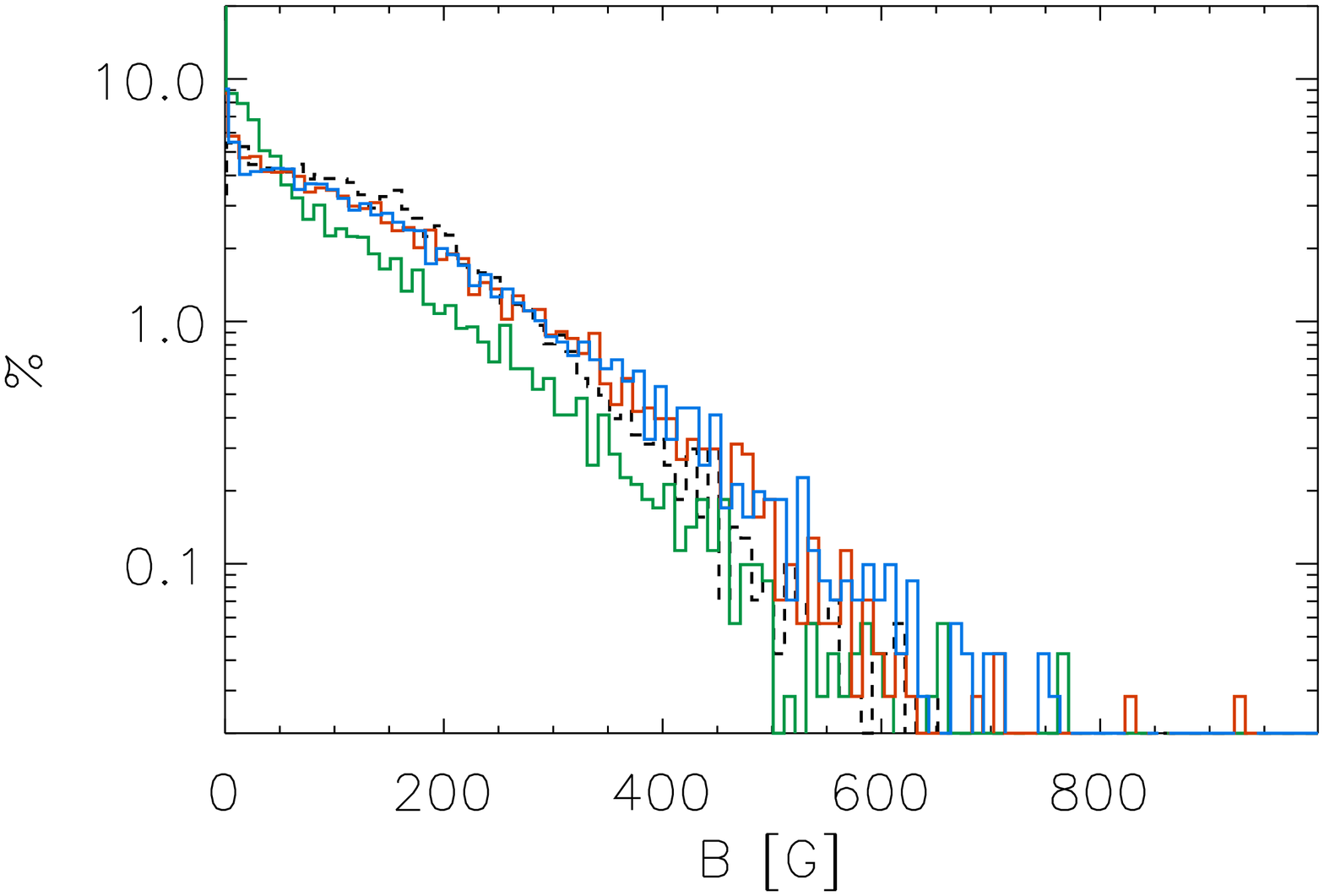}
  \includegraphics[angle=90,width=0.45\linewidth ,trim=0cm 1cm 0.5cm 1.3cm,clip=true]{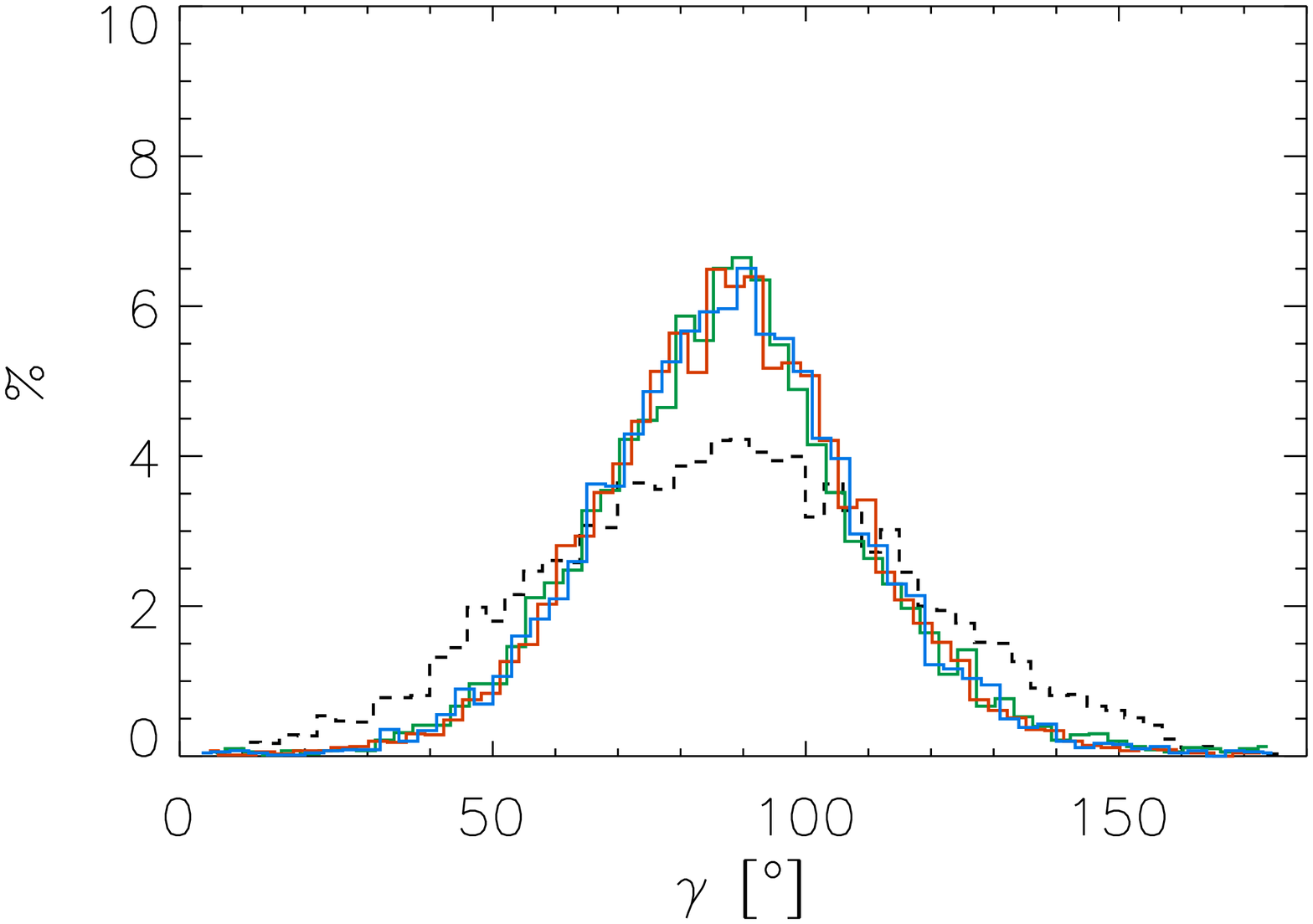}
  \caption{Results of 2D inversions applied to simulations - effects of different noise level: Histograms of magnetic field strength (left) and inclination (right) at $\log  \tau =0$ for Sim~1 retrieved with 2D inversions. Green, red, blue lines correspond to noise levels of $0$, $8\times10^{-4}$~I$_{c}$ and $1.2\times10^{-3}$~I$_{c}$, respectively. Black dashed lines mark the original distributions after the highest spatial frequencies were removed. Bin sizes are $10$~G and $3^{o}$.}
\label{hist_diffnoise}
\end{figure*}

\begin{figure*}
  \centering
  \includegraphics[angle=90,width=0.45\linewidth ,trim=0cm 1cm 0.5cm 1.3cm,clip=true]{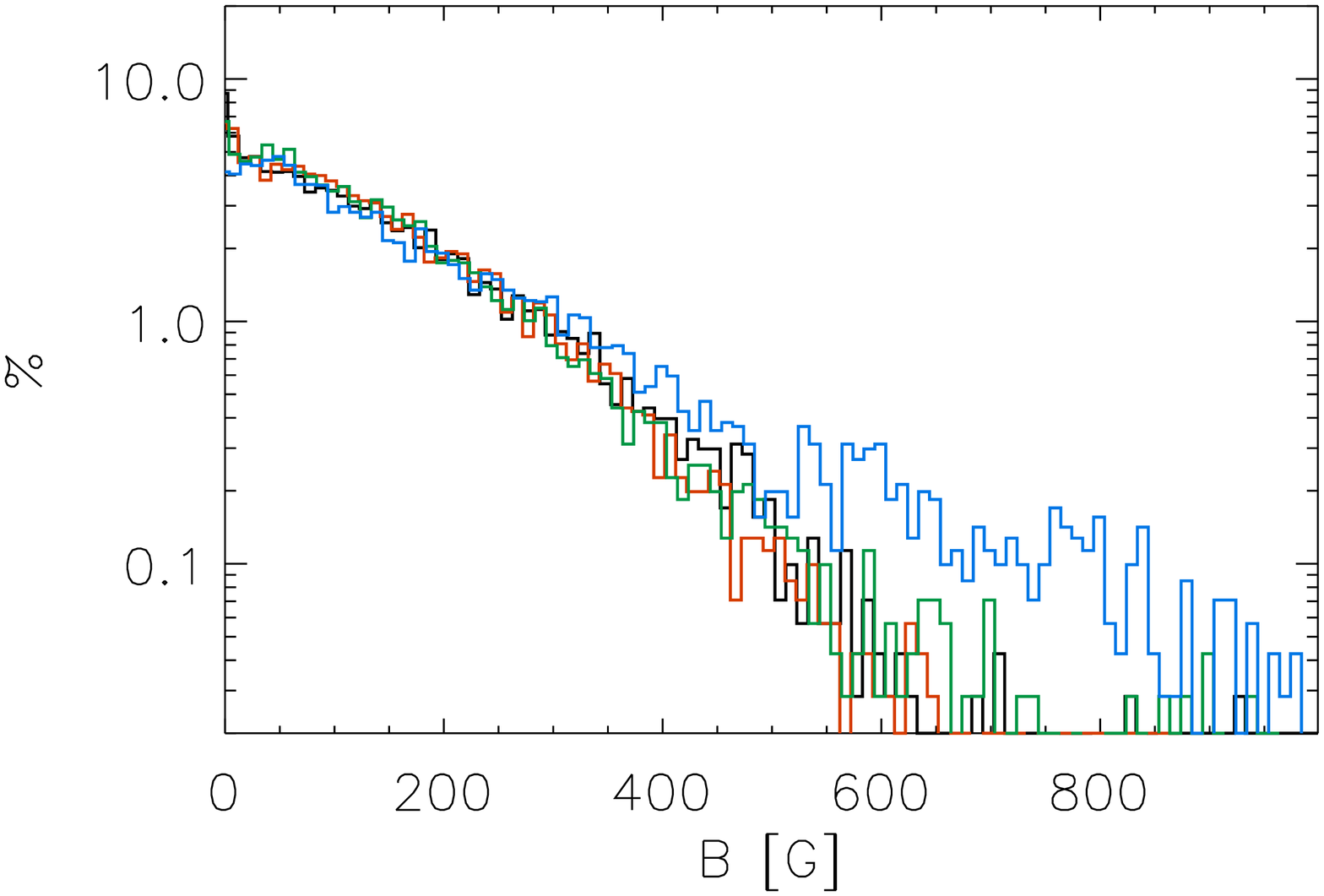}
  \includegraphics[angle=90,width=0.45\linewidth ,trim=0cm 1cm 0.5cm 1.3cm,clip=true]{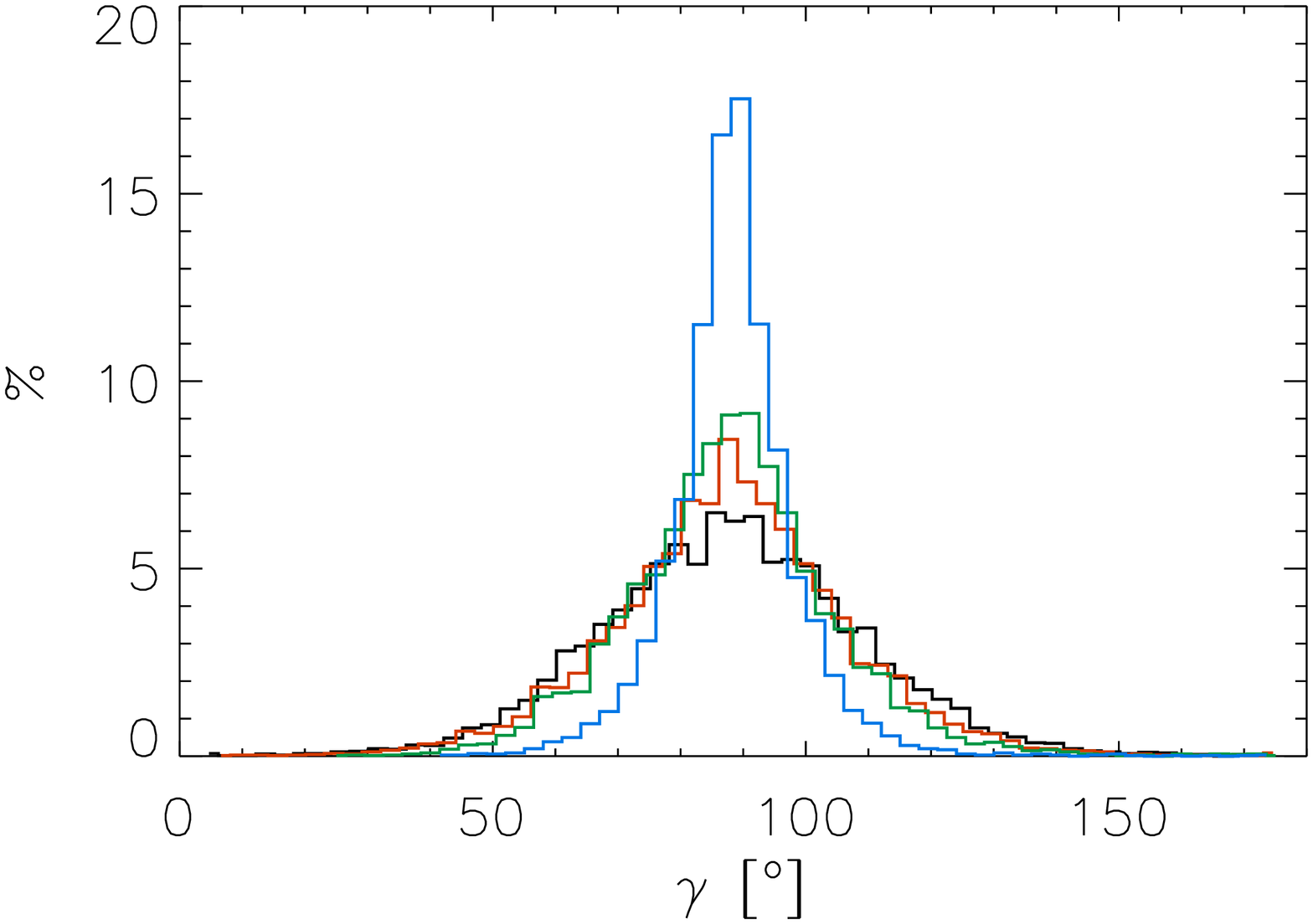}
  \caption{Results of 2D inversions applied to simulations - effect of temporal integration: Histograms of magnetic field strength (left) and inclination (right) at $\log  \tau =0$ for Sim~1 retrieved with 2D inversions. Black, red, green and blue lines correspond to cases where we intergrated over 0~s, 20~s, 2~min and 9~min, respectively. Bin sizes are $10$~G and $3^{o}$.}
\label{hist_diffevol}
\end{figure*}

\subsection{Choice of PSF}

Due to the Hinode/SOT design, the Hinode/SP observations were often taken with a significant departure from the best focus for the instrument \citep{lites2011,David2013}. The combination of this somewhat random shift with the long-term drift of the best focus position makes the determination of the precise focus departure troublesome. This is why we test here how much the error in the defocus of the PSF used in our inversion, effects the final result. We use three PSFs with different amount of defocus. The first one, where defocus of 7 focus steps is included \citep{danilovic08}, is also used for degrading the synthetic images. In the other two PSFs, the defocus is underestimated or overestimated by 4 steps.  

The test shows that the disagreement between the cases is hardly visible, so we omit the figure with the parameter differences as a function of optical depth. The systematic error in the velocity is slightly higher for the PSF where the defocus is underestimated, but the mean error, even then, is not larger than $30$~m/s. When the PSF with too little defocus is used, the mean field strength at the solar surface is underestimated by less than 10 G. Too much defocus on the other hand overestimates the field by only a few G more than the 'right' PSF. While the systematic error does not go over $10$~G when the proper PSF is used, too much defocus increases the mean error to $20$~G in the upper atmosphere.

Figure~\ref{hist_diffpsf} shows how small the differences in the final inverted distributions really are. The difference is the largest in the weak field part of the field strength distribution. The PSF with underestimated amount of defocus tends to also underestimate the contribution of the pixels with a weak field. On the other hand, assuming that defocus was larger than it really was will slightly increase the number of pixels with weak, horizontal field. But, all in all, the impact is statistically negligible.
    
\subsection{Noise level}

High-resolution Hinode/SP scans are taken also in deep magnetogram mode which shifts the level of noise from  $1.2\times10^{-3}$~I$_{c}$ (for normal mode) to $8\times10^{-4}$~I$_{c}$. To check if this brings any improvement, we simulate both cases. Figure~\ref{hist_diffnoise} shows the resulting distributions of magnetic field strength and inclination retrieved with 2D inversions. There seems to be no significant difference. The result stays almost the same even if no noise is included. The figure suggests that impact of noise is in redistributing the pixels with very weak field to hG field bins and generally making the field stronger. The distribution of inclination also does not benefit from lack of noise. The problem is intrinsically in a simple model we assume which tries to match very asymmetric Stokes profiles produced by complex MHD atmospheres.

\subsection{Solar photospheric evolution}

For all the previous tests, we used one snapshot to simulate the observations taken with a finite exposure time. Here, we test how much the results change if we take into account the actual evolution of magnetic features during this time. Because the simulations where emergence takes place would lead to an amount of horizontal field that increases with time, we take Sim~1, where the flux stays almost constant. Simulating a proper Hinode/SP scan requires a large set of snapshots, produced with very high cadence. Given that these simulation cover only $6\arcsec \times 6\arcsec$ that Hinode/SP scans in only 3 min in the normal mode, we find that integrating for every pixel over the whole box would be close enough approximation that would produce similar effect as averaging for every slit position separately. We choose to integrate over 20~s, 2~min and 9~min. Every snapshot within this period is then treated the same. Spatial smearing and adding the noise of  $8\times10^{-4}$~I$_{c}$ is applied first to each of them and then temporal integration of the resulting signal is performed for every rebinned pixel and every wavelength point separately. In this way, the intensity contrast is reduced by 16~\% and apparent longitudinal flux density decreased by 37~\%.

Figure~\ref{hist_diffevol} shows the resulting distributions of the magnetic field strength and inclination. An obvious trend is that the contribution of the more inclined field is larger as we integrate longer. The difference is not great for integration up to 2~min, especially for the field strength, but it becomes significant if we integrate longer. If we integrate over 9~min, the distribution of field inclination becomes very narrow with most pixels harbouring horizontal field, while the retrieved mean field strength at $\log  \tau =0$ increases by $34$~\%.

\begin{figure*}
  \centering
  \includegraphics[angle=90,width=\linewidth ,trim=1.7cm 0.8cm 1.75cm 1.5cm,clip=true]{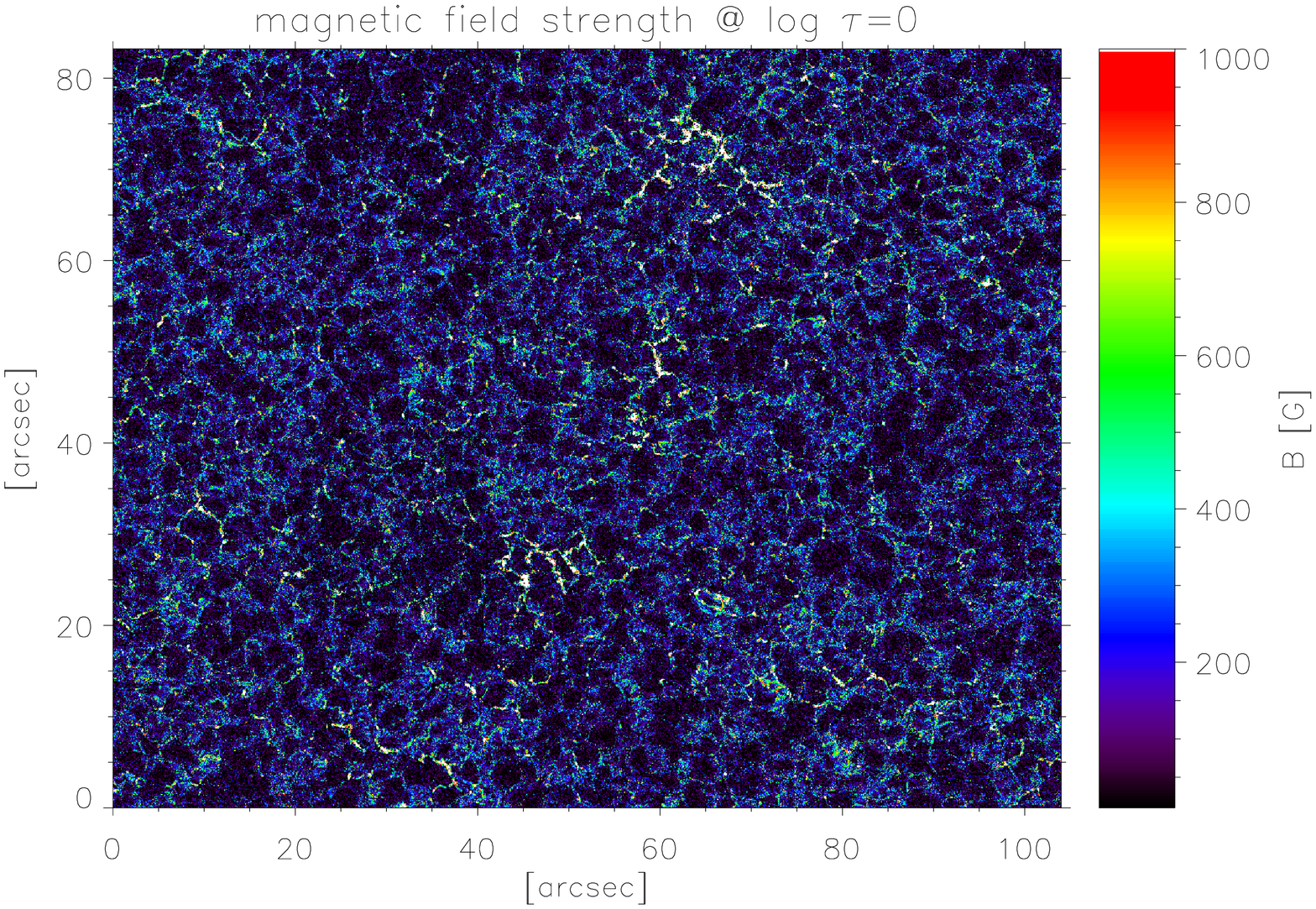}
  \includegraphics[angle=90,width=\linewidth ,trim=0.3cm 0.8cm 1.75cm 1.5cm,clip=true]{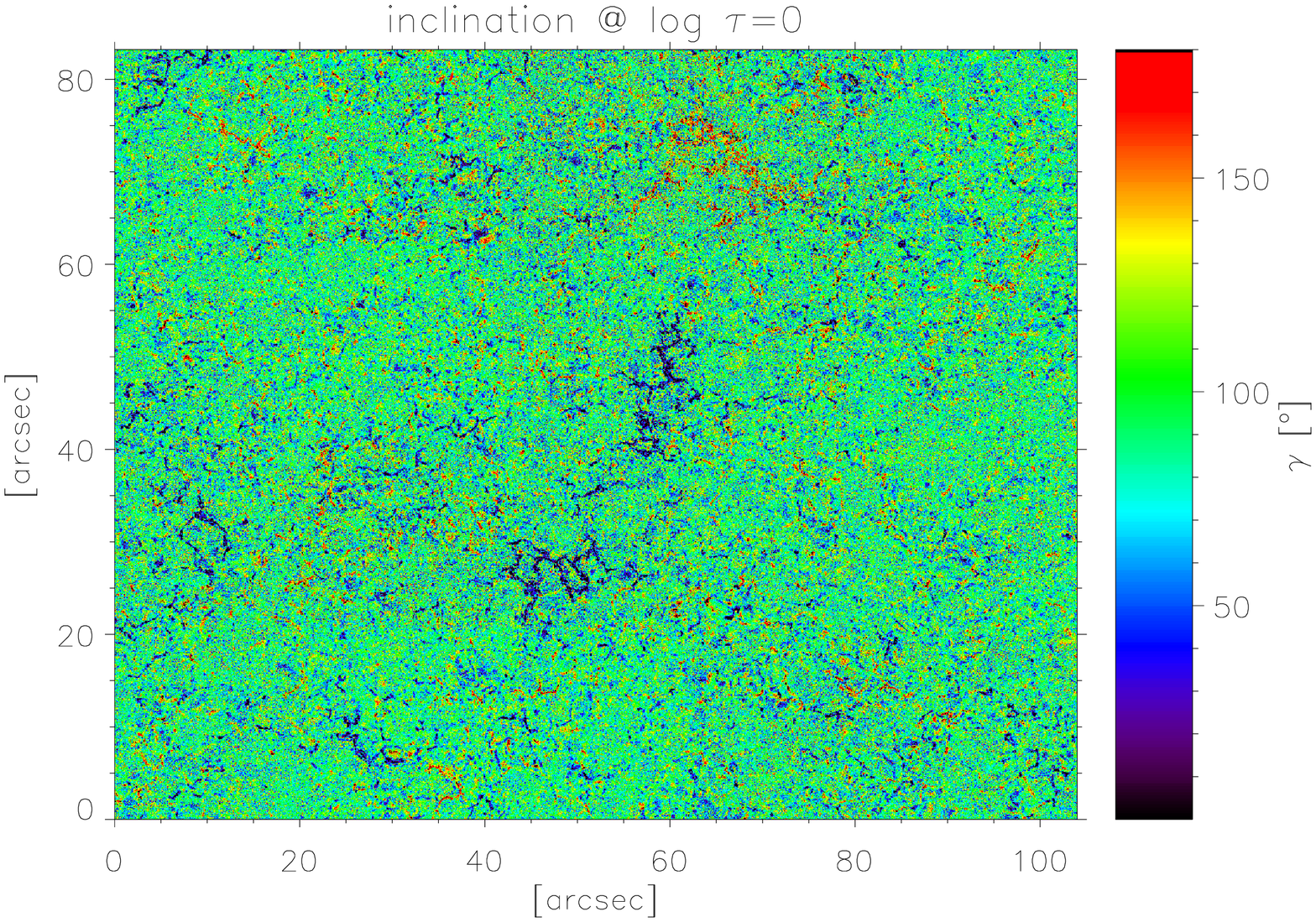}
  \caption{Results of 2D inversions applied to Hinode/SP observations: maps of magnetic field strength (top) and inclination (bottom) at $\log  \tau =0$. }
\label{maps_full_fov}
\end{figure*}

\begin{figure*}
  \centering
  \includegraphics[angle=90,width=0.37\linewidth ,trim=3cm 2cm 1.75cm 1cm,clip=true]{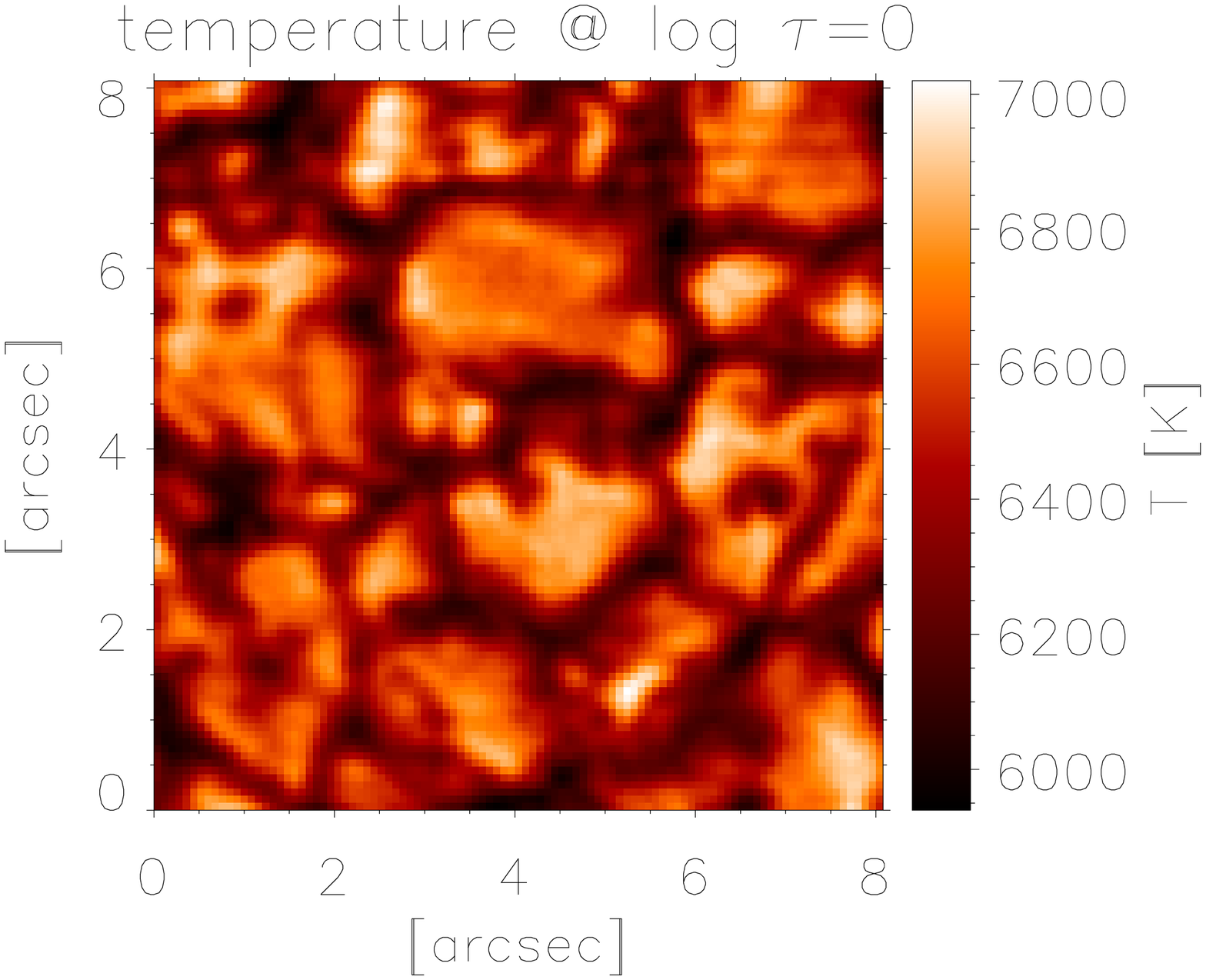}
  \includegraphics[angle=90,width=0.3\linewidth ,trim=3cm 2.2cm 1.75cm 5.1cm,clip=true]{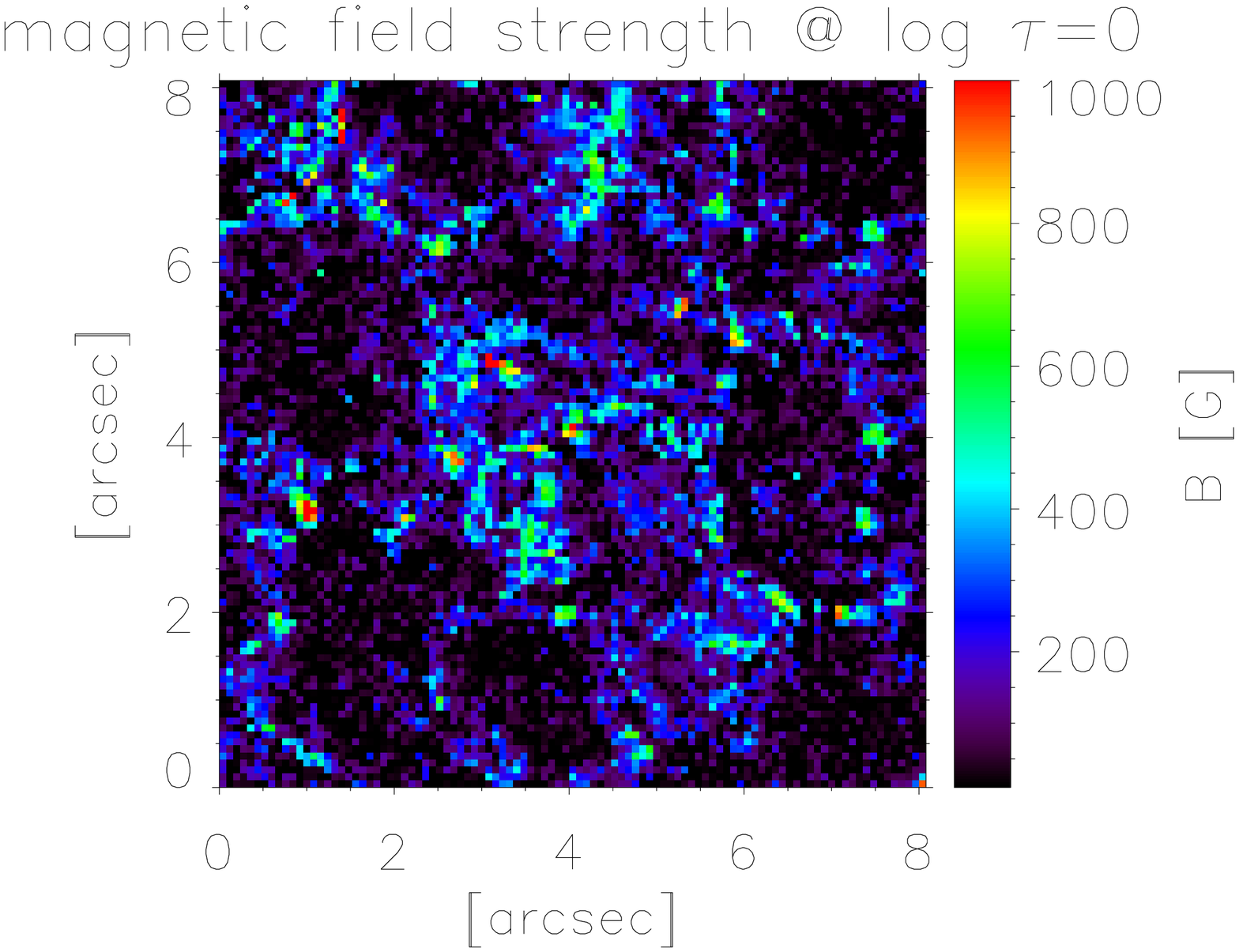}
  \includegraphics[angle=90,width=0.3\linewidth ,trim=3cm 2cm 1.75cm 5.1cm,clip=true]{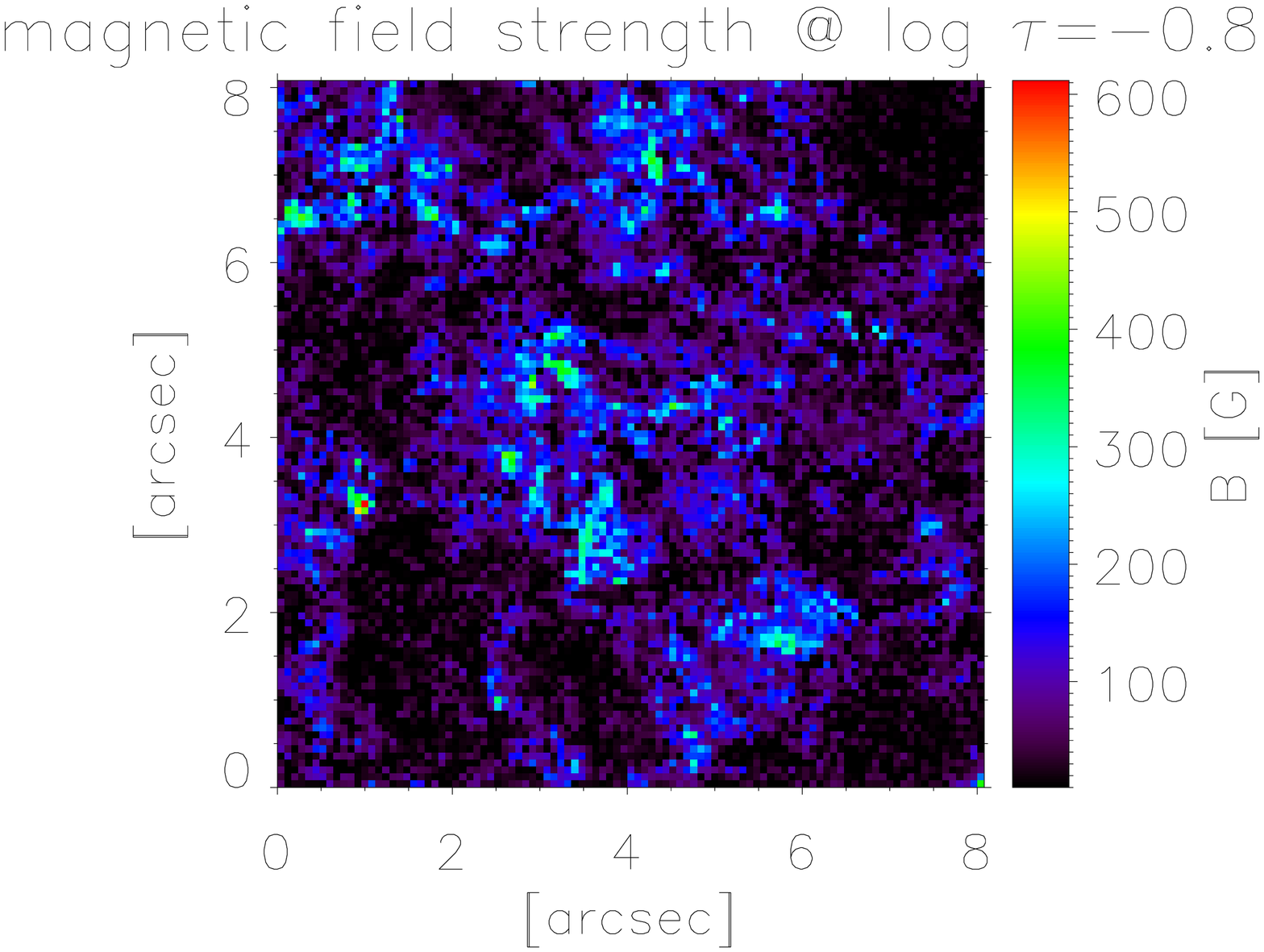}
  \includegraphics[angle=90,width=0.37\linewidth ,trim=0.8cm 2cm 1.75cm 1cm,clip=true]{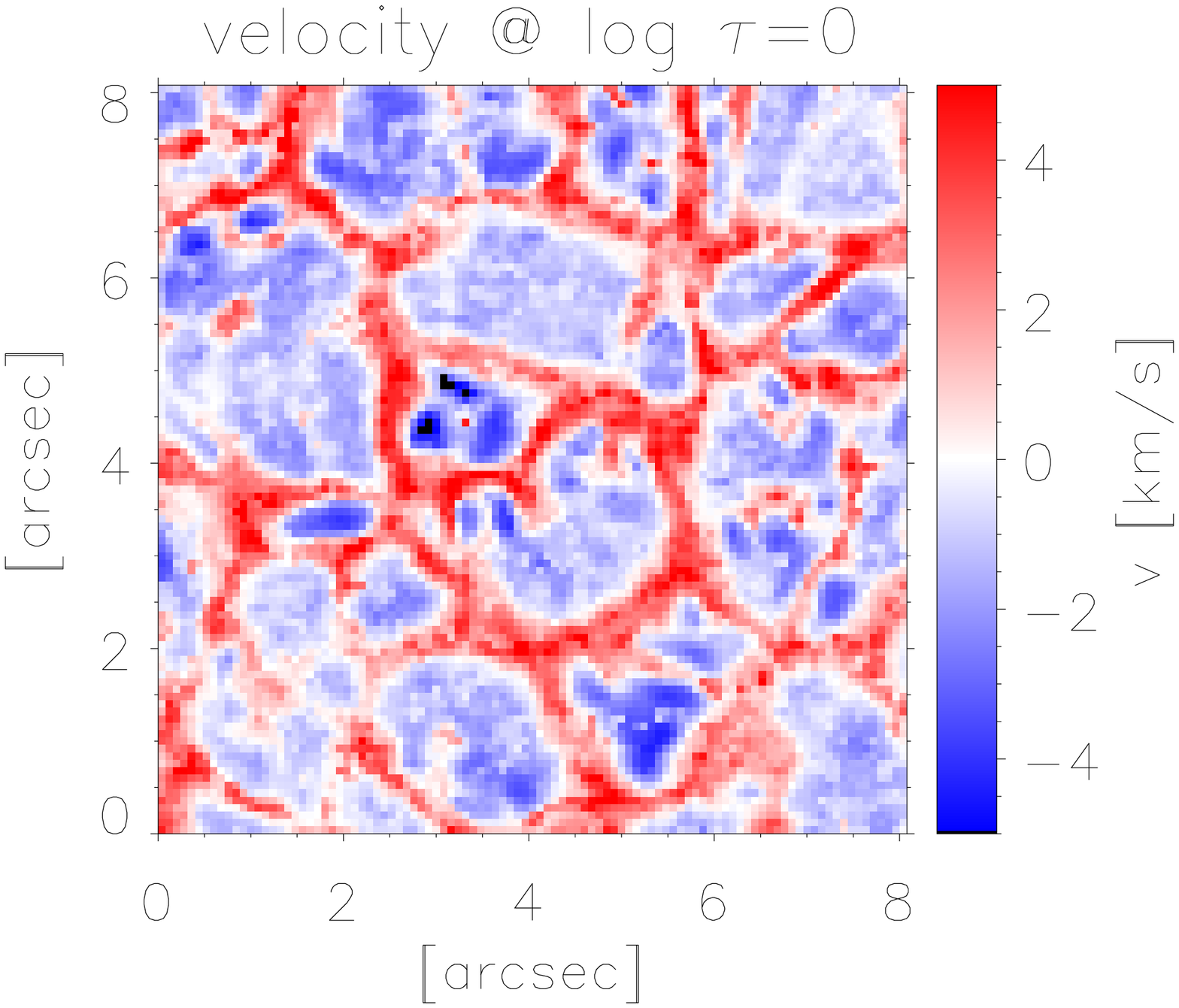}
  \includegraphics[angle=90,width=0.3\linewidth ,trim=0.8cm 2.2cm 1.75cm 5.1cm,clip=true]{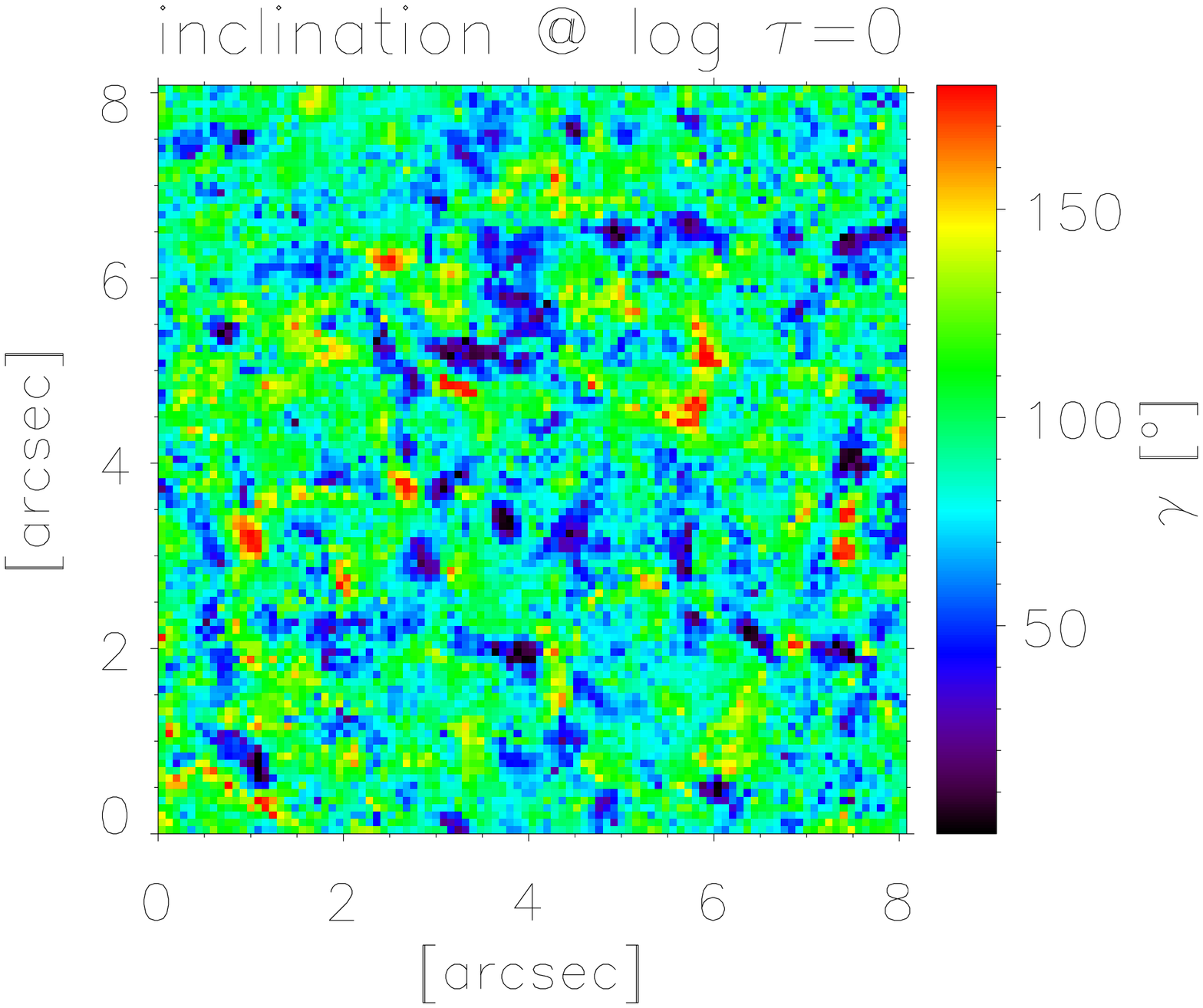}
  \includegraphics[angle=90,width=0.3\linewidth ,trim=0.8cm 2cm 1.75cm 5.1cm,clip=true]{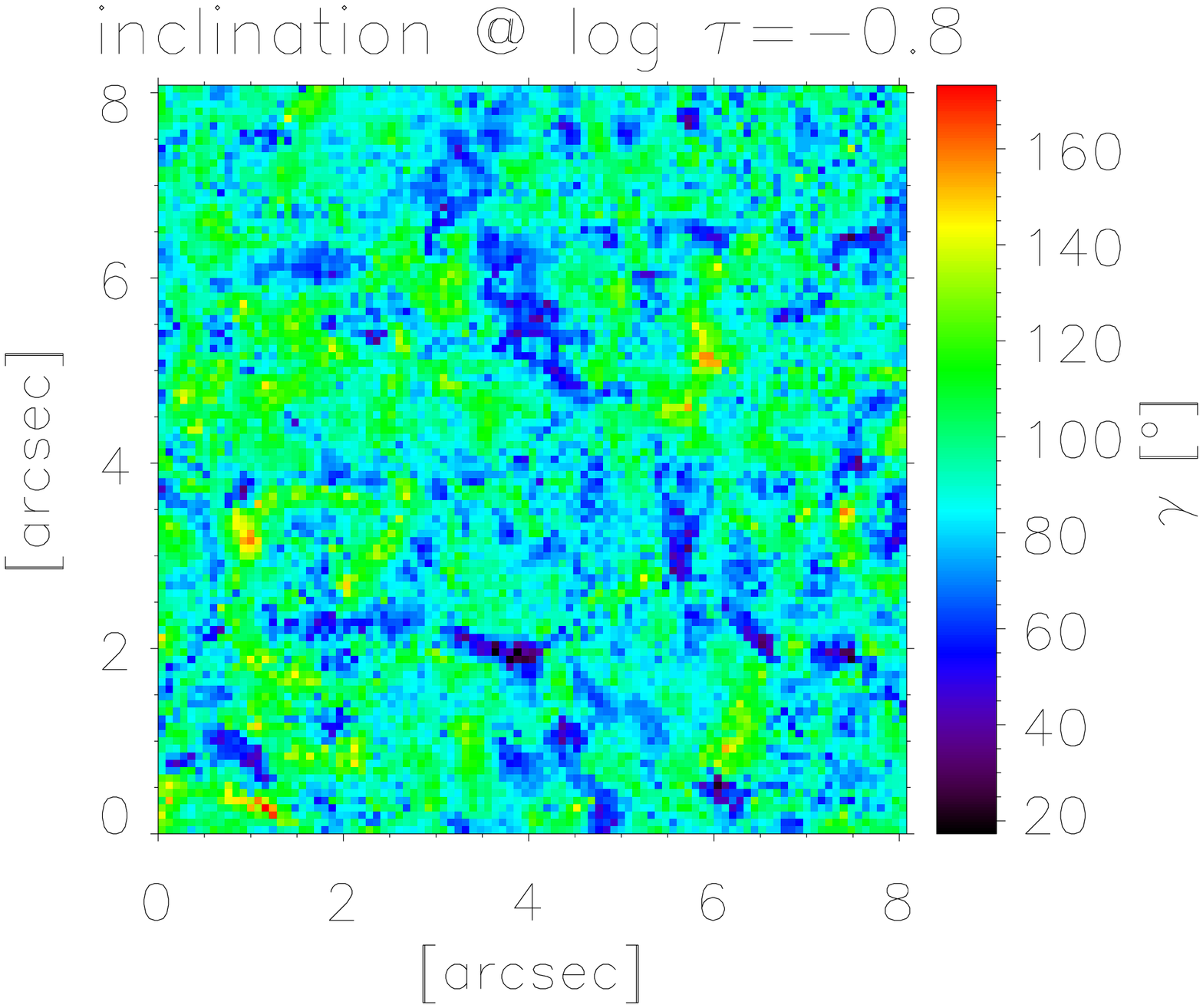}
  \caption{Results of 2D inversions applied to Hinode/SP observations: zoom in to a small field of view (see the text). \textit{Top row, from left to right}: temperature  at $\log  \tau =0$ and magnetic field strength at $\log  \tau =0$ and $-0.8$. \textit{Bottom row, from left to right}: line of sight velocity at $\log  \tau =0$ and  magnetic field inclination at $\log  \tau =0$ and $-0.8$. }
\label{maps_small_fov}
\end{figure*}

\section{Inversions of observed Hinode/SP maps}

Previous tests convinced us that the combination of node positions at $\log  \tau =0,-0.8$ and $-2.0$ gives statistically good results, irrespective of the field configurations, so we use it for inverting real solar observations. Furthermore, we found out that choosing a PSF with a slightly different amount of defocus does not make a large difference so we use the PSF where a defocus of 7 focus steps is included. In order to compare our results to the results of others, we take a standard scan made in normal mode. 

Figure~\ref{maps_full_fov} shows an inverted part of the normal mode scan used in various studies starting with \citep{lites08} and \citep{david07a,david07b}. Comparison with these studies confirms that the 2D inversions give inverted maps that are similar to the ones retrieved with other codes, but show much more detail. The network features are finer and the voids containing almost no field are smaller - not bigger than $5$\arcsec. 

In Fig.~\ref{maps_small_fov}, we zoom in on the field of view shown in Fig. 2 of \cite{david07b} to demonstrate what our code gives on the smallest scales. Magnetic field strength and inclination at $\log  \tau =-0.8$ is similar to the maps in \cite{david07b} and \cite{asensio2009}. Fields are predominantly weak with a few stronger features in the intergranular lanes that are still hG. Our maps at $\log  \tau =0$, on the other hand, reveal many more features of which some harbour kG field. The inclination map, at the same time, reveals densely packed mixed polarities over the whole map. Generally, oversampling to pixels size of 0.08\arcsec, in case of our 2D inversions, reveals much more fine structure in all the maps.    

\begin{figure*}
  \centering
  \includegraphics[angle=90,width=0.45\linewidth]{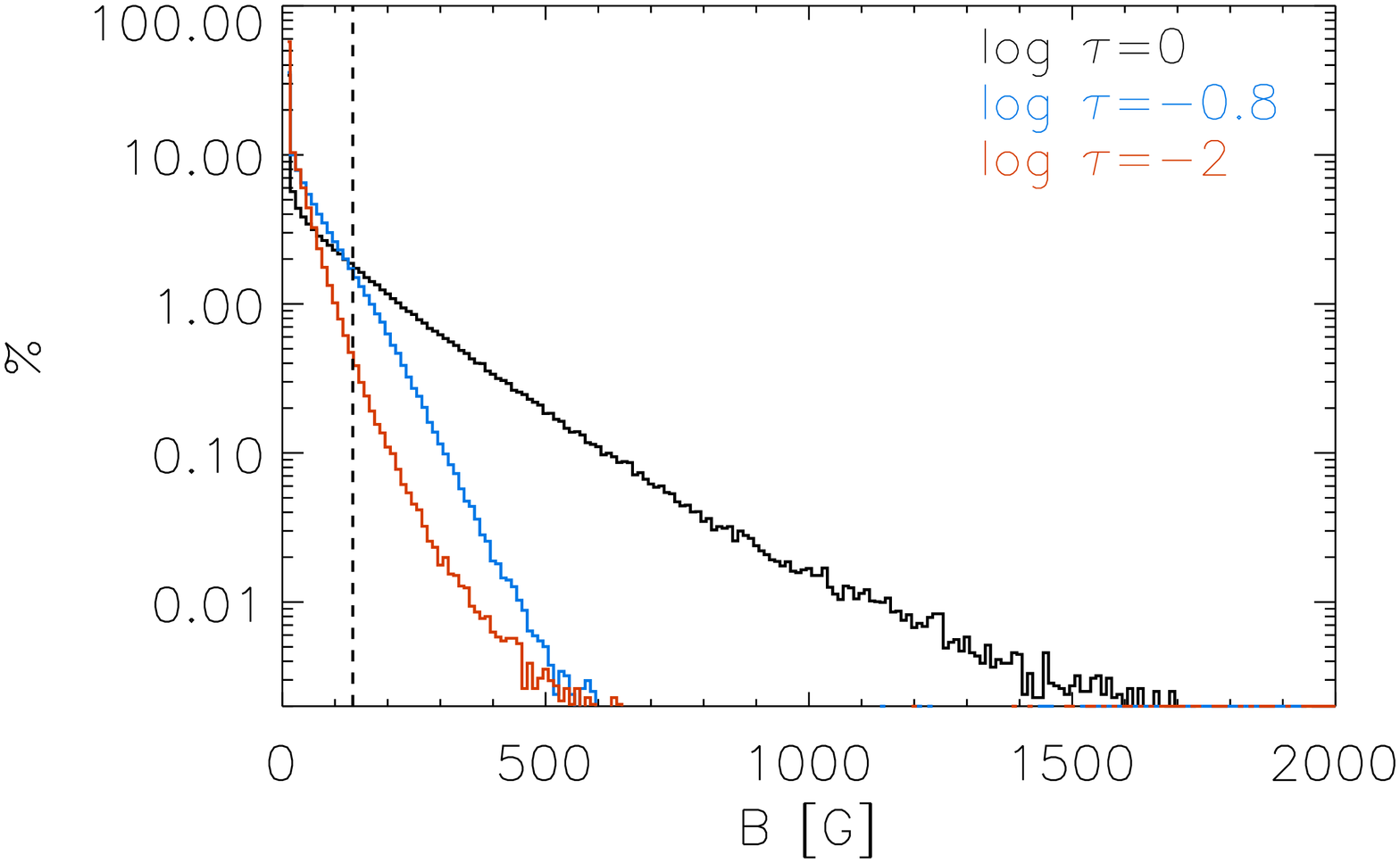}
   \includegraphics[angle=90,width=0.45\linewidth]{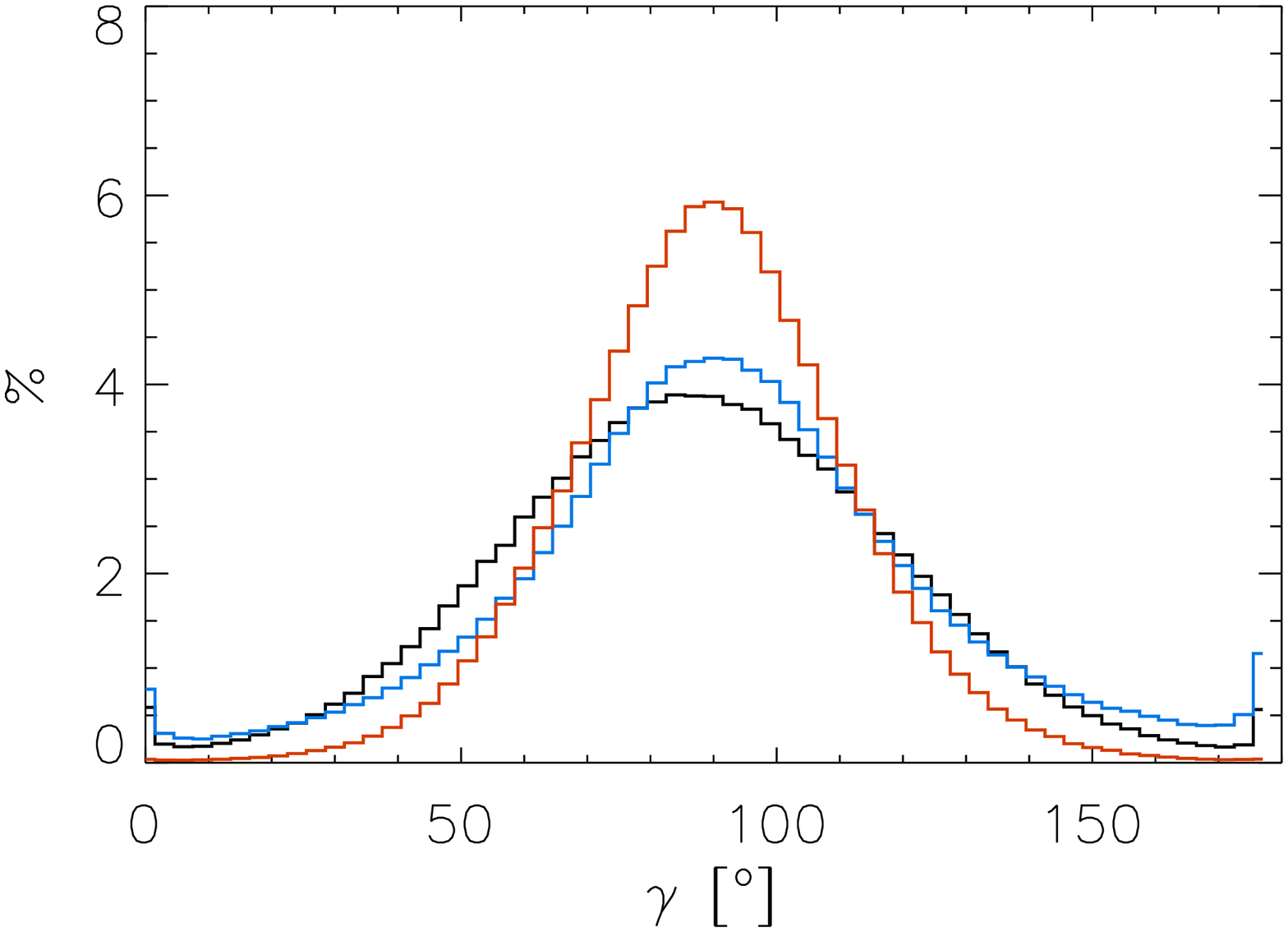} 

  \caption{Results of 2D inversions applied to Hinode/SP observations: histograms of magnetic field strength and field inclinations at 3 nodes. Black, blue and red denote heights of $\log  \tau =0, -0.8$ and $-2.0$ respectively. Bin sizes are $10$~G and $3^{o}$. Vertical line marks the position of the mean field strength at $\log  \tau =0$.}
\label{hist_obs_normal}
\end{figure*}

Figure~\ref{hist_obs_normal} shows final distributions of magnetic field strength and inclination of internetwork. We filtered  out the network contribution by following the procedure from \cite{lites2011}. All the regions where longitudinal apparent flux density exceeds $100$~G together with its 2\arcsec\ surroundings are excluded from the statistics. The retrieved mean field strength at $\log  \tau =0$ is around $130$~G. At the higher layers, the field strength drops as the field becomes more horizontal. 

\section{Conclusions and discussion}

An inversion code that self-consistently accounts for the effects that the instrumental PSF has on data, was applied, for the first time, to quiet Sun observations. Extensive testing on MHD simulations confirmed that the inverted results make sense and also provided an inversion strategy which we then applied to the solar observations. 

In studies like the one presented here, typically only pixels are selected that contain significant signals in Q,U and/or V, to ensure that the result returned by the inversion code used to recover the magnetic field strength is reliable. However, this selection disregards signals that in individual pixels are below the noise level, but averaged over a large number of pixels are statistically significant. This not only results in loss of signal, it may also introduce a bias in the results towards the properties that are particular of strong magnetic fields only. The spatially coupled inversion method used here is able to constrain a result using such signals, and can therefore make use of {\em all} pixels in the FOV.

The inversions return a distribution with mainly weak field, without any secondary peak at kG field strengths, as in \cite{stenflo10} and \cite{lites2011}. It shows no peak either at hG values as detected by \cite{david07a}, but monotonously increases towards the smallest field strengths. This however does not exclude the possibility that real distribution of the field strength does not have a peak at $5-10$~G as local dynamo simulations show. The tests on the simulations demonstrate that the code tends to set the field strength to zero when signals are too weak. The code also tends to retrieve mostly horizontal field in the regions which harbour very weak field. This will then produce large differences between the original and retrieved distribution of the field inclinations in the case of local dynamo simulations which show salt and pepper pattern even in these regions.

Due to the noise, the code tends to overestimate the hG field which results in a slight overestimation of the mean field strength. Nevertheless the retrieved value comes close to the original. The mean field strength $>100$~G at optical depth unity retrieved from the observations, puts ours results closer to the results based on the Hanle effect \citep{trujillo04}, although we cannot confirm that the field strength in the upper photosphere is also over 100~G, as found by \cite{Shchukina2011}. 

A mean field strength at the solar surface of the same magnitude was also retrieved by \cite{david2012a} and \cite{Luis2012}. Their results, however, show much more horizontal field. Our tests show that this can be explained as an artefact, produced by prolonged temporal averaging, since increasing the integration time beyond the evolution time scale of the solar scene (2-3~min.), artificially increases the apparent contribution of the horizontal fields significantly. In the case that the distribution of magnetic field is isotropic, this can be easily understood, since in that case integration beyond the evolution timescale will superpose a statistically independent realization of the magnetic field distribution, which will decrease the measured polarimetric signal at the same rate as the photon noise, so that no nett improvement of the S/N ratio can be obtained by continued integration. The scaling properties of the noise equivalent horizontal and magnetic fields, however, continue to favour a more and more horizontally inclined field configuration. At the same time, the granular motions produce wider line profiles which results in larger retrieved field strengths.  

The distribution of the field inclination has a maximum at $90^{o}$, which confirms the results given by \cite{david07b} and \cite{lites08}. However, we cannot interpret this as evidence for a predominantly horizontal field, nor can we state that this is in agreement with the results obtained by \cite{asensio2014}, who claim that the distribution is quasi-isotropic. As shown in Fig.~\ref{hist_diffsim}, similar results can be obtained from both predominantly horizontal and a quasi-isotropic distribution of the magnetic field. Re-examining the lower left panel of Fig.~\ref{hist_diffsim}, one can conclude that even if the photospheric magnetic field is isotropic, it might not be possible to recover it as such. We are currently limited, not only by our inversion tools, but also by the resolution limit of our instruments. Tests of the 2D inversion technique on the simulated Hinode/SP data at the disc center suggest that the information for discerning between the two distributions is just not there. 


\begin{acknowledgements}
We thank M. Sch\"ussler for snapshots of local dynamo simulations (Sim~1). The National Center for Atmospheric Research (NCAR) is sponsored by the National Science Foundation. We would like to acknowledge high-performance computing support from Yellowstone (http://n2t.net/ark:/85065/d7wd3xhc) provided by NCAR's Computational and Information Systems Laboratory, sponsored by the National Science Foundation. Hinode is a Japanese mission developed and launched by ISAS/JAXA, with NAOJ as domestic partner and NASA and STFC (UK) as international partners. It is operated by these agencies in co-operation with ESA and NSC (Norway). 
\end{acknowledgements}

\end{document}